\documentclass[english,12pt]{article}
\usepackage[T1]{fontenc}
\usepackage[latin9]{inputenc}
\usepackage{xcolor}
\usepackage{colortbl}
\usepackage{babel}
\usepackage{array}
\usepackage{float}
\usepackage{booktabs}
\usepackage{mathtools}
\usepackage{url}
\usepackage{multirow}
\usepackage{amsmath}
\usepackage{amsthm}
\usepackage{amssymb}
\usepackage{graphicx}
\usepackage{geometry}
\usepackage{setspace}
\usepackage[authoryear]{natbib}
\usepackage[pdfusetitle,
 bookmarks=true,bookmarksnumbered=false,bookmarksopen=false,
 breaklinks=true,pdfborder={0 0 1},backref=false,colorlinks=true]
 {hyperref}
\hypersetup{
 citecolor=blue}

\makeatletter

\providecommand{\tabularnewline}{\\}
\floatstyle{ruled}
\newfloat{algorithm}{tbp}{loa}
\providecommand{\algorithmname}{Algorithm}
\floatname{algorithm}{\protect\algorithmname}

\usepackage{algorithmic}

\date{}

\ifdefined\showcaptionsetup
 \PassOptionsToPackage{caption=false}{subfig}
\fi
\usepackage{subfig}
\makeatother

\theoremstyle{plain}
\newtheorem{thm}{\protect\theoremname}[section]
\providecommand{\theoremname}{Theorem}

\begin{document}
\title{\textbf{cuRegOT: A GPU-Accelerated Solver for Entropic-Regularized
Optimal Transport}}
\author{Yixuan Qiu\\
{\normalsize School of Statistics and Data Science \& Institute of
Big Data Research}\\
{\normalsize Shanghai University of Finance and Economics}\\
{\normalsize\texttt{qiuyixuan@sufe.edu.cn}}}
\maketitle
\begin{abstract}
Optimal transport (OT) has emerged as a fundamental tool in modern
machine learning, yet its computational cost remains a significant
bottleneck for large-scale applications. While harnessing the massive
parallelism of modern GPU hardware is critical for efficiency, the
\emph{de facto} standard Sinkhorn algorithm, despite its ease of parallelization,
often suffers from slow convergence in challenging problems. More
recently, the sparse-plus-low-rank quasi-Newton method offers a balance
between convergence rate and per-iteration complexity; however, its
efficiency on GPUs is severely hindered by the serial nature of sparse
matrix symbolic analysis and irregular memory access patterns. To
bridge this gap, we present cuRegOT, a high-performance GPU solver
tailored for entropic-regularized OT. We introduce a suite of algorithmic
and architectural optimizations, including an amortized symbolic analysis
strategy to mitigate CPU bottlenecks, an asynchronous Sinkhorn iterates
generation mechanism, and a fused kernel for bandwidth-efficient gradient
evaluation. These strategies are backed by rigorous theoretical guarantees
ensuring algorithmic convergence. Extensive numerical experiments
demonstrate that cuRegOT achieves significant speedups over state-of-the-art
GPU-based solvers across a variety of benchmark tasks.
\end{abstract}

\section{Introduction}

\label{sec:introduction}

Optimal transport (OT) provides a principled way to compare probability
measures by seeking the minimum cost of coupling two distributions
under a prescribed ground cost \citep{villani2009optimal}. The resulting
Wasserstein distance endows the space of probability measures with
a rich geometry, and has found widespread success in statistical machine
learning, computer vision, natural language processing, and modern
data science \citep{peyre2019computational}.

In machine learning, OT has been widely adopted as a loss function
or regularizer for tasks that require aligning probability distributions.
Representative examples include unsupervised domain adaptation \citep{courty2017optimal},
generative modeling via Wasserstein GAN \citep{arjovsky2017wasserstein},
computer graphics \citep{solomon2015convolutional}, and single-cell
biology \citep{schiebinger2019optimal}, among many others. These
applications often require solving OT problems repeatedly, making
the computational efficiency and hardware scalability of OT solvers
a prioritized concern.

The discrete OT problem is a linear programming problem:
\begin{equation}
\min_{P\in\Pi(a,b)}\langle P,M\rangle,\label{eq:ot}
\end{equation}
where $M\in\mathbb{R}^{n\times m}$ is a given cost matrix,
\[
\Pi(a,b)=\{P\in\mathbb{R}^{n\times m}:P\mathbf{1}_{m}=a,P^{T}\mathbf{1}_{n}=b,P\ge0\},
\]
$a$ and $b$ are two probability vectors satisfying $a>0$, $b>0$,
and $\sum^{n}_{i=1}a_{i}=\sum^{m}_{j=1}b_{j}=1$, and all inequality
signs applied to vectors and matrices are elementwise.

Despite its elegance, solving (\ref{eq:ot}) at scale is challenging.
Standard linear programming solvers typically incur a super-cubic
complexity of $O(n^{3}\log(n))$ for $n\approx m$ \citep{pele2009fast}.
This computational burden becomes prohibitive when dealing with high-dimensional
data or large support sizes (e.g., $n,m>10^{4}$), necessitating the
development of more efficient approximation schemes.

To overcome the computational difficulty of solving (\ref{eq:ot}),
\citet{cuturi2013sinkhorn} proposes the entropic-regularized OT problem
as an approximation to the original OT problem, which adds an entropic
regularization to the objective function of (\ref{eq:ot}):
\begin{equation}
\min_{P\in\Pi(a,b)}\langle P,M\rangle-\eta\cdot h(P),\label{eq:entropic}
\end{equation}
where $h(P)=\sum_{i,j}P_{ij}(1-\log(P_{ij}))$ is the entropy term.

At first glance, problem (\ref{eq:entropic}) is no simpler than (\ref{eq:ot}),
but its benefit is clear by studying the dual problem of (\ref{eq:entropic}):
\begin{equation}
\max_{\alpha\in\mathbb{R}^{n},\beta\in\mathbb{R}^{m}}\ \mathcal{L}(\alpha,\beta),\label{eq:entropic_dual}
\end{equation}
where
\[
\mathcal{L}(\alpha,\beta)=-\eta\sum^{n}_{i=1}\sum^{m}_{j=1}\exp\{\eta^{-1}(\alpha_{i}+\beta_{j}-M_{ij})\}+\alpha^{T}a+\beta^{T}b.
\]
Clearly, (\ref{eq:entropic_dual}) is a concave, smooth, and unconstrained
maximization problem, which allows for many first- and second-order
optimization techniques. Let $T^{*}$ and $(\alpha^{*},\beta^{*})$
be the primal and dual optima of the problems (\ref{eq:entropic})
and (\ref{eq:entropic_dual}), respectively. Then they are connected
by the relation $T^{*}_{ij}=\exp\{(\alpha^{*}_{i}+\beta^{*}_{j}-M_{ij})/\eta\}$.

\citet{cuturi2013sinkhorn} uses the well-known Sinkhorn algorithm
\citep{yule1912methods,sinkhorn1964relationship} to solve (\ref{eq:entropic}),
which is equivalent to applying the block coordinate ascent method
to the dual problem (\ref{eq:entropic_dual}). One major advantage
of the Sinkhorn algorithm is that it can be highly parallelized, thus
suitable for GPU implementation. This has led to the development
of several widely-used GPU-based OT solvers. For instance, POT (Python
optimal transport, \citealp{flamary2021pot}) can use the CuPy package
\citep{cupy_learningsys2017} as a backend to support GPU computing,
while OTT-JAX \citep{cuturi2022optimal} leverages the JAX framework
to offer just-in-time compilation on accelerators.

However, empirical results show that the Sinkhorn algorithm may demonstrate
slow convergence on challenging problems. More recently, there is
an increasing interest in applying second-order or quasi-Newton method
to solving (\ref{eq:entropic_dual}). As a preliminary, note that
$(\alpha,\beta)$ has one redundant degree of freedom, as $\mathcal{L}(\alpha,\beta)\equiv\mathcal{L}(\alpha+c\mathbf{1}_{n},\beta-c\mathbf{1}_{m})$
for all $c\in\mathbb{R}$. Therefore, by fixing $\beta_{m}=0$ and
defining the free variable $x=(\alpha,\beta_{-m})$, where $\beta_{-m}=(\beta_{1},\ldots,\beta_{m-1})^{T}$,
our goal reduces to the smooth convex optimization problem
\begin{equation}
\min_{x\in\mathbb{R}^{n+m-1}}f(x),\quad f(x)=-\mathcal{L}(\alpha,\beta).\label{eq:dual_minimization}
\end{equation}

Several advanced Newton-type methods have been proposed to solve
(\ref{eq:dual_minimization}). For instance, the SNS \citep{tang2024accelerating}
and SSNS \citep{tang2024safe} algorithms use second-order search
directions to accelerate convergence, and exploit the approximate
sparsity in the Hessian matrices to control the per-iteration cost.
Additionally, the sparse-plus-low-rank algorithm (SPLR, \citealp{wang2025sparse})
further incorporates the specific structure of the Hessian matrices
in entropic-regularized OT, and develops a quasi-Newton method that
improves robustness when the transport plan is dense.

While these Newton-type methods demonstrate promising convergence
speeds compared to the Sinkhorn algorithm, their practical efficiency
on GPUs remains an open challenge. The primary bottleneck lies in
solving large sparse linear systems required for Newton steps. Unlike
the dense matrix-vector operations in Sinkhorn, sparse factorizations
typically rely on symbolic analysis and reordering steps that are
inherently sequential and difficult to parallelize on SIMT (single
instruction, multiple threads) architectures. Consequently, existing
implementations of these advanced algorithms are often confined to
CPUs or suffer from severe GPU under-utilization.

\begin{figure}[h]
\begin{centering}
\includegraphics[width=0.99\textwidth]{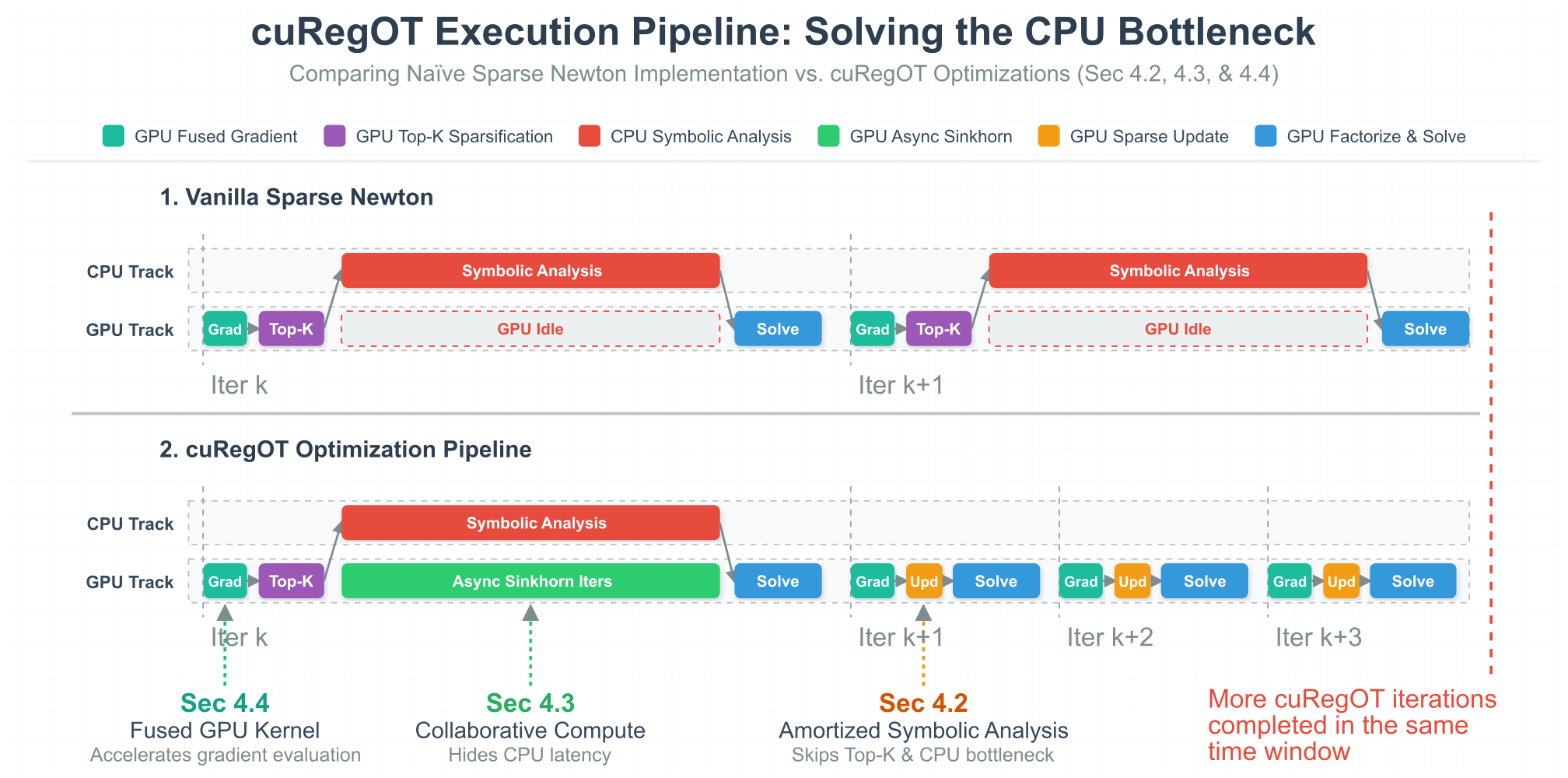}
\par\end{centering}
\caption{\label{fig:pipeline}An overview of the cuRegOT pipeline.}
\end{figure}

To address this gap, this article presents the cuRegOT library (CUDA-accelerated
regularized optimal transport), a high-performance GPU solver tailored
for entropic-regularized OT. cuRegOT adds important improvements to
the SPLR algorithm, and is programmed on the CUDA parallel computing
platform. Specifically, we focus on GPU-oriented algorithm and system
designs to mitigate the bottlenecks in solving sparse Newton-type
linear systems, including: (1) amortizing symbolic analysis by reusing
sparsity patterns across multiple iterations; (2) overlapping CPU-side
symbolic analysis with auxiliary GPU-side candidate iterate updates
to better utilize accelerator resources; and (3) developing efficient
fused CUDA kernels for gradient evaluation. A preview of the cuRegOT
execution pipeline is illustrated in Figure \ref{fig:pipeline}, with
details elaborated in subsequent sections. The source code of cuRegOT
is available at \url{https://github.com/yixuan/regot-cuda}.

\section{Related Work}

\paragraph{Entropic-regularized OT}

Entropic regularization turns discrete OT into a strictly convex problem
whose dual is smooth and unconstrained, enabling scalable iterative
solvers \citep{cuturi2013sinkhorn,peyre2019computational}. The dominant
approach is Sinkhorn matrix scaling \citep{sinkhorn1964relationship},
together with stabilization and $\varepsilon$-scaling heuristics
for small regularization \citep{schmitzer2019stabilized}. Recent
works revisit complexity and acceleration of Sinkhorn-type schemes,
e.g., greedy coordinate variants and improved analyses \citep{altschuler2017nearlinear,lin2022efficiency}.
Despite these progresses, widely-used GPU solvers for entropic OT
still largely center on Sinkhorn-style iterations.

\paragraph{GPU-based OT solvers}

Because Sinkhorn iterations are dominated by dense kernel evaluations
and elementwise operations, most practical GPU OT solvers focus on
Sinkhorn and its variants, including POT \citep{flamary2021pot} and
OTT-JAX \citep{cuturi2022optimal}. Beyond entropic-regularized OT,
PDOT \citep{lu2024pdot} provides a GPU solver aimed at high-accuracy
solutions for the \emph{unregularized} OT linear program. In addition,
Douglas--Rachford splitting has been adapted to deliver an efficient
GPU implementation for a range of \emph{non-entropic} regularizers
\citep{lindback2023bringing}, explicitly contrasting with entropic
regularization.

\paragraph{Quasi-Newton methods}

Quasi-Newton methods approximate curvature using gradient information
and are widely used in large-scale optimization due to their strong
practical performance \citep{nocedal2006numerical}. In entropic-regularized
OT, recent Newton-type solvers exploit the structure in the dual curvature,
notably through Hessian sparsification and low-rank approximation
\citep{tang2024accelerating,tang2024safe,wang2025sparse,ouyang2026sparsification}.
However, realizing these methods efficiently on GPUs is challenging
because each outer step typically requires solving large sparse linear
systems. This motivates our GPU-oriented algorithm and system designs
that attempt to overcome this difficulty.

\section{Motivation and Preliminaries}

\subsection{The SPLR Algorithm}

Our proposed solver, cuRegOT, is based on the SPLR algorithm for entropic-regularized
OT, with special designs tailored for modern GPU hardware. In this
section we provide a brief and self-contained introduction to the
SPLR algorithm. Recall that our goal is to solve the smooth convex
optimization problem (\ref{eq:dual_minimization}), and SPLR can be
viewed as a quasi-Newton method, in the sense that it uses both the
gradient $\nabla f(x)$ and an approximation to the Hessian $\nabla^{2}f(x)$
to generate a sequence of iterates $\{x_{k}\}$.

Let $T=T(x)$ be an $n\times m$ matrix with elements $T_{ij}=\exp\{(\alpha_{i}+\beta_{j}-M_{ij})/\eta\}$,
given the iterate variable $x=(\alpha,\beta_{-m})$. Then it can be
shown that the gradient function and the Hessian matrix have closed-form
expressions:
\begin{align*}
\nabla f(x) & =\begin{bmatrix}T\mathbf{1}_{m}-a\\
T^{T}_{-m}\mathbf{1}_{n}-b_{-m}
\end{bmatrix},\\
\nabla^{2}f(x) & =\eta^{-1}\begin{bmatrix}\mathbf{diag}(T\mathbf{1}_{m}) & T_{-m}\\
T^{T}_{-m} & \mathbf{diag}(T^{T}_{-m}\mathbf{1}_{n})
\end{bmatrix},
\end{align*}
where $T_{-m}$ means removing the $m$-th column from $T$.

The SPLR algorithm generates iterates using the update rule $x^{+}=x-\gamma B^{-1}g$,
where $x$ and $g=\nabla f(x)$ represent the current iterate and
gradient, respectively, $x^{+}$ is the next iterate, $\gamma$ is
a step size that may vary with the iteration, and the matrix $B$
is an approximation to the current Hessian matrix, $B\approx H=\nabla^{2}f(x)$.
$B$ is chosen such that it contains useful information about $H$
and meanwhile admits faster computation of $B^{-1}g$ than $H^{-1}g$.
Unlike the well-known BFGS method that constructs $B$ using purely
diagonal and low-rank matrices, the SPLR algorithm defines $B$ as
\begin{equation}
B=H_{\Omega}+(\xi uu^{T}+\zeta vv^{T})+\tau I,\label{eq:approx_hessian}
\end{equation}
where $H_{\Omega}$ is a sparse matrix, $(\xi uu^{T}+\zeta vv^{T})$
is a rank-two matrix, and $\tau I$ is diagonal. The rank-two term
inherits the structure of the BFGS rule, and is determined by the
following procedure. Let $x^{-}$ and $g^{-}=\nabla f(x^{-})$ be
the previous iterate and gradient, respectively, and define $y^{-}=g-g^{-}$
and $s^{-}=x-x^{-}$. Then we set
\begin{equation}
\begin{aligned}u & =y^{-}, & v & =(H_{\Omega}+\tau I)s^{-},\\
\xi & =\frac{1}{(y^{-})^{T}s^{-}}, & \zeta & =-\frac{1}{v^{T}s^{-}}.
\end{aligned}
\label{eq:low_rank}
\end{equation}
The diagonal term can be simply set to $\tau=\min\{\tau_{\max},\Vert g\Vert\}$
for some fixed constant $\tau_{\max}>0$.

The sparse term $H_{\Omega}$ is the core component of the SPLR algorithm,
which has the following structure:
\begin{equation}
H_{\Omega}=\eta^{-1}\begin{bmatrix}\mathbf{diag}(T\mathbf{1}_{m}) & T^{\Omega}_{-m}\\
(T^{\Omega}_{-m})^{T} & \mathbf{diag}(T^{T}_{-m}\mathbf{1}_{n})
\end{bmatrix},\label{eq:sparse_term}
\end{equation}
where $T^{\Omega}_{-m}$ is an $n\times(m-1)$ matrix with entries
\[
(T^{\Omega}_{-m})_{ij}=\begin{cases}
T_{ij}, & (i,j)\in\Omega,\\
0, & (i,j)\notin\Omega,
\end{cases}
\]
and $\Omega$ is a subset of the matrix coordinates, $\Omega\subseteq\{(i,j):1\le i\le n,1\le j\le m-1\}$.
Since $H_{\Omega}$ is primarily determined by the index set $\Omega$,
we call $\Omega$ a \emph{sparsification scheme} for clarity. For
theoretical analysis, we assume that $\Omega$ contains a minimum
set of indices: $\Omega\supseteq\Omega^{*}=\{(i,j):i=1\text{ or }j=1,1\le i\le n,1\le j\le m-1\}$.

Finally, the step size $\gamma>0$ is typically determined by a line
search procedure \citep{more1994line} that guarantees the Wolfe conditions:
\begin{equation}
\begin{aligned}f(x^{+}) & \le f(x)+c_{1}\gamma g^{T}d,\\{}
[\nabla f(x^{+})]^{T}d & \ge c_{2}g^{T}d,
\end{aligned}
\label{eq:wolfe}
\end{equation}
where $d=-B^{-1}g$, and $0<c_{1}<1/2$ and $c_{1}<c_{2}<1$ are pre-specified
constants. The overall structure of the SPLR algorithm, after omitting
some minor details such as the iteration loop and the convergence
test, is summarized in Algorithm \ref{alg:splr}. In our implementation
of cuRegOT, the sparsification scheme $\Omega$ is based on a simple
top-$k$ rule: we select the largest $k$ entries of $T$, forming
$T^{\Omega}$ by zeroing out all other entries. Then $H_{\Omega}$
is assembled using formula (\ref{eq:sparse_term}). Readers are referred
to \citet{wang2025sparse} for more details of choosing $\Omega$
in each iteration.

\begin{algorithm}[h]
\caption{\label{alg:splr}Overview of the SPLR algorithm for solving entropic-regularized
OT.}


\begin{algorithmic}[1]

\REQUIRE Previous iterate and gradient $(x^{-},g^{-})$, current
iterate $x$, constant $\tau_{\max}>0$

\ENSURE Next iterate $x^{+}$

\STATE Compute $T=T(x)$, $g=\nabla f(x)$, $H=\nabla^{2}f(x)$

\STATE Determine $\Omega$ and compute $H_{\Omega}$

\STATE Compute $s^{-}=x-x^{-}$ and $y^{-}=g-g^{-}$

\STATE Compute $\xi,\zeta,u,v$ according to (\ref{eq:low_rank})

\STATE Let $R=\begin{cases}
\xi uu^{T}+\zeta vv^{T}, & \text{if }(y^{-})^{T}s^{-}{>}10^{-6}\Vert y^{-}\Vert^{2}\\
O, & \text{otherwise}
\end{cases}$

\STATE Set $\tau=\min\{\tau_{\max},\Vert g\Vert\}$

\STATE Compute $d=-B^{-1}g$, where $B=H_{\Omega}+R+\tau I$

\RETURN $x^{+}=x+\gamma d$ with $\gamma$ selected by line search

\end{algorithmic}

\end{algorithm}

\subsection{Challenges on GPU Implementation}

\label{subsec:challenges}

The SPLR algorithm has achieved promising computational efficiency
on the CPU-based implementation, compared with well-known reference
algorithms including the Sinkhorn algorithm. However, it is non-trivial
to adapt it to GPU due to its major computational bottleneck on computing
the quasi-Newton search direction. Specifically, one of the most critical
part of the algorithm is computing the sparse Cholesky decomposition
of the sparsified Hessian matrix $H_{\Omega}$, which mainly consists
of three steps:
\begin{enumerate}
\item \textbf{Symbolic analysis}: analyzing the sparsity pattern of $H_{\Omega}$,
and determining the reordering method for $H_{\Omega}$. Reordering
the rows and columns of $H_{\Omega}$ is critical to achieve a sparse
decomposition result, in the sense that $L$ may be sparser than $L_{0}$
with $P^{T}H_{\Omega}P=LL^{T}$ and $H_{\Omega}=L_{0}L^{T}_{0}$,
for some permutation matrix $P$.
\item \textbf{Factorization}: computing the sparse lower-triangular $L$
matrix given $H_{\Omega}$ and the reordering method $P$.
\item \textbf{Solving}: obtaining $H^{-1}_{\Omega}g=P(L^{T})^{-1}L^{-1}P^{T}g$
by computing the forward substitution $L^{-1}u$ and the backward
substitution $(L^{T})^{-1}v$ for some vectors $u$ and $v$.
\end{enumerate}
For most existing GPU-based sparse Cholesky solvers, the factorization
and solving steps can well support parallel computing, whereas the
symbolic analysis part primarily runs on CPU, and typically takes
up a significant amount of computing time. This fact becomes the major
obstacle to adapting the SPLR algorithm to efficient GPU computing.

\section{Method: The cuRegOT Solver}

\label{sec:method}

\subsection{Overview}

To overcome the issues introduced in Section \ref{subsec:challenges}
and to facilitate efficient OT solvers, we propose three algorithm
and system designs that are tailored for modern GPU. Below we provide
an overview of the methods we develop, with details of these designs
elaborated in subsequent sections:
\begin{itemize}
\item We show with theoretical guarantees that the CPU computing bottleneck
can be amortized, in the sense that the symbolic analysis result can
be reused by multiple iterations. On average, the cost resulted from
pure CPU computing can be reduced by a large factor.
\item We design an asynchronous algorithm that allow GPU to compute useful
information when CPU is working on the symbolic analysis. This additional
information can potentially accelerate the convergence of the algorithm,
thus reducing the total runtime.
\item We design a fused CUDA kernel for efficient gradient evaluation, aiming
at reducing overhead and memory IO.
\end{itemize}

\subsection{Amortized Symbolic Analysis of Sparsity Pattern}

\label{subsec:amortized}

The standard SPLR algorithm re-evaluates the sparsity pattern of $H_{\Omega}$
at every iteration, which depends on the value of the current iterate
$x$. However, performing symbolic analysis and matrix reordering
(e.g., via nested dissection) to minimize fill-in during sparse Cholesky
decomposition is a highly serial process, typically executed on the
CPU. This creates a significant synchronization bottleneck, leaving
the GPU idle between numerical steps.

We observe that the sparsity pattern, typically determined by the
top-$k$ entries in $T$, evolves slowly across iterations (an illustration
of this phenomenon is given in Appendix \ref{subsec:evolution_plan}).
To this end, we introduce an amortized symbolic analysis strategy.
We perform the expensive CPU-based symbolic analysis only once every
$S$ iterations (e.g., $S=10$). For the subsequent $S-1$ iterations,
we reuse the established symbolic pattern structure, updating only
the numerical values of the non-zero elements. This dramatically reduces
the CPU overhead and improves the overall GPU duty cycle.

This method is simple and easy to implement, but we shall emphasize
that this is not a trivial engineering modification. Clearly, reusing
the sparsity pattern will change the quasi-Newton search direction
in each iteration, and we must show that it does not compromise the
convergence properties of the algorithm. A formal proof is presented
in Section \ref{subsec:convergence}, and here we provide some insights.
The validity of this modification is grounded on Theorem \ref{thm:eigenvalues},
which is derived from Corollary 3.4 of \citet{wang2025sparse}:
\begin{thm}[Rephrased Corollary 3.4 of \citealp{wang2025sparse}]
 \label{thm:eigenvalues}For any sparsification scheme $\Omega\supseteq\Omega^{*}$,
the sparsified Hessian matrix $H_{\Omega}$ satisfies
\[
0<\lambda_{\min}(H)\le\lambda_{\min}(H_{\Omega})\le\lambda_{\max}(H_{\Omega})\le\lambda_{\max}(H),
\]
where $\lambda_{\min}(A)$ and $\lambda_{\max}(A)$ stand for the
smallest and largest eigenvalues of a matrix $A$. 
\end{thm}

In other words, any sparsification made to the Hessian matrix $H$
does not expand the range of its eigenvalues, so no matter we reuse
the sparsity pattern or not, the condition number of $H_{\Omega}$
is always bounded by $\lambda_{\max}(H)/\lambda_{\min}(H)$. In theoretical
analysis, the condition number of the $B$ matrix is crucial to the
convergence of quasi-Newton methods. Since $B$ is strongly connected
to $H_{\Omega}$, a control of $\lambda_{\max}(B)/\lambda_{\min}(B)$
can also be anticipated. In summary, the amortized symbolic analysis
provides significant computational benefits, and meanwhile is a safe
algorithmic change that possesses a solid theoretical guarantee.

\subsection{Collaborative CPU-GPU Computing for Candidate Iterate Generation}

\label{subsec:async}

\begin{figure}[h]
\begin{centering}
\includegraphics[width=0.8\textwidth]{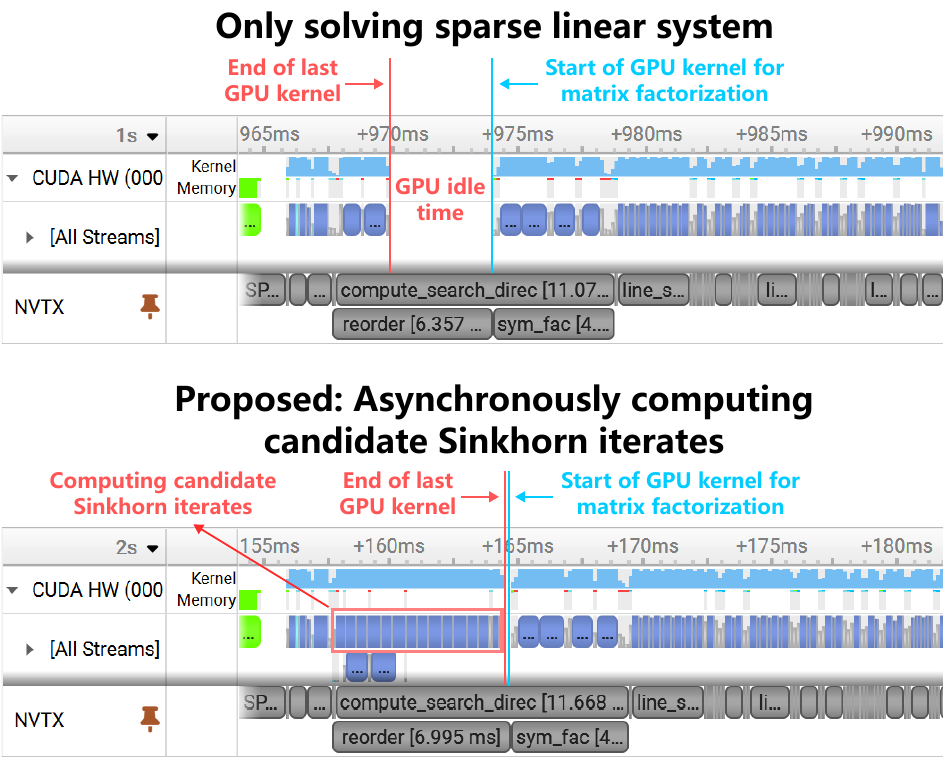}
\par\end{centering}
\caption{\label{fig:kernel_occupancy}Visualization of GPU kernel occupancy.}
\end{figure}

Even with amortization, the GPU must wait for the CPU during the
symbolic analysis phase every $S$-th iteration, and there is still
massive computing time spent on the CPU-only part. During this period,
GPU is almost idle, which potentially results in a waste of computing
power. The upper part of Figure \ref{fig:kernel_occupancy} visualizes
this phenomenon by profiling on the compiled code of a vanilla version
of the proposed solver that implements the method in Section \ref{subsec:amortized}.
In most of the time, the occupacy of GPU kernels keeps staying at
a high level. However, when the sparsity pattern is re-computed, there
is a visible gap between kernel function calls, indicating that only
CPU is working during this time.

To make the best use of this residual idle time, we propose a collaborative
CPU-GPU computing mechanism. The core idea is to generate candidate
iterates on GPU during the symbolic analysis part that merely utilizes
CPU. Recall that the Sinkhorn algorithm solves the same dual problem
(\ref{eq:entropic_dual}). While generally slower to converge than
Newton-type methods, its iterations are computationally inexpensive
and highly parallelizable on the GPU.

Therefore, when the host CPU begins the serial symbolic analysis task,
we propose to simultaneously trigger an asynchronous CUDA stream on
the GPU, which performs several lightweight Sinkhorn iterations on
the current dual variables, generating an alternative candidate solution
$x^{s}_{k}$. Once the CPU completes the analysis and the GPU finishes
the subsequent quasi-Newton step to generate the primary candidate
$x^{q}_{k}$, cuRegOT evaluates the objective function and gradient
norms for both candidates. It then selects the one yielding greater
improvement. This strategy effectively masks the CPU latency and provides
more explorations of the solution space, thus potentially resulting
in faster overall convergence without incurring additional wall-clock
time. The lower part of Figure \ref{fig:kernel_occupancy} shows the
profiling result on the improved algorithm. It can be seen that multiple
Sinkhorn iterations have been computed during the symbolic analysis
stage, and the runtime for this stage is almost unaffected.

We shall clarify that inserting external iteration steps into an existing
optimization algorithm does not automatically guarantee the convergence
of the modified procedure, especially when the insertion occurs possibly
infinite times. This is because the inserted steps may disrupt the
dependence structure between the original consecutive iterates, which
the convergence guarantee may rely on. In the case of the proposed
algorithm, however, we provide a rigorous theoretical guarantee in
Section \ref{subsec:convergence}, showing that adding such candidate
selection step preserves the convergence properties of the original
SPLR algorithm.

\subsection{Fused Kernel Design for Gradient Evaluation}

\label{subsec:fused_kernel}

A naive implementation of the gradient calculation involves multiple
passes over the large cost matrix $M$ and the intermediate transport
plan matrix $T$ in global memory: one pass to compute $T_{ij}$ elements,
and subsequent passes to reduce rows to compute $\nabla_{\alpha}f=T\mathbf{1}_{m}-a$
and columns to compute $\nabla_{\beta_{-m}}f=T^{T}_{-m}\mathbf{1}_{n}-b_{-m}$.
On GPUs, this is severely bound by memory bandwidth.

We propose a fused CUDA kernel that performs element-wise computation,
row reduction, and column reduction in a single pass over the data,
maximizing data reuse in the fastest levels of the memory hierarchy
(registers and shared memory). The kernel operates on the matrix
$M$ using a grid-stride loop to handle arbitrary matrix dimensions.
We deploy thread blocks of size $32\times D$, where $32$ corresponds
to the CUDA warp size, and $D$ is a tunable parameter (empirically
set to $D=8$ in our implementation). Each thread block processes
a $D\times32$ tile of the matrix.

\paragraph{Register-level computation}

The thread loads $\alpha_{i}$, $\beta_{j}$, and $M_{ij}$ from global
memory and computes $T_{ij}=\exp\{(\alpha_{i}+\beta_{j}-M_{ij})/\eta\}$.
This value is held locally in a register.

\paragraph{Warp-level row reduction}

Threads within a single warp (processing the same row $i$) cooperate
to sum their register values of $T_{ij}$. We utilize highly efficient
warp shuffle intrinsics \texttt{\_\_shfl\_down\_sync()} to perform
a parallel reduction tree directly in registers, requiring zero shared
memory usage. The thread at lane 0 of the warp writes the partial
row sum to the global memory.

\paragraph{Shared memory column reduction}

Threads within the same block column (processing the same column $j$)
sum their values, utilizing the fast shared memory (L1 cache). Each
thread atomically adds its $T_{ij}$ value to a shared memory buffer
dedicated to its column index within the block. By choosing a small
dimension $D=8$, we ensure low bank conflict pressure during these
atomic operations. Once the block finishes the grid-stride loop,
it adds the partial column sums to the global memory.

This fused kernel ensures that each $M_{ij}$ is read exactly once,
significantly alleviating the bandwidth bottleneck.

\section{Theoretical Analysis}

\label{sec:theoretical_analysis}

\subsection{Computational Complexity}

In this section we focus on analyzing the computational complexity
of the fused kernel proposed in Section \ref{subsec:fused_kernel},
and most of the other parts of the algorithm involve standard matrix
and vector operations that have well-established complexity analyses.
Suppose that there are $K$ GPU threads that can work in parallel.
In our setting, every $32D$ threads form a block of size $D\times32$
that each time processes a tile of the $T$ matrix, and we assume
that the total $K$ threads form a grid of blocks with $g_{r}$ rows
and $g_{c}$ columns. In other words, a grid consists of $g_{r}\times g_{c}$
blocks, and each block contains $D\times32$ threads. Clearly, we
have $K=32Dg_{r}g_{c}$. We summarize the computational costs for
major operations in gradient evaluation in Table \ref{tab:complexity},
with calculation details given in Appendix \ref{subsec:complexity_details}.
The memory-write terms are conservative estimates that account for
possible serialization caused by atomic additions to the same output
row/column accumulator.

\begin{table}[h]
\caption{\label{tab:complexity}Computational costs for major operations in
the fused kernel of gradient evaluation.}

\centering{}%
\begin{tabular}{>{\centering}m{0.2\textwidth}>{\centering}m{0.37\textwidth}>{\centering}m{0.25\textwidth}}
\toprule 
\textbf{Operation} & \textbf{Hardware Level} & \textbf{Running Time on $K$ Processors}\tabularnewline
\midrule
\midrule 
\centering{}$T$ matrix elements & Register & $O(nm/K)$\tabularnewline
\midrule 
\multirow{2}{0.2\textwidth}[-0.25em]{\centering{}Row sum reduction} & Register & $O(5nm/K)$\tabularnewline
\cmidrule{2-3}
 & Global memory write & $O(g_{c}nm/K)$\tabularnewline
\midrule 
\multirow{3}{0.2\textwidth}[-0.45em]{\centering{}Column sum reduction} & Register & $O(mn/K)$\tabularnewline
\cmidrule{2-3}
 & Shared memory write & $O(Dm/(32g_{c}))$\tabularnewline
\cmidrule{2-3}
 & Global memory write & $O(g_{r}m/(32g_{c}))$\tabularnewline
\bottomrule
\end{tabular}
\end{table}

\subsection{Convergence of the Algorithm}

\label{subsec:convergence}

As explained in Sections \ref{subsec:amortized} and \ref{subsec:async},
the algorithmic improvements we have made to the SPLR algorithm do
not automatically preserve the convergence properties. Therefore,
we formally establish the theoretical guarantee for the modified algorithm
via the following two theorems.
\begin{thm}
\label{thm:global_convergence}Let $\{x_{k}\}$ be a sequence of iterates
such that $x_{k}$ is generated by the SPLR algorithm (Algorithm \ref{alg:splr})
for $k\neq iS$, $i=1,2,\ldots$, where $S\ge2$ is a pre-specified
integer. For $k=iS$, $i=1,2,\ldots,$ let $x^{q}_{k}$ be the next
iterate of $x_{k-1}$ using Algorithm \ref{alg:splr}, and $x^{s}_{k}$
be an arbitrary point. We set $x_{k}=x^{s}_{k}$ if $f(x^{s}_{k})\le f(x^{q}_{k})$,
and take $x_{k}=x^{q}_{k}$ otherwise. Then we have
\[
\lim_{k\rightarrow\infty}\Vert\nabla f(x_{k})\Vert=0.
\]
\end{thm}

In other words, we have shown that the sequence $\{x_{k}\}$ is globally
convergent to an optimal point of $f(x)$. Moreover, the modified
algorithm still has a linear convergence rate.
\begin{thm}
\label{thm:linear_convergence}Under the same setting as Theorem \ref{thm:global_convergence},
the iterates $\{x_{k}\}$ at least have a linear convergence rate.
That is, there exists a constant $0<r<1$ such that
\[
f(x_{k})-f^{*}\le r\left[f(x_{k-1})-f^{*}\right]
\]
for all $k\ge1$, where $f^{*}$ is the optimal objective function
value of $f(x)$.
\end{thm}

The expression of $r$ is given in Appendix \ref{subsec:proof_rate},
and we note that it depends on the problem parameters $(M,a,b,\eta,x_{0})$.

Equipped with the two theorems, the algorithmic designs in Section
\ref{sec:method} not only provide practical engineering optimizations,
but also enjoy rigorous theoretical guarantees.

\section{Numerical Experiments}

\label{sec:experiments}

\subsection{Overview}

In this section, we evaluate the performance of different GPU-based
OT solvers on a variety of benchmark problems. For each problem instance,
we generate the cost matrix $M\in\mathbb{R}^{n\times m}$, the source
distribution vector $a\in\mathbb{R}^{n}_{+}$, and the target distribution
vector $b\in\mathbb{R}^{m}_{+}$, where $\sum^{n}_{i=1}a_{i}=\sum^{m}_{j=1}b_{j}=1$.
Following standard practice in entropic-regularized OT literature,
we normalize the cost matrix by its maximum value, \emph{i.e.}, $M\leftarrow M/\max_{i,j}M_{ij}$,
to make the regularization parameter $\eta$ comparable across test
examples. We fix $\eta=0.001$ in the main experiments, and explore
the impact of different $\eta$ values in Appendix \ref{subsec:different_eta}.

The GPU solvers included in our experiments are the POT package \citep{flamary2021pot}
with two algorithms, Sinkhorn and Greenkhorn, the OTT-JAX package
\citep{cuturi2022optimal} with the Sinkhorn algorithm and its Anderson
acceleration version, a CuPy implementation of the accelerated Sinkhorn
algorithm (AccSinkhorn) proposed in \citet{lin2022efficiency}, and
the proposed cuRegOT solver. The memory size for Anderson acceleration
is set to 5, and we fix the hyperparameter $S$ in cuRegOT to $S=10$
for the experiments. In Appendix \ref{subsec:hyperparameter}, we
additionally explore the impact of different $S$ values in the cuRegOT
solver.

For each solver, we set the convergence tolerance to zero and run
for some fixed numbers of iterations (typically 10, 20, 50, 100, etc.)
to obtain detailed convergence trajectories. This approach allows
us to compare the efficiency of different solvers without being confounded
by their different stopping criteria. For each test problem and solver
configuration, we first run the solver for a few iterations to warm
up before measuring their actual runtime, and we repeat the experiment
ten times to account for runtime variability. We report the median
runtime and median optimization error across these runs, where the
optimization error is measured by the marginal error of the computed
transport plan $T$: $\text{Error}=\Vert T\mathbf{1}_{m}-a\Vert_{1}+\Vert T^{T}\mathbf{1}_{n}-b\Vert_{1}$.
The rationale of using the marginal error to quantify optimality
in our setting is explained in Appendix \ref{subsec:duality_gap},
where we also consider the duality gap as an alternative metric. Finally,
we visualize the results by plotting the optimization error (on a
logarithmic scale) against runtime for each solver, which reveals
both the convergence speed and final accuracy of different algorithms.
Lower curves indicate better performance, achieving lower errors in
less time.

\subsection{Synthetic Data}

\label{subsec:synthetic}

We construct three types of synthetic benchmark problems to evaluate
the solver performance across different geometric structures and distribution
properties. For each type, multiple problem sizes of $(n,m)=(1600,1200),(3200,2400),(6400,4800)$
are tested to assess scalability.

\paragraph{Synthetic I: Gaussian point clouds}

We generate two samples of Gaussian distributions to create cost matrices
based on squared Euclidean distances. We consider two variants:
\begin{itemize}
\item \emph{Same distribution (iid)}: entries of both source and target
samples are independently drawn from $N(0,1)$, thus testing solvers
on symmetric, well-conditioned problems.
\item \emph{Different distributions (diff)}: source and target samples have
$N(0,1)$ and $N(1,0.25)$ entries, respectively. This setting creates
asymmetric problems that require more substantial mass transport.
\end{itemize}
For both variants, the source and target distribution vectors $a$
and $b$ are uniform. 

\paragraph{Synthetic II: Exponential to mixture Gaussian transport}

This benchmark models transport between (discretized) continuous probability
distributions. The source distribution is an exponential distribution
with mean one, admitting the density function $a(x)=e^{-x}$, while
the target distribution is a mixture of two Gaussians: $b(x)=0.2\cdot N(x;1,0.04)+0.8\cdot N(x;3,0.25)$.
We discretize both distributions on a uniform grid of points $x_{1},\ldots,x_{n}$
and $y_{1},\ldots,y_{m}$ on $[0,5]$, respectively, and compute the
squared Euclidean distance $|x_{i}-y_{j}|^{2}$ as the $(i,j)$ entry
of the cost matrix. This problem tests solvers on problems where the
OT transport plan exhibits a multi-modal structure.

\subsection{CIFAR-10 Data}

To evaluate solvers on real-world computer vision tasks, we construct
OT problems using the CIFAR-10 image dataset \citep{krizhevsky2009learning}.
CIFAR-10 contains 60000 color images of size $32\times32\times3$
across 10 classes, with 5000 training images per class. For each benchmark
instance, we select two classes from CIFAR-10 as the source and target
distributions, where each image is a discrete point in the distribution.
This creates problems with $n=m=5000$. For the cost matrix, we do
not directly use raw pixel values, but instead extract deep features
using a pretrained ResNet-18 model \citep{he2016deep}. The preprocessing
details are described in Appendix \ref{subsec:cifar10_preprocess}.
For each pair of source and target images, we compute the squared
Euclidean distance between their feature vectors to construct the
cost matrix $M\in\mathbb{R}^{5000\times5000}$, and both marginal
distributions $a$ and $b$ are set to uniform.

This benchmark tests solvers on large-scale, real-world problems where
the cost structure reflects high-level semantic relationships rather
than simple geometric distances.

\begin{figure}[h]
\centering
\subfloat[Synthetic data I with i.i.d. source and target distributions.]{\begin{centering}
\includegraphics[width=0.33\textwidth]{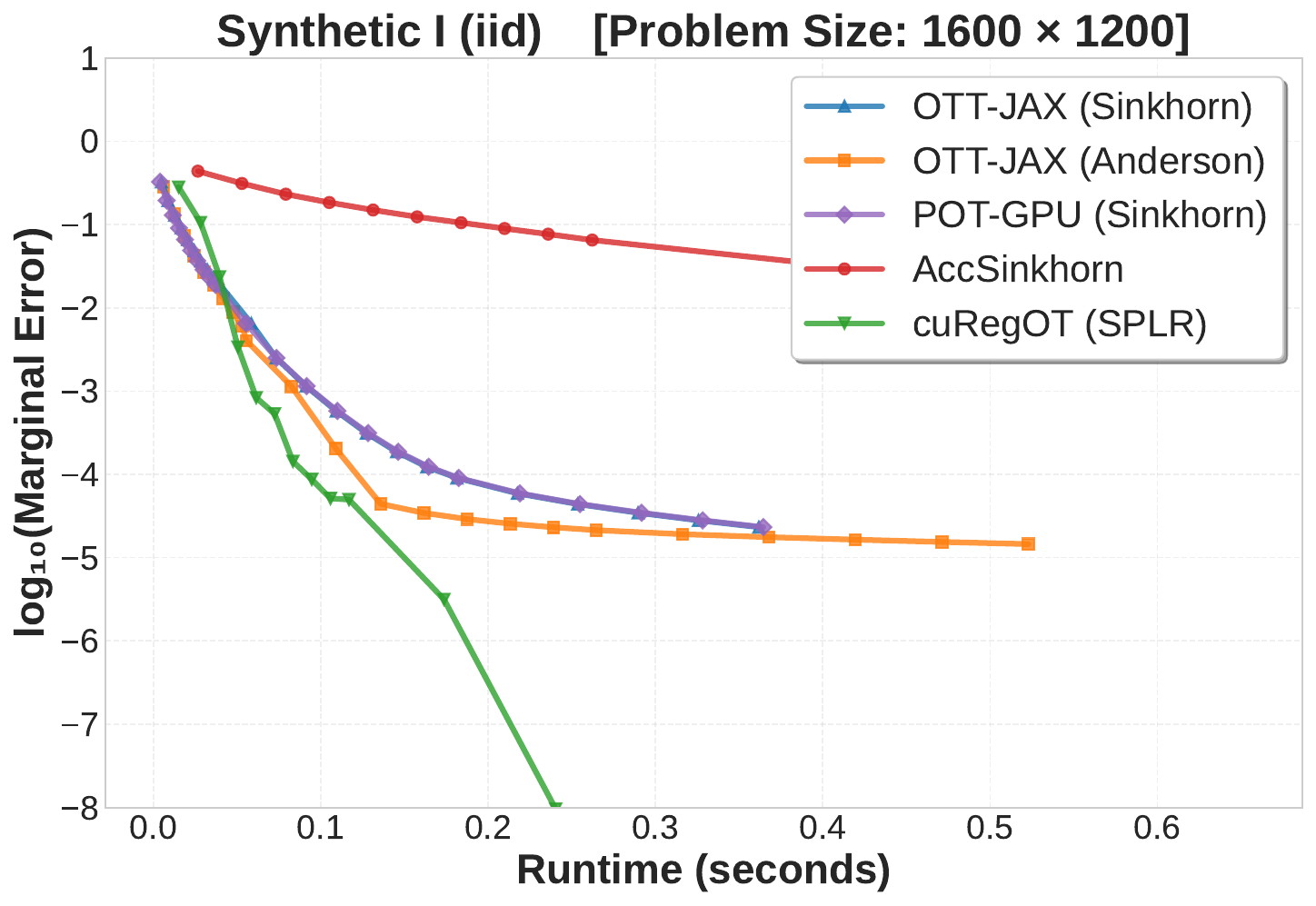}
\includegraphics[width=0.33\textwidth]{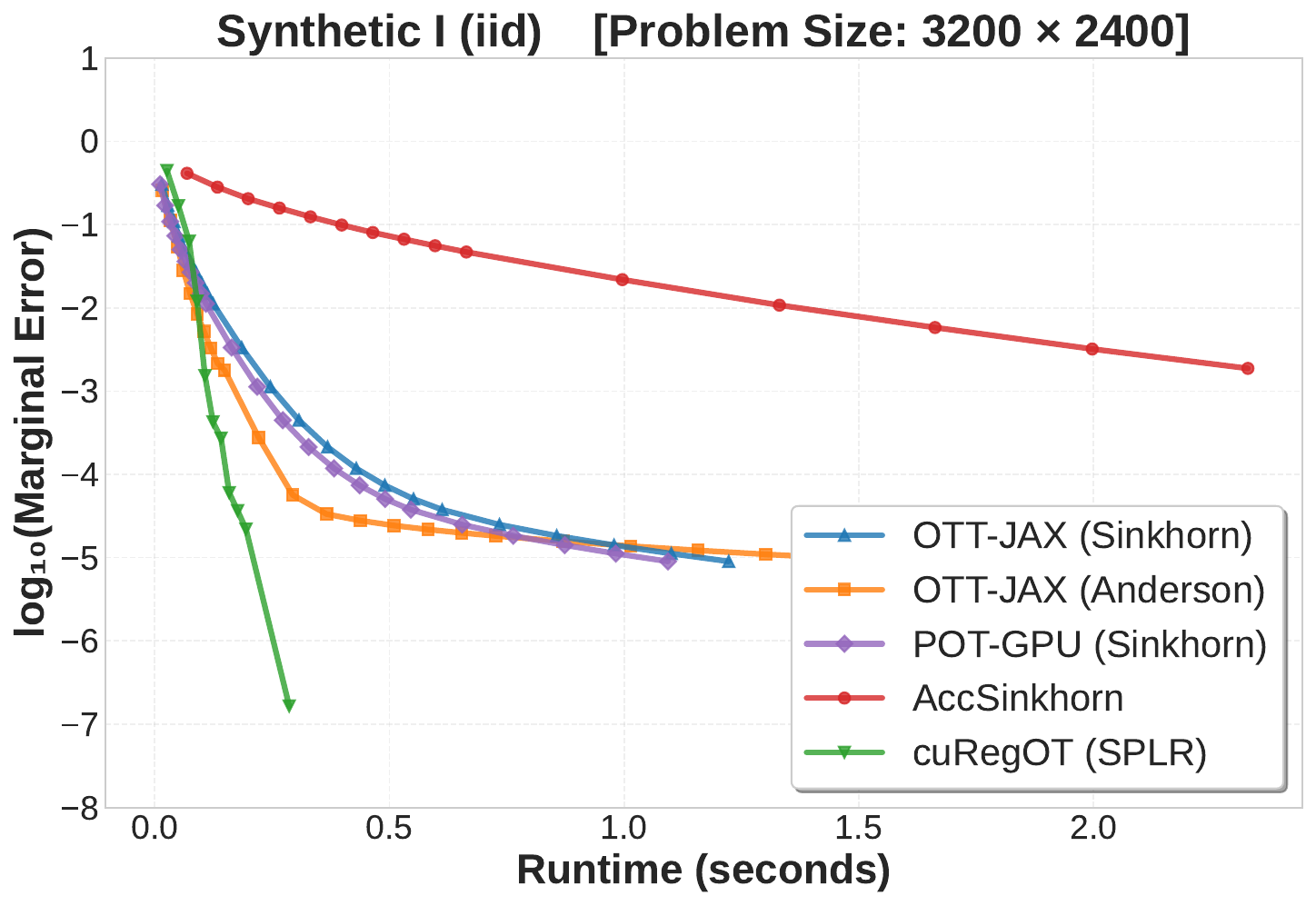}
\includegraphics[width=0.33\textwidth]{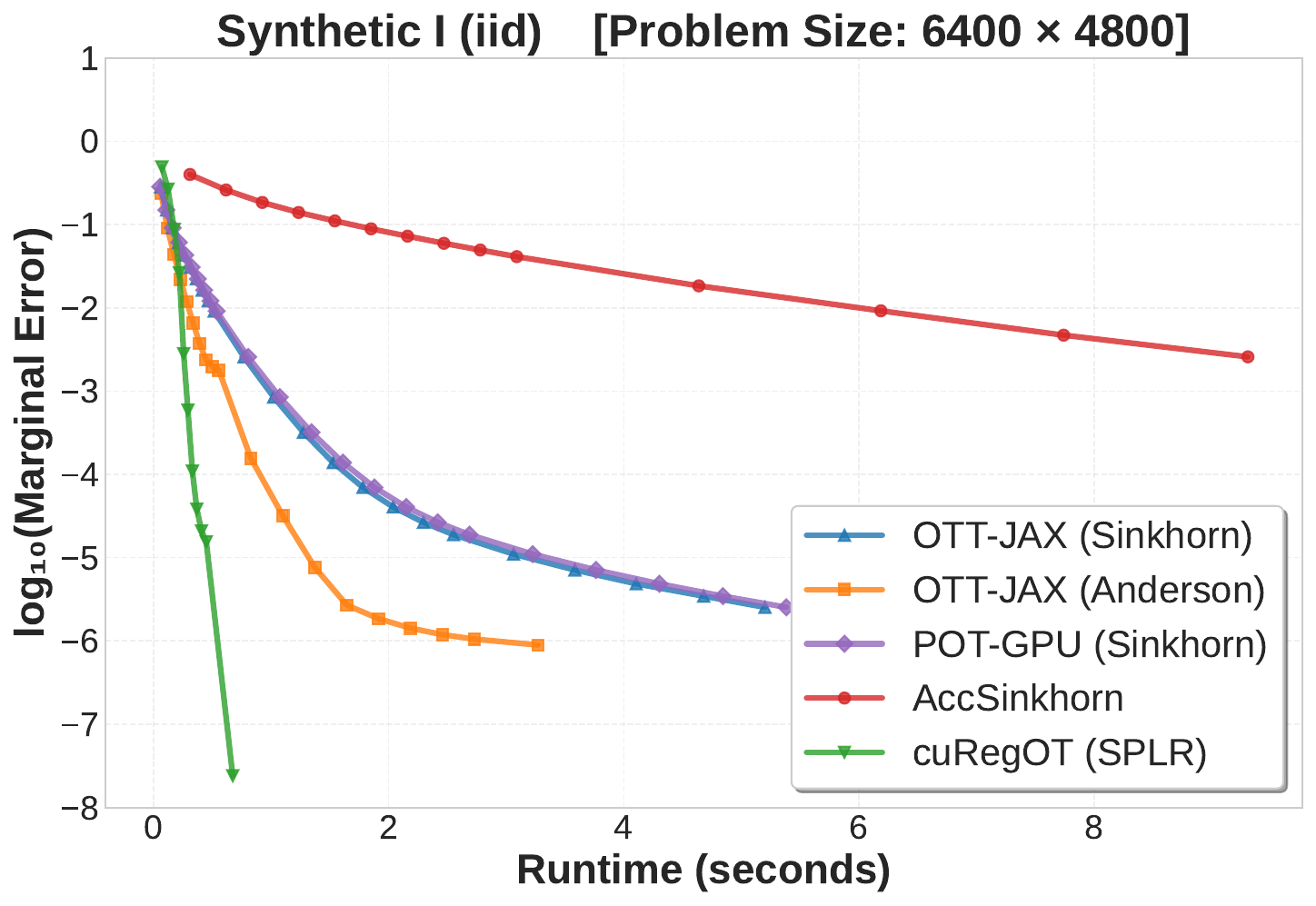}
\par\end{centering}
}

\subfloat[Synthetic data I with difference source and target distributions.]{\begin{centering}
\includegraphics[width=0.33\textwidth]{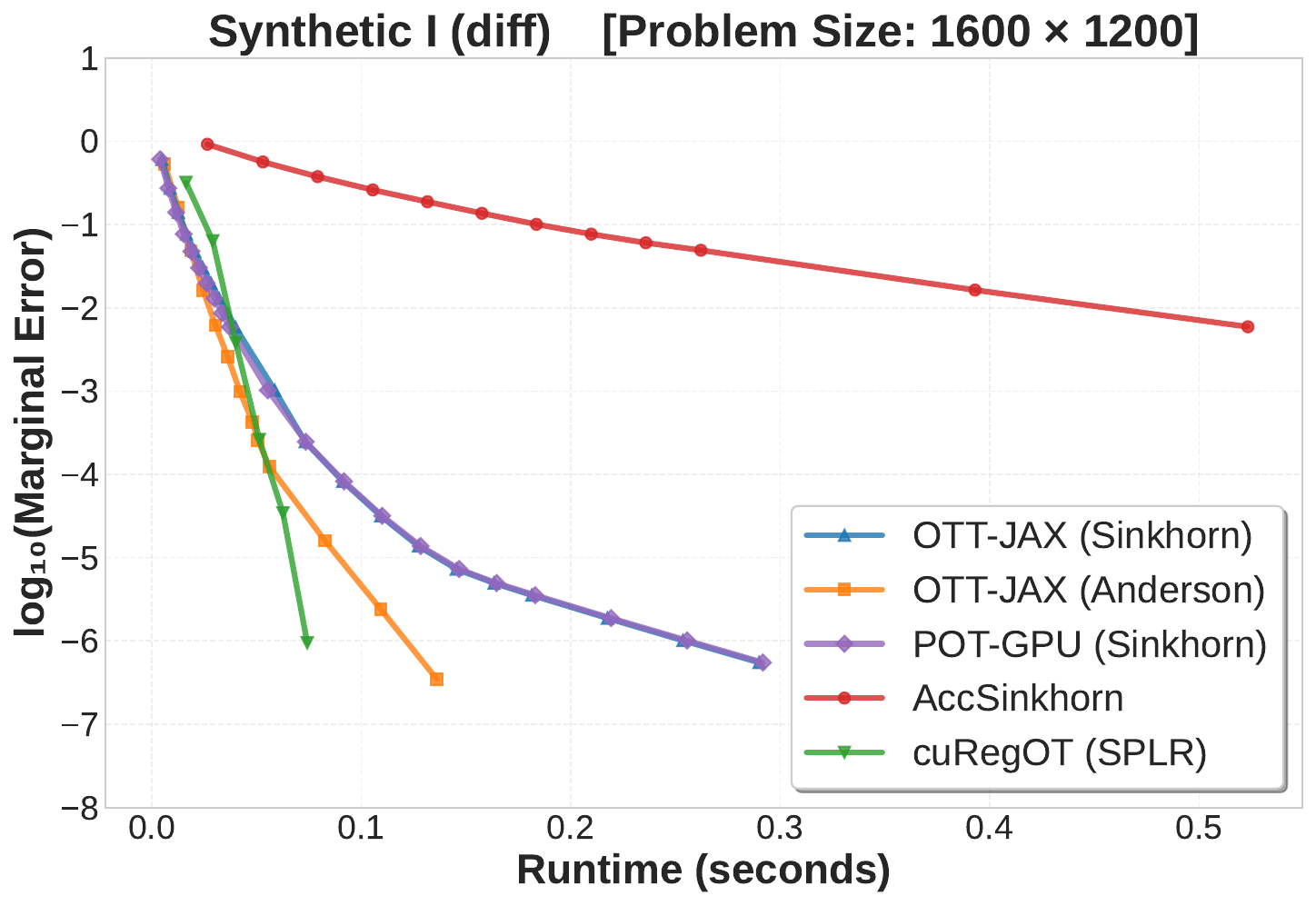}
\includegraphics[width=0.33\textwidth]{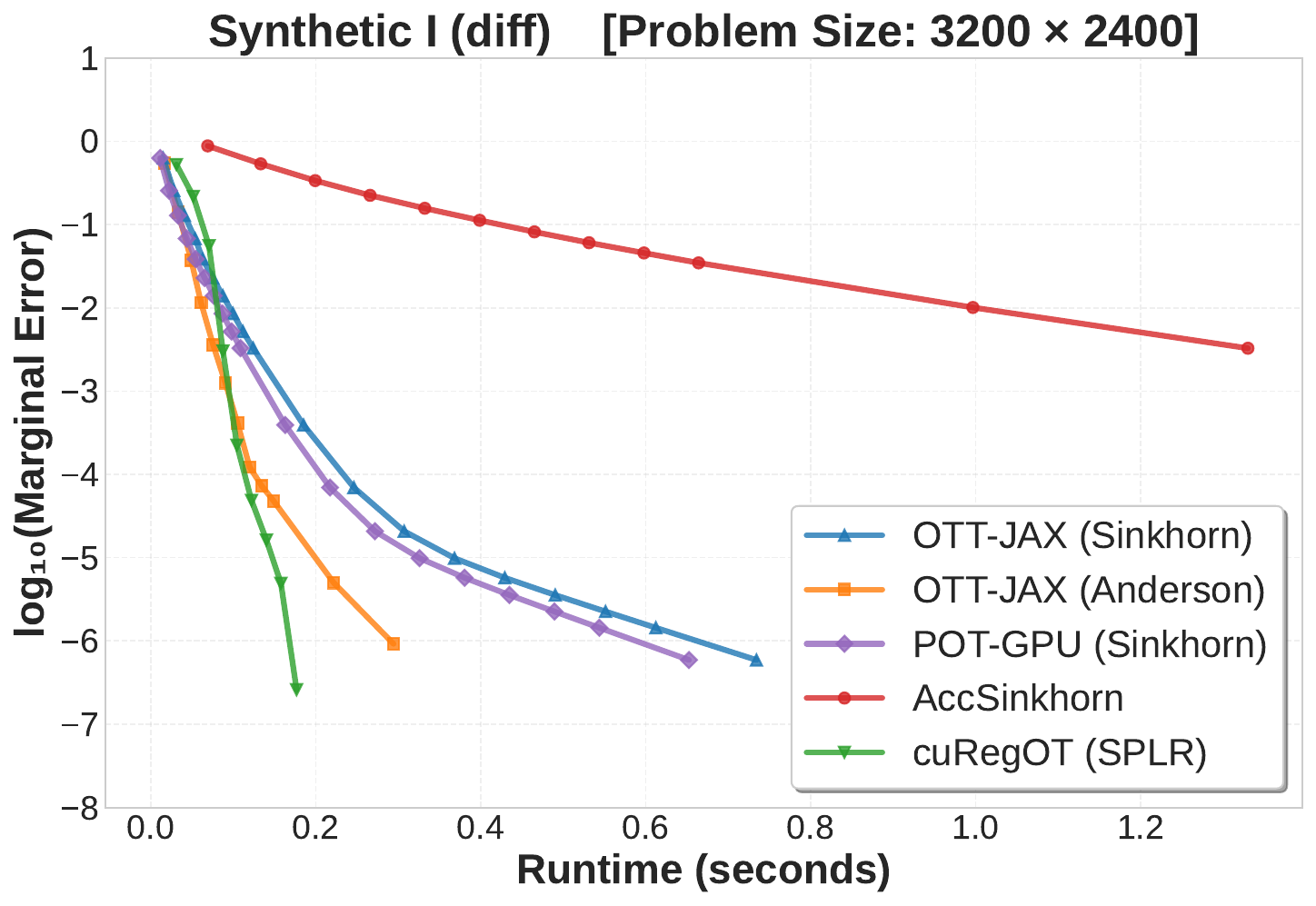}
\includegraphics[width=0.33\textwidth]{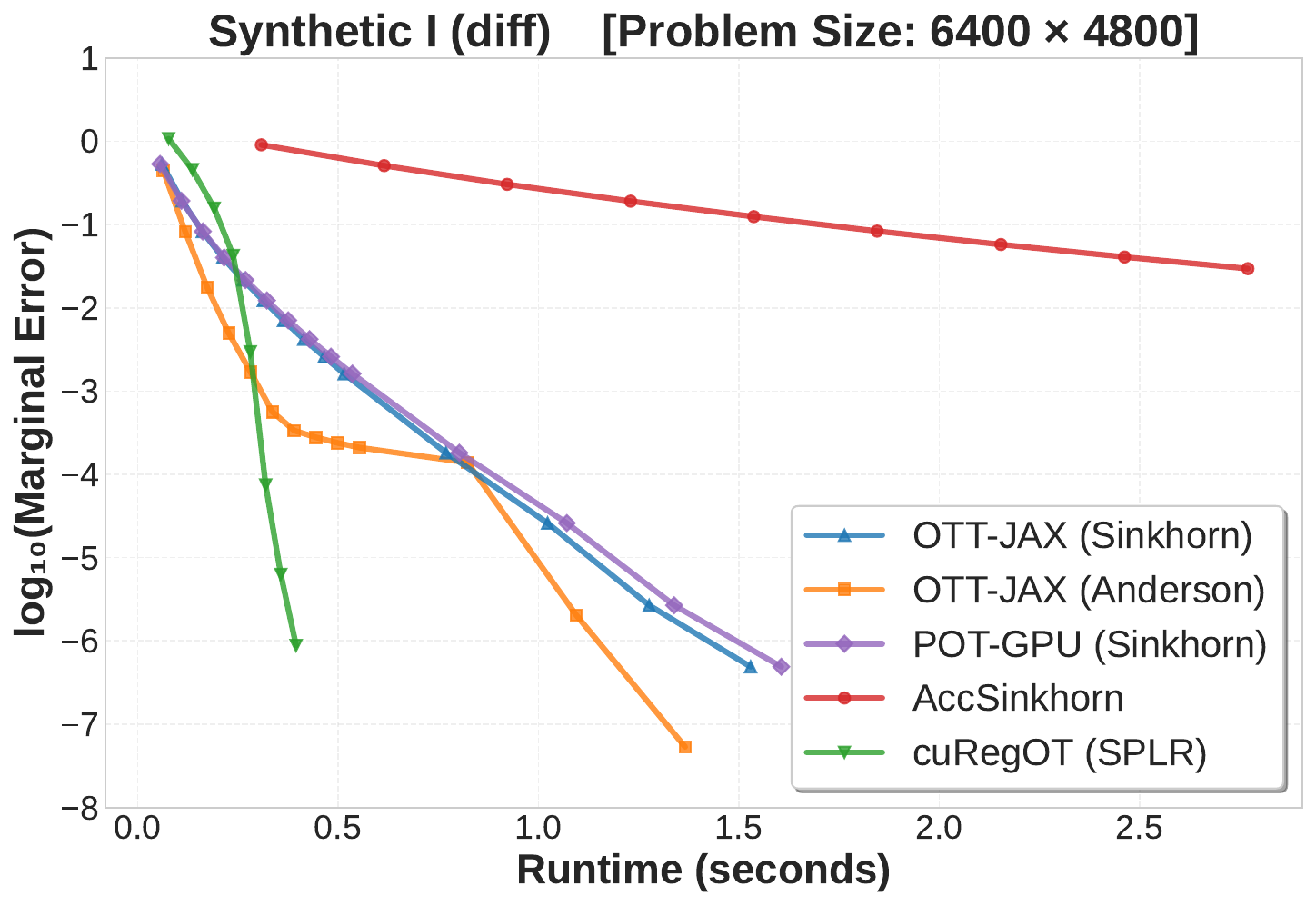}
\par\end{centering}
}

\subfloat[Synthetic data II.]{\begin{centering}
\includegraphics[width=0.33\textwidth]{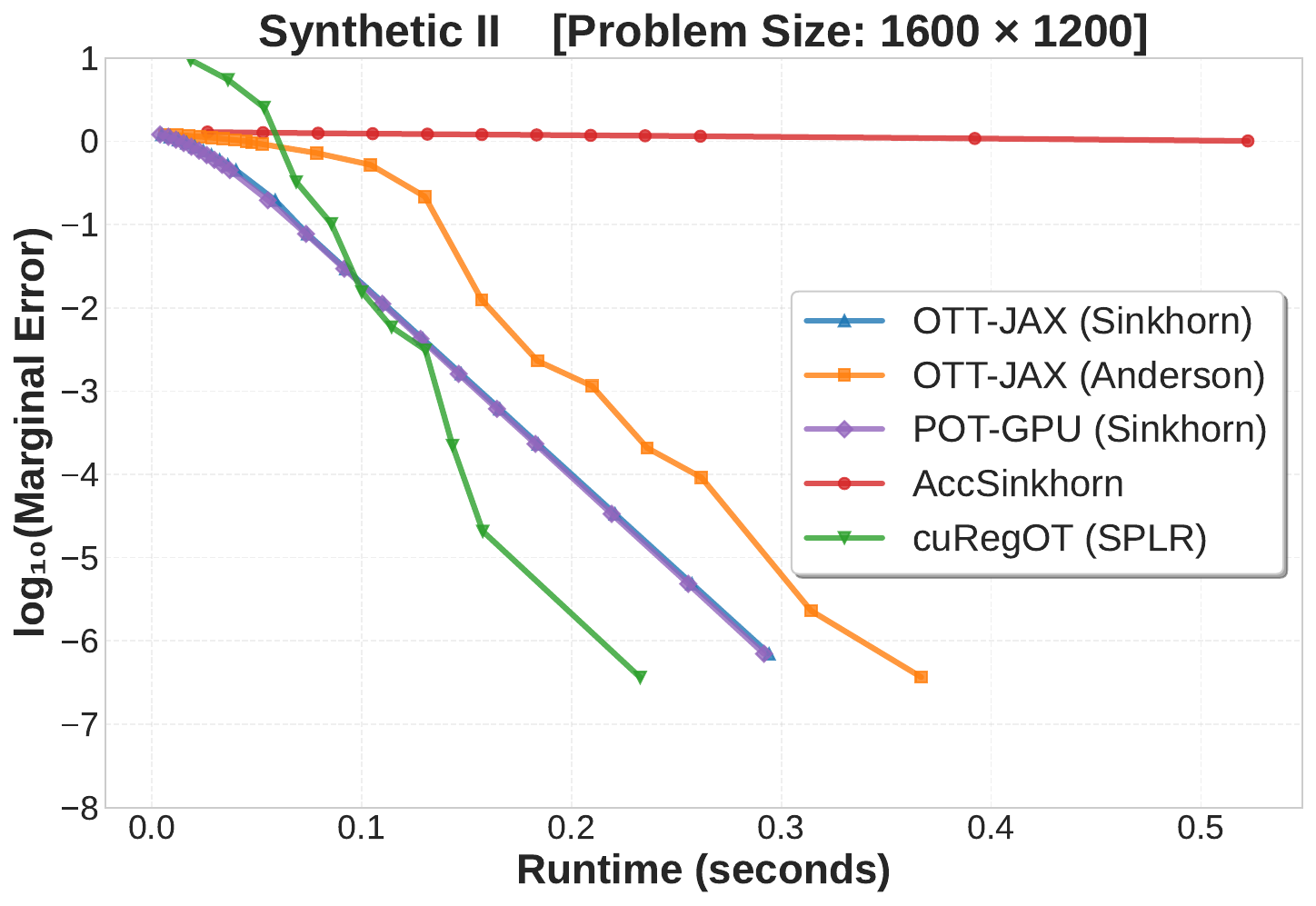}
\includegraphics[width=0.33\textwidth]{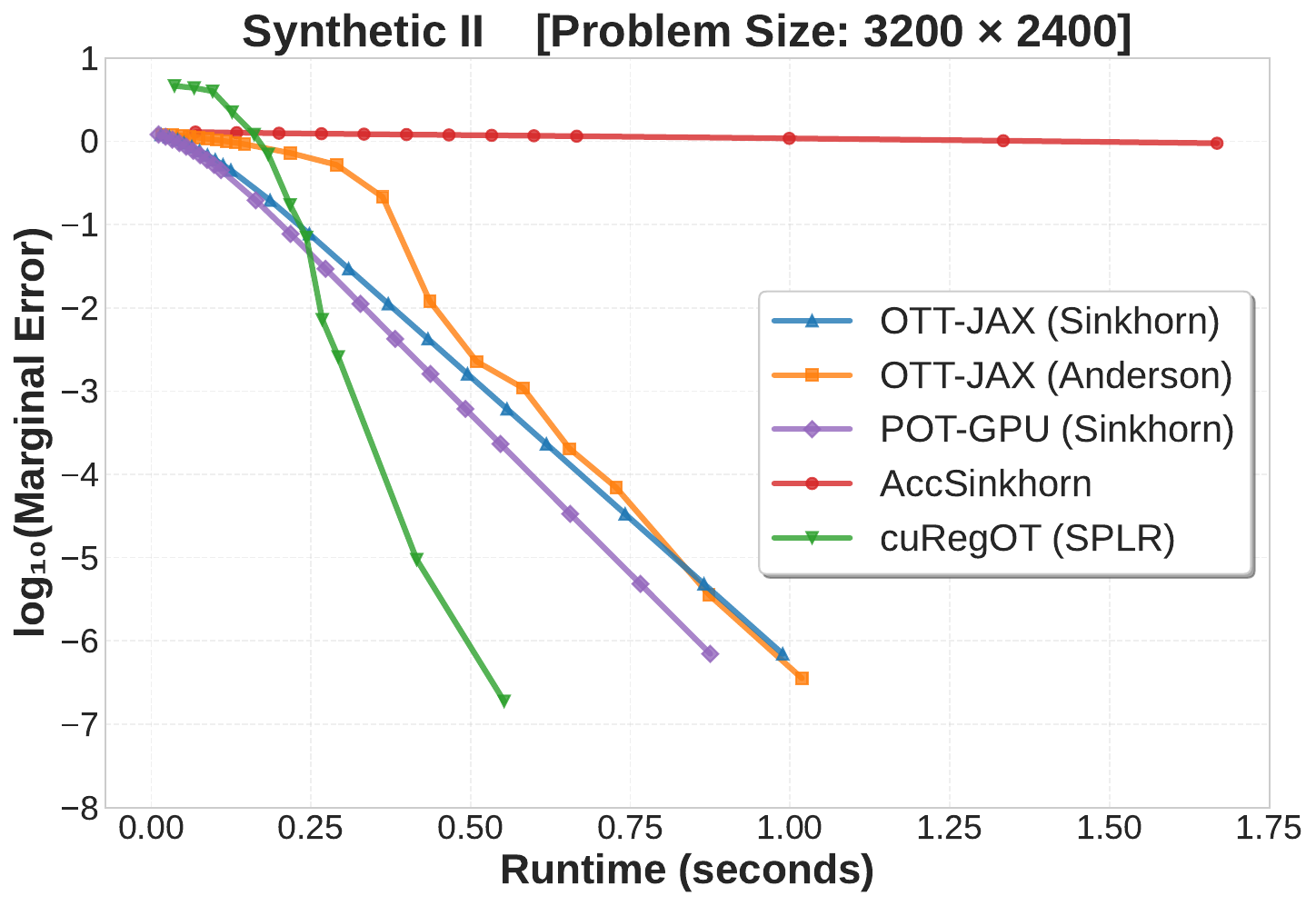}
\includegraphics[width=0.33\textwidth]{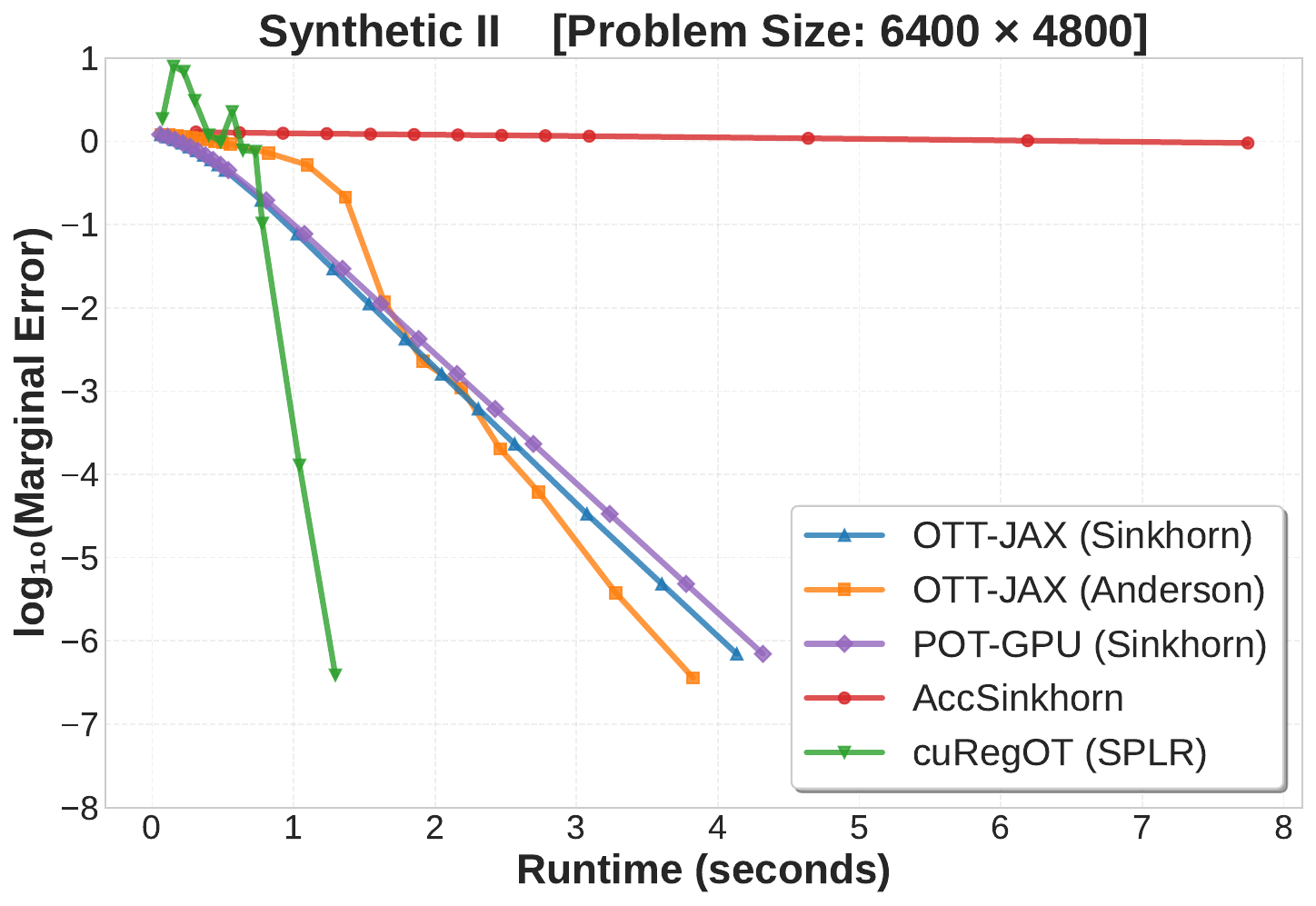}
\par\end{centering}
}

\caption{\label{fig:synthetic}Benchmark result for synthetic datasets. The
horizontal axis represents the elapsed wall time, and the vertical
axis is the optimization error on a logarithmic scale.}
\end{figure}

\begin{figure}[h]
\centering
\begin{centering}
\includegraphics[width=0.328\textwidth]{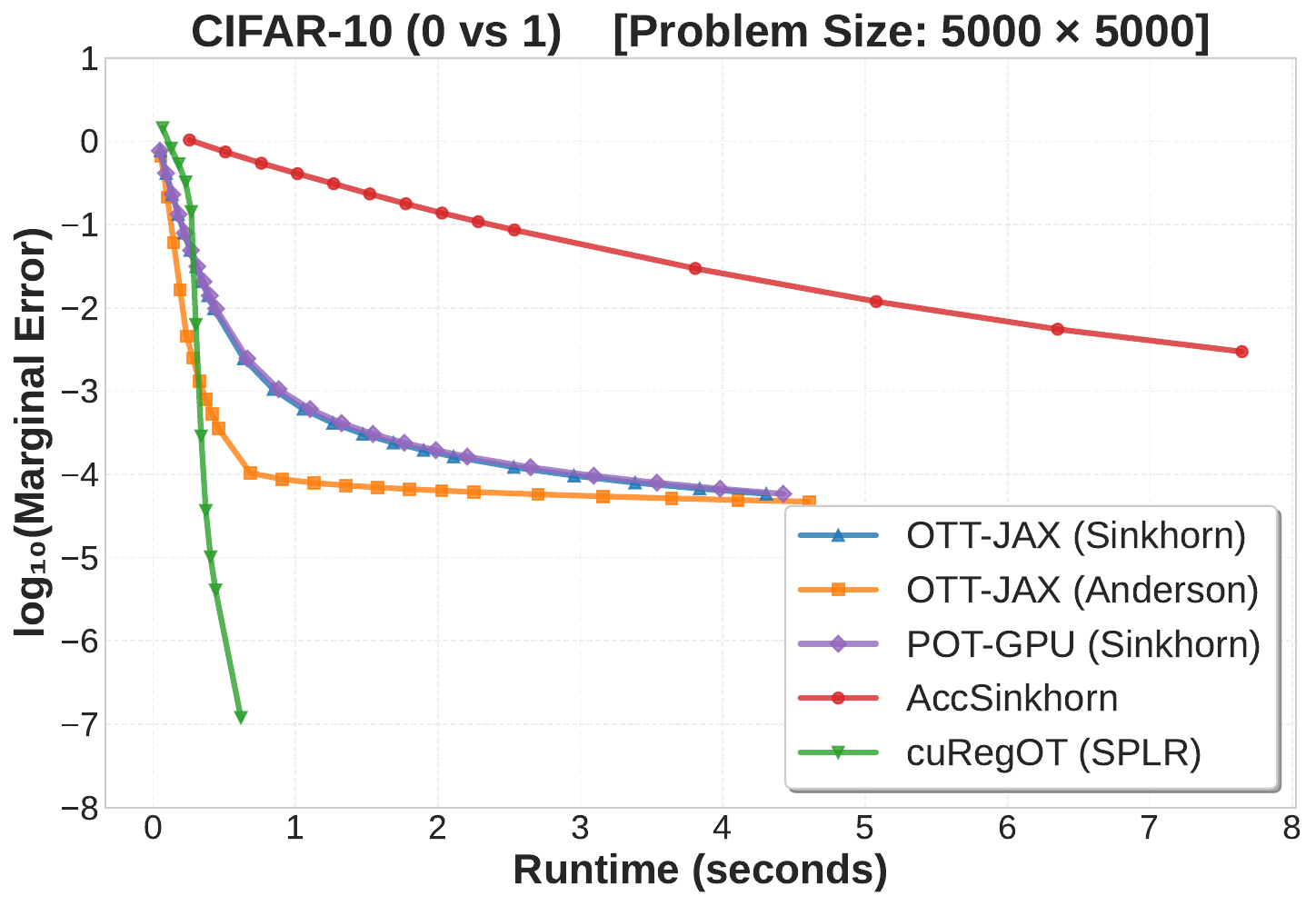}
\includegraphics[width=0.328\textwidth]{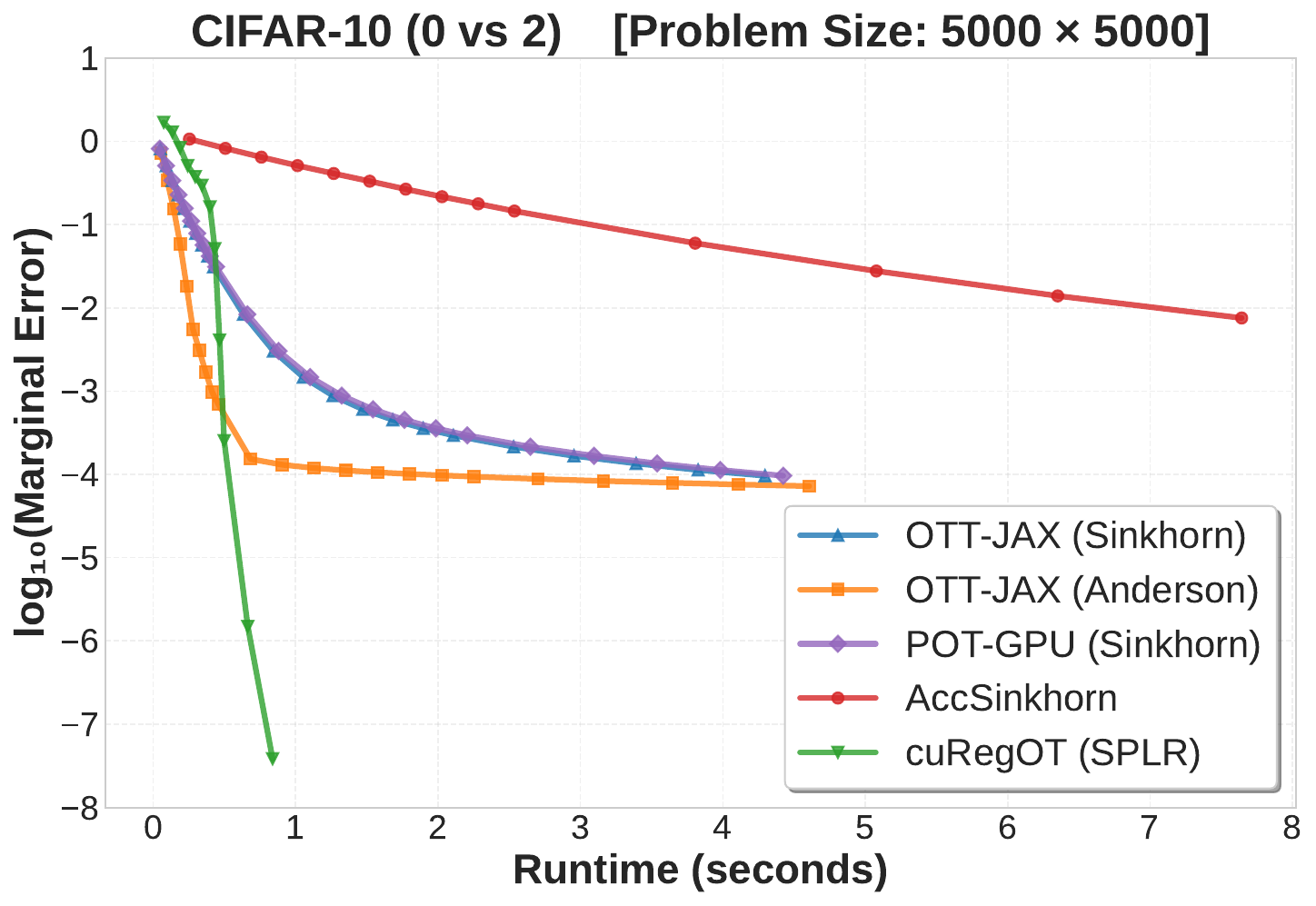}
\includegraphics[width=0.328\textwidth]{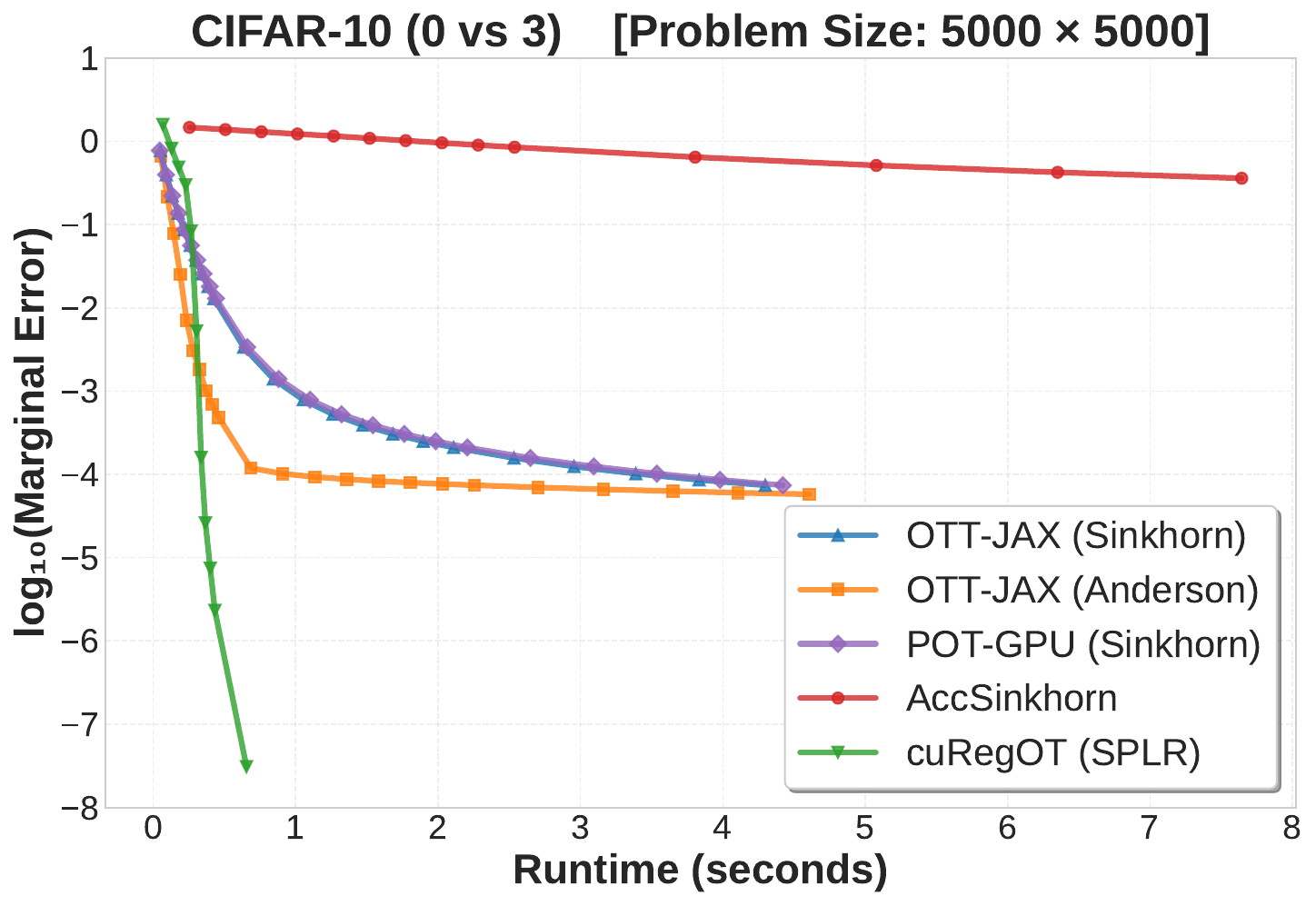}
\par\end{centering}
\begin{centering}
\includegraphics[width=0.328\textwidth]{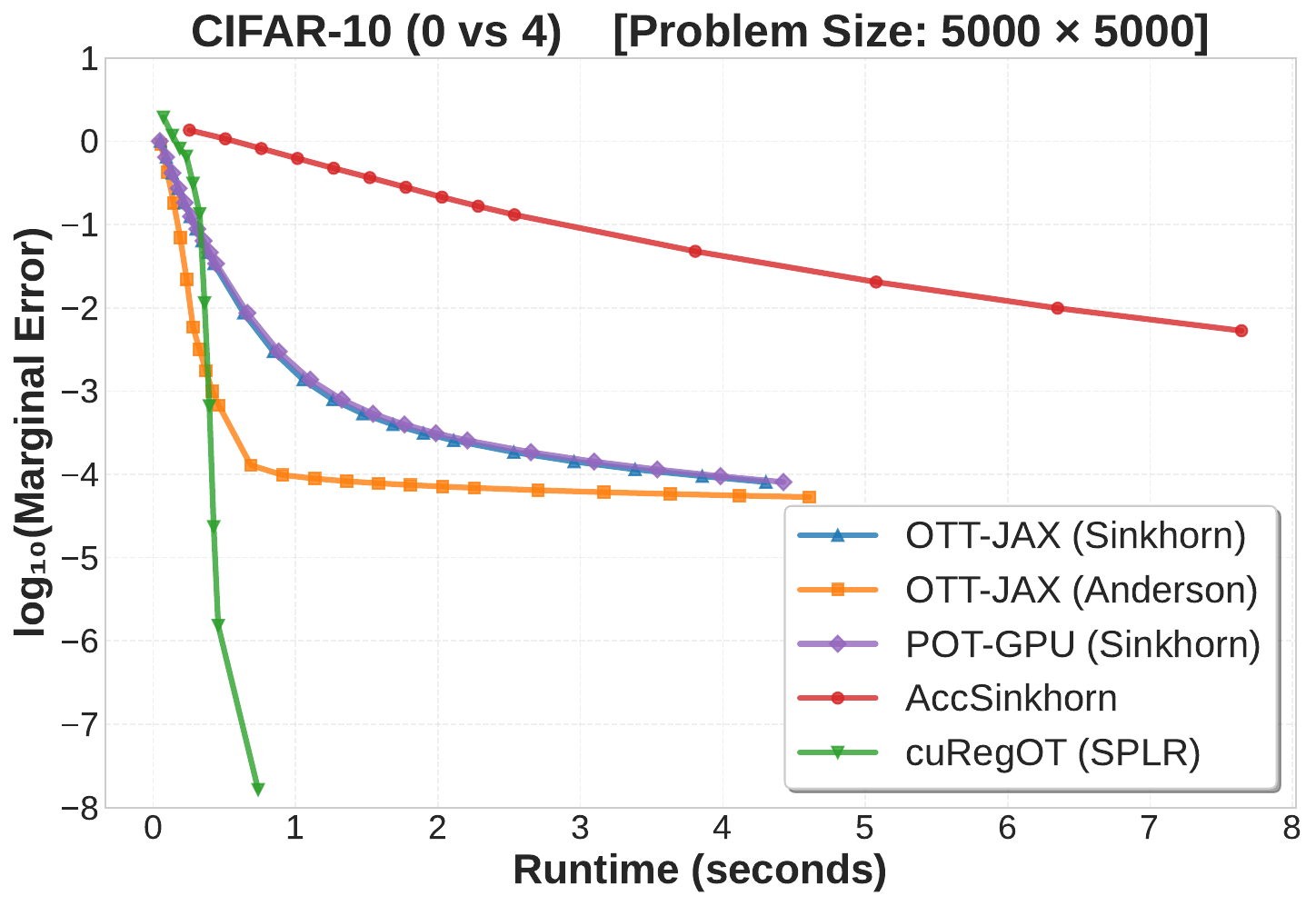}
\includegraphics[width=0.328\textwidth]{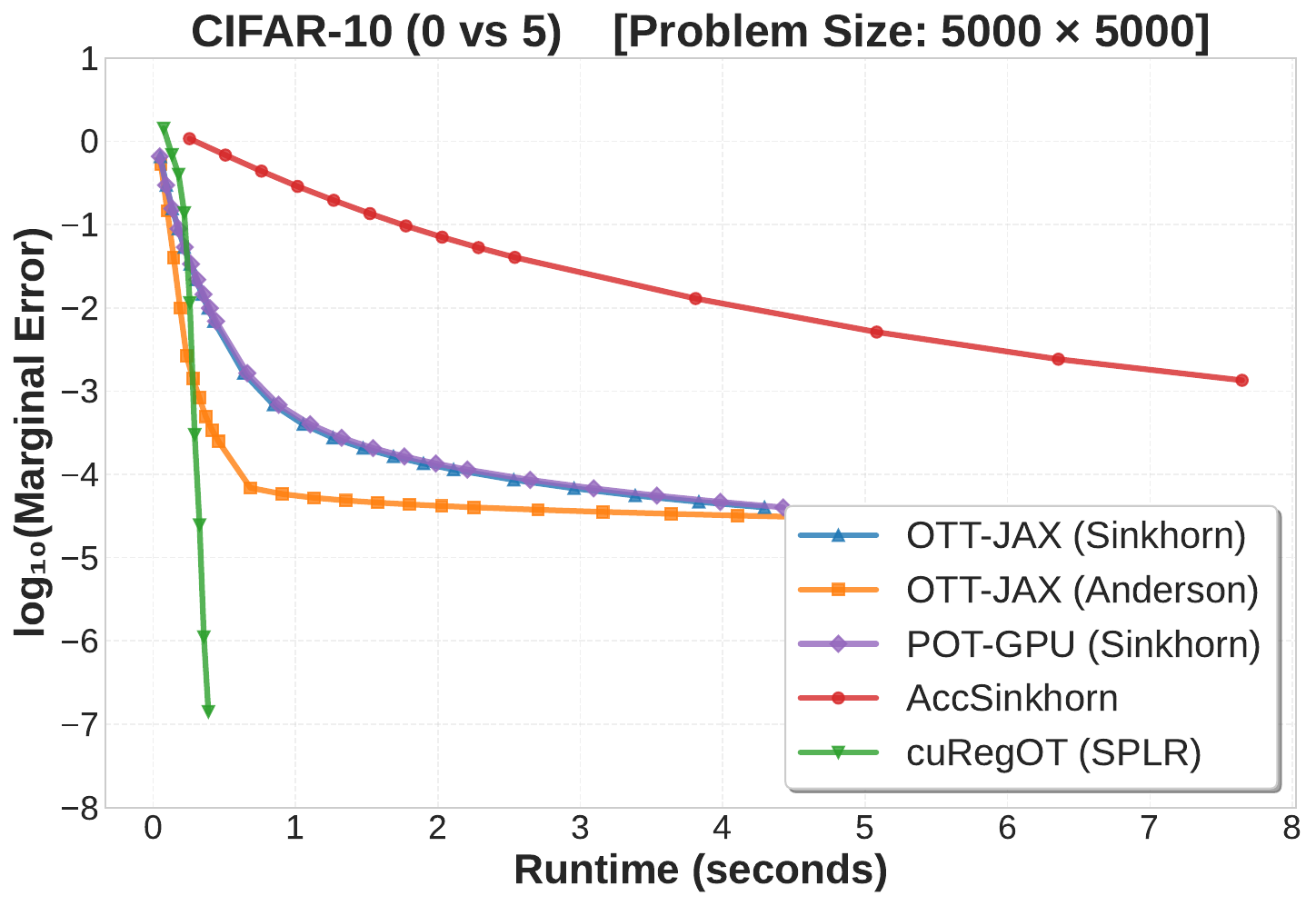}
\includegraphics[width=0.328\textwidth]{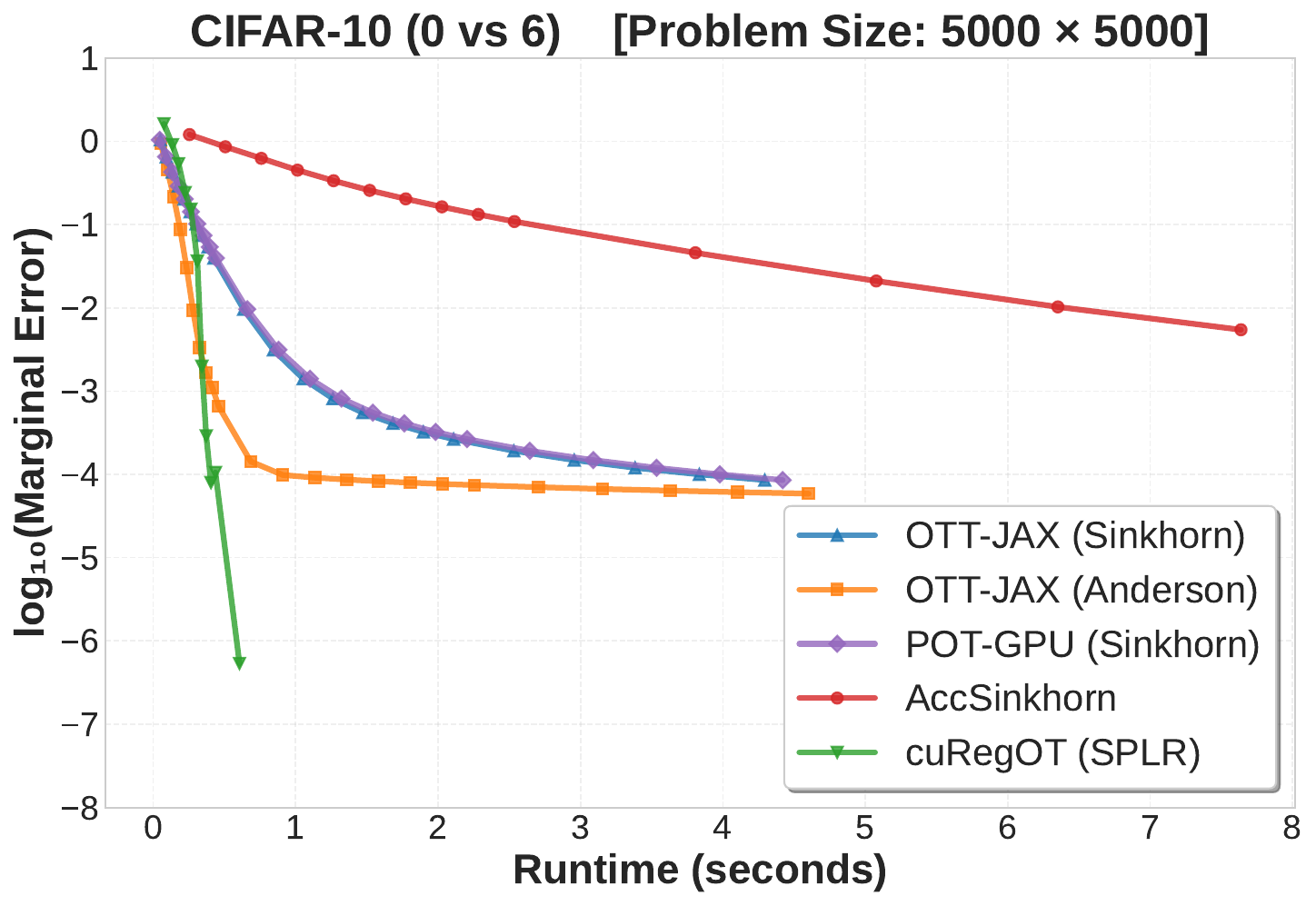}
\par\end{centering}
\begin{centering}
\includegraphics[width=0.328\textwidth]{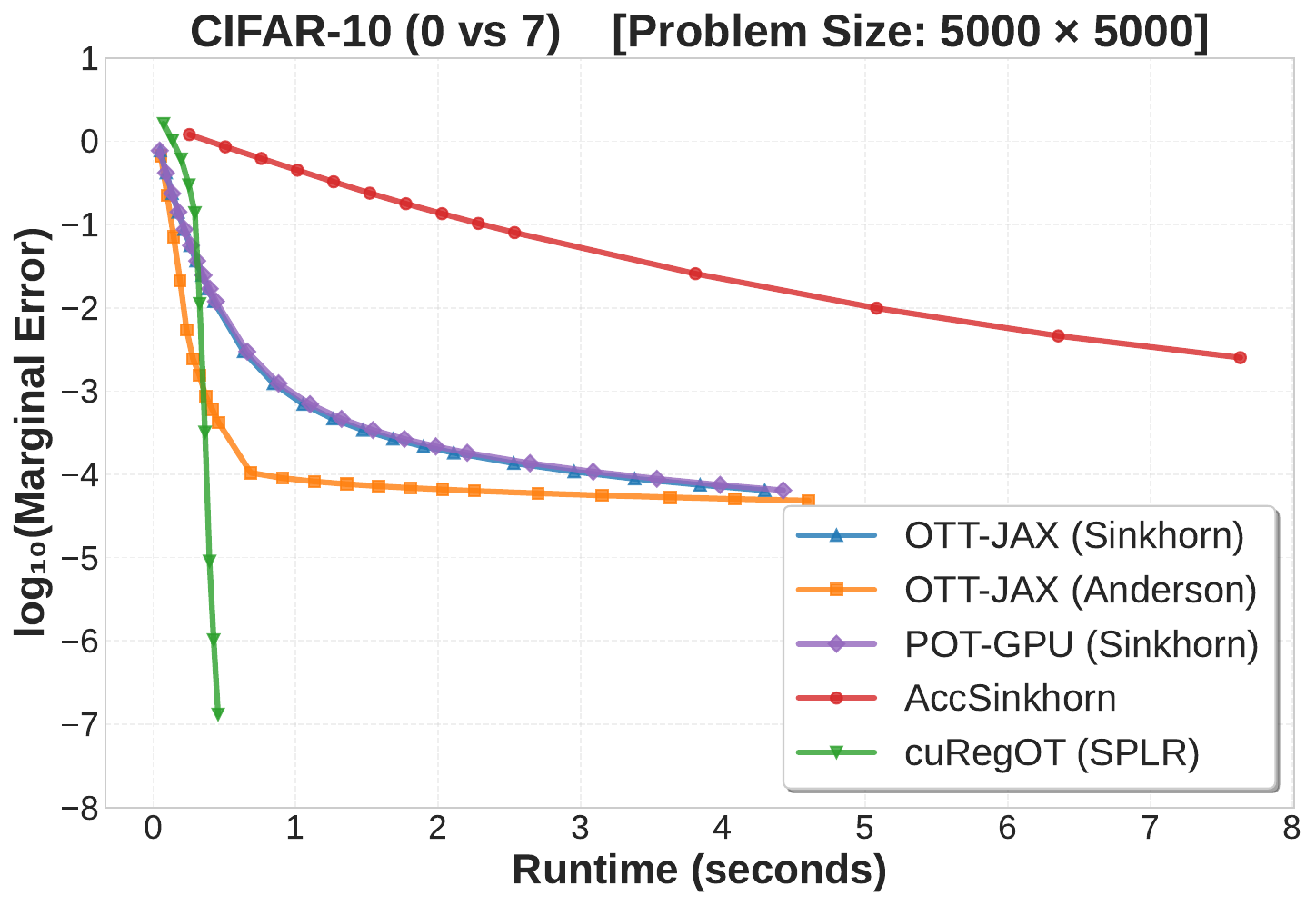}
\includegraphics[width=0.328\textwidth]{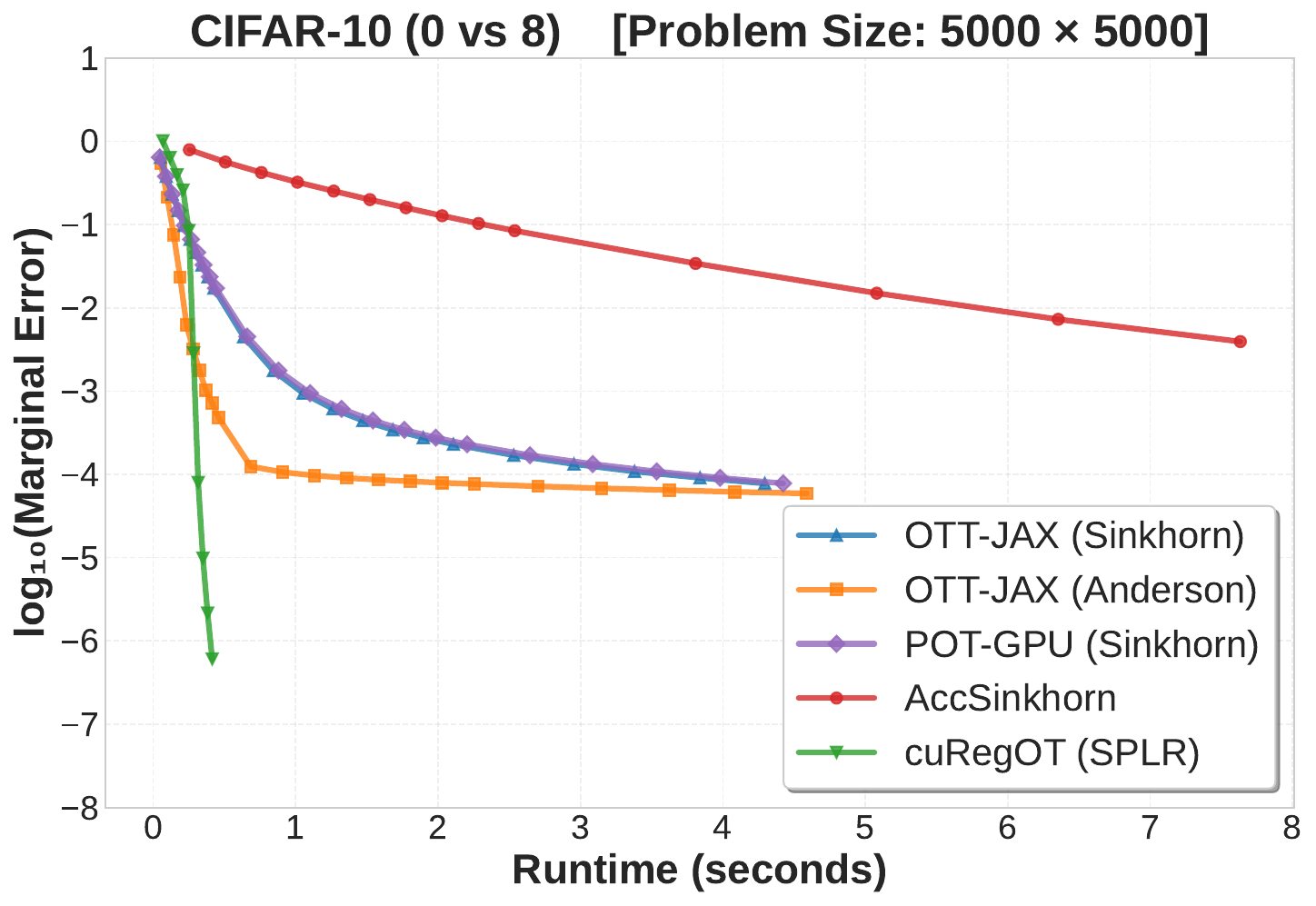}
\includegraphics[width=0.328\textwidth]{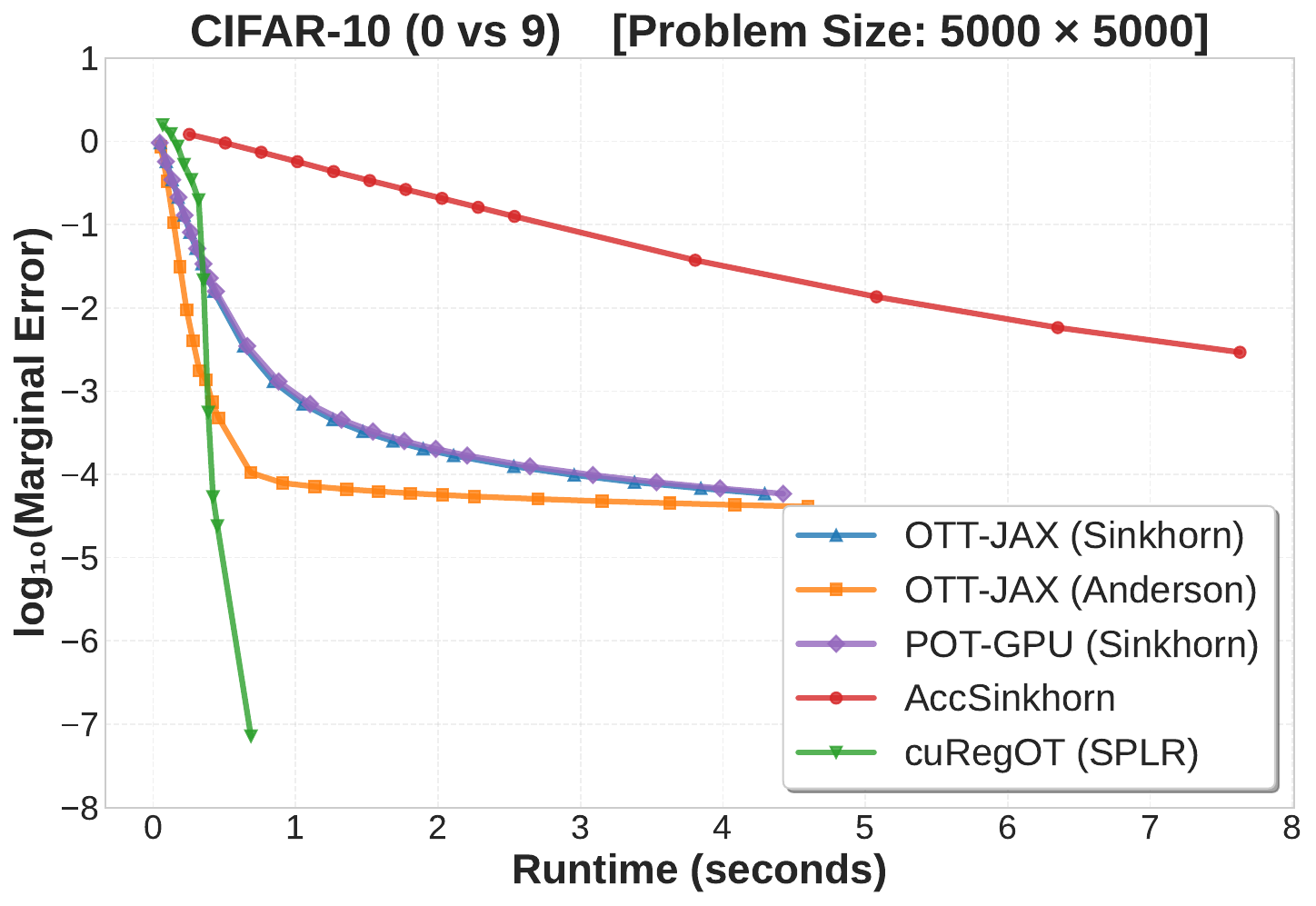}
\par\end{centering}
\caption{\label{fig:cifar10}Benchmark result for the CIFAR-10 dataset.}
\end{figure}

\subsection{Result}

In all experiments, we have found that the Greenkhorn algorithm has
a quite slow convergence speed, exhibiting almost flat curves, so
we exclude it in all the plots. The experiment results for the synthetic
data and the CIFAR-10 data are visualized in Figures \ref{fig:synthetic}
and \ref{fig:cifar10}, respectively. First of all, we can observe
that the two implementations of the Sinkhorn algorithm are close to
each other, suggesting that their performance are representative.
For the Anderson acceleration, it provides visible speedup to the
standard Sinkhorn algorithm in the Synthetic I and CIFAR-10 cases,
but still generally faces a slow convergence to a high precision.
On the other hand, it can be even slower in the Synthetic II case.

For the proposed cuRegOT solver, it is clear that it exhibits a significantly
faster convergence in almost all cases, especially when a relatively
high precision is requested. Moreover, the advantage of cuRegOT tends
to be larger for bigger problem sizes, further validating its scalability
for large-scale and real-word problems.

\subsection{Ablation Study}

\begin{figure}[h]
\centering
\begin{centering}
\includegraphics[width=0.328\textwidth]{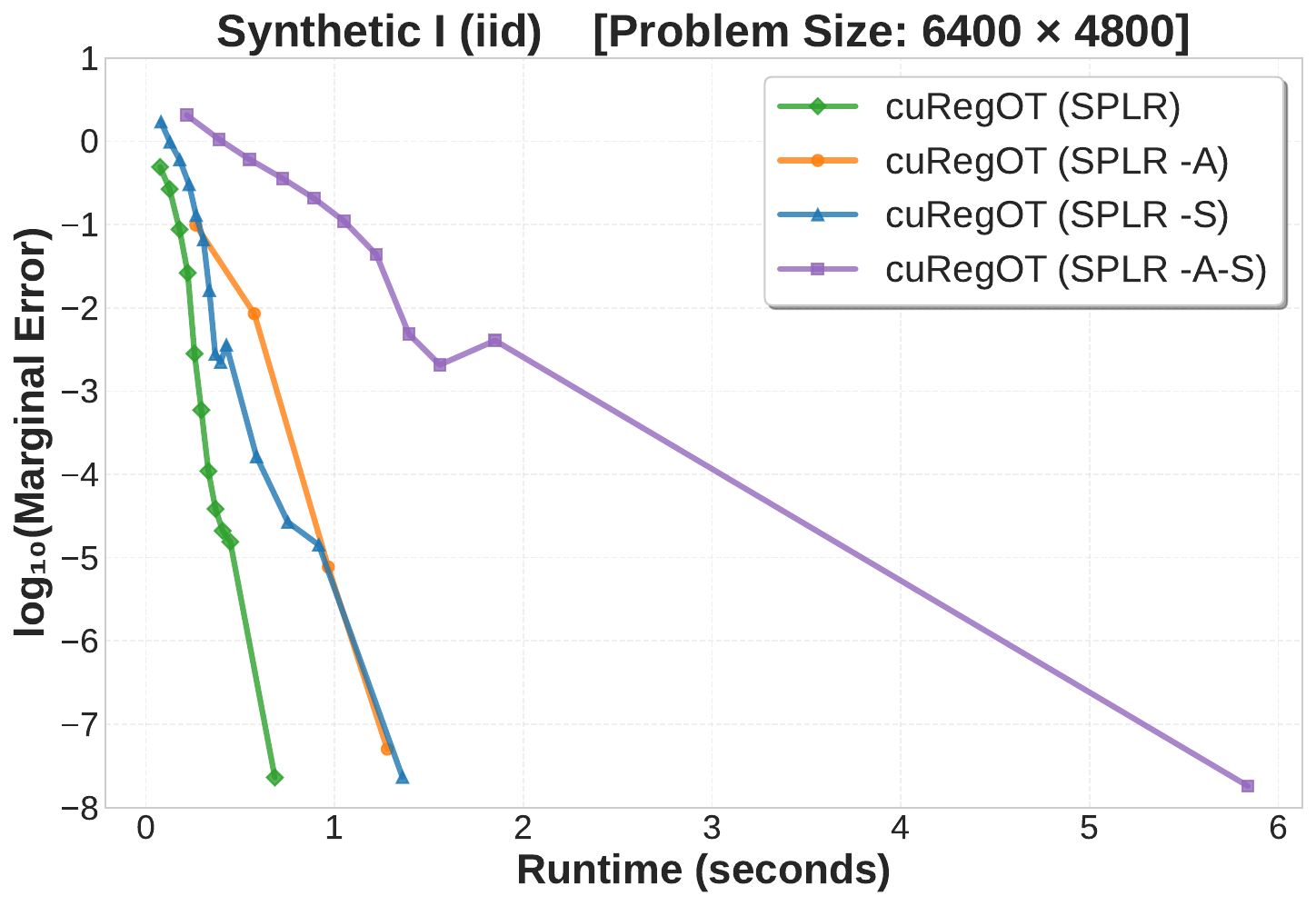}
\includegraphics[width=0.328\columnwidth]{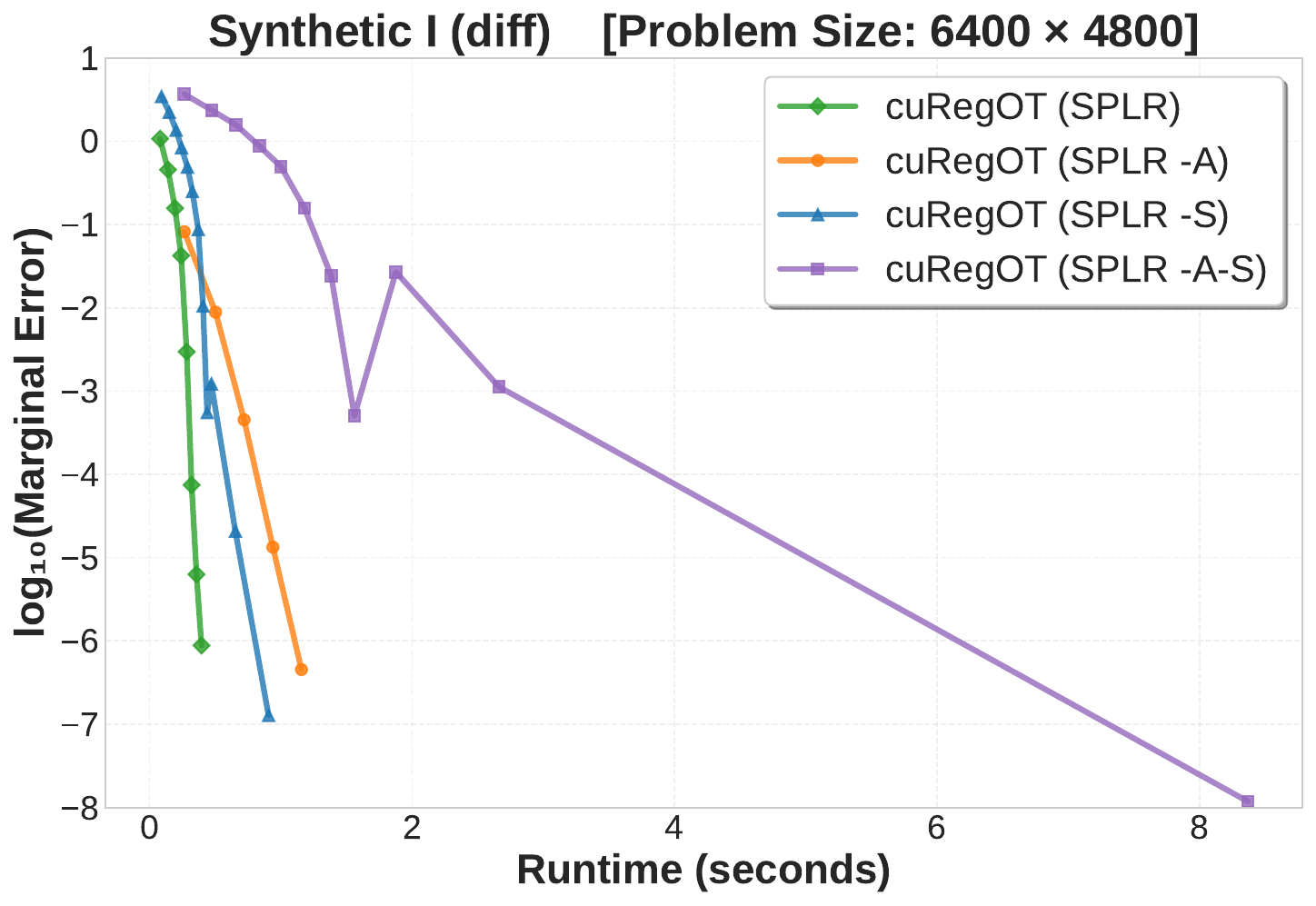}
\includegraphics[width=0.328\columnwidth]{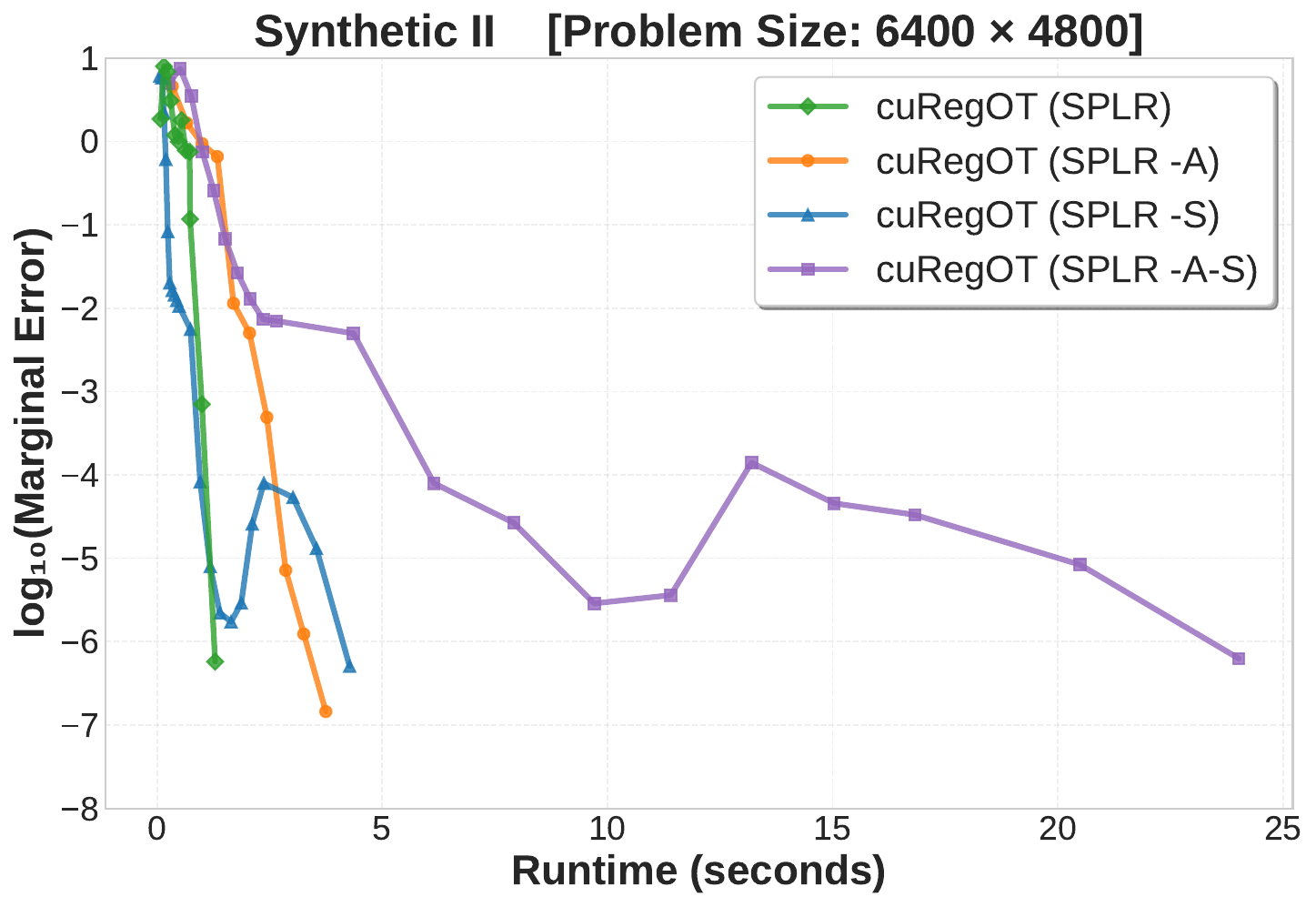}
\par\end{centering}
\begin{centering}
\includegraphics[width=0.328\columnwidth]{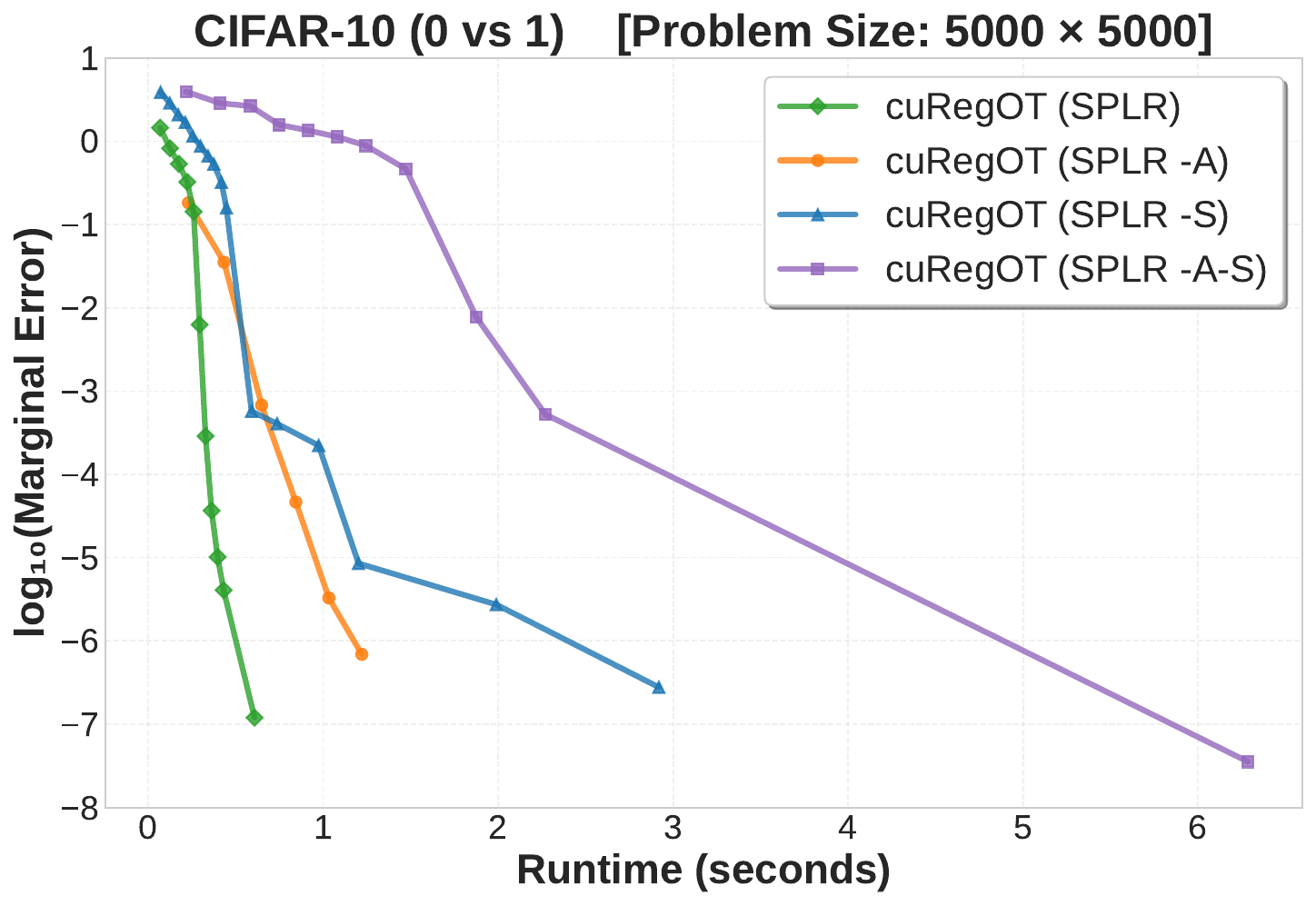}
\includegraphics[width=0.328\columnwidth]{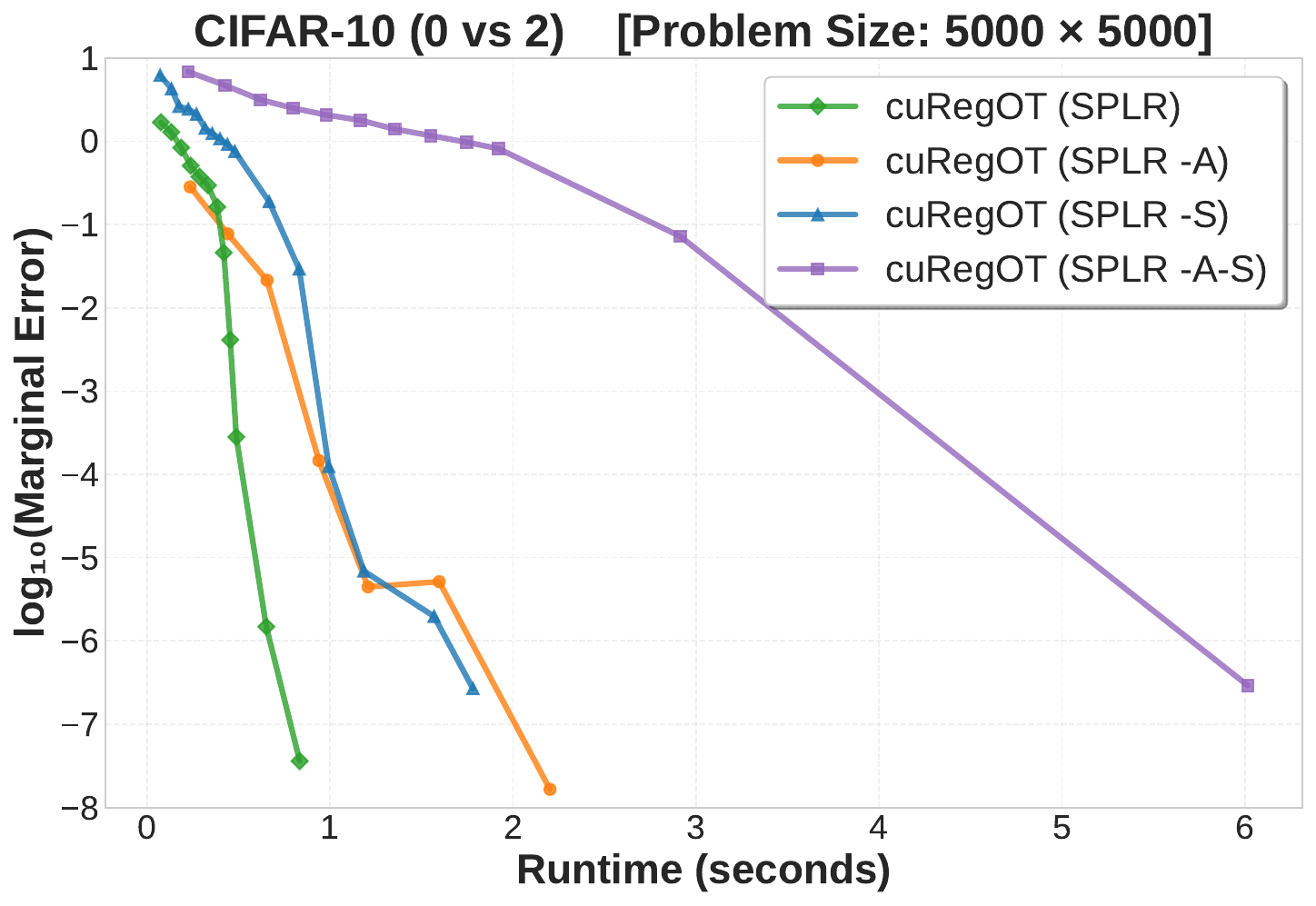}
\includegraphics[width=0.328\columnwidth]{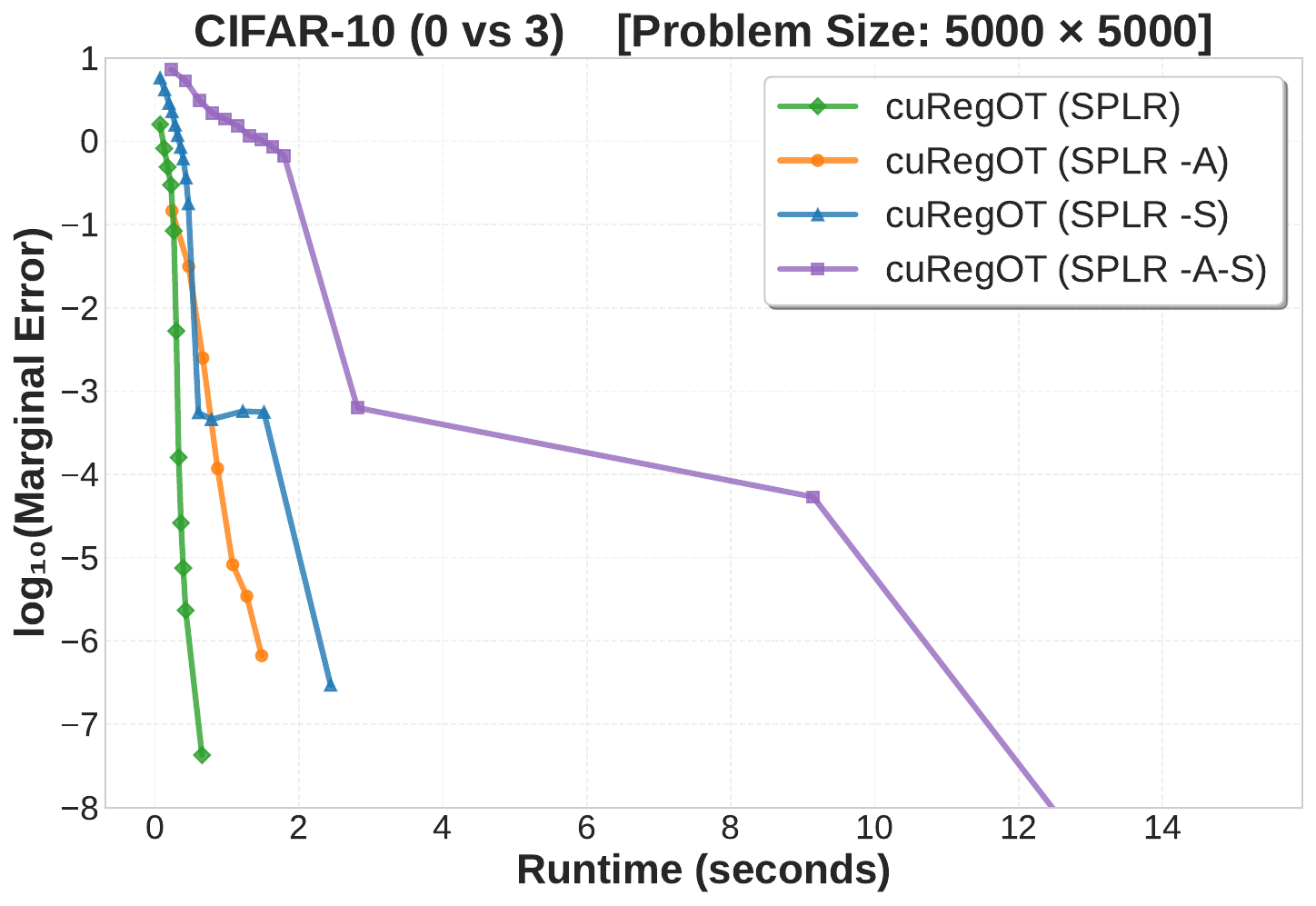}
\par\end{centering}
\caption{\label{fig:ablation}Ablation study on the algorithm design.}
\end{figure}

To study the effectiveness and necessity of the algorithm improvements
introduced in Section \ref{sec:method}, we conduct an ablation study
by testing the solver performance after removing one or more components
of the algorithm design. The results are shown in Figure \ref{fig:ablation},
in which ``-A'' means removing the amortized symbolic analysis strategy,
``-S'' means removing the candidate Sinkhorn iterate generation,
and ``-A-S'' means removing both. Clearly, the plots show that removing
either of the algorithmic improvement may slow down the solver. This
suggests that both the amortized symbolic analysis and the Sinkhorn
iteration generation contribute to the performance of the cuRegOT
solver.

\section{Conclusion}

This paper presented cuRegOT, a high-performance GPU solver for entropic-regularized
OT, built upon the SPLR quasi-Newton method. The main obstacle in
bringing Newton-type methods to GPUs are sparse linear systems: the
symbolic analysis and reordering needed by the sparse Cholesky decomposition
are difficult to parallelize and are often executed on the CPU. To
address this bottleneck, cuRegOT introduces three GPU-oriented designs:
(1) \emph{amortized symbolic analysis} by reusing sparsity patterns
across multiple iterations; (2) \emph{collaborative CPU--GPU computing}
that overlaps host-side analysis with GPU Sinkhorn-based candidate
generation; and (3) \emph{fused CUDA kernel} for efficient gradient
evaluation to reduce memory traffic.

We established rigorous guarantees showing that these modifications
preserve the global convergence of the original SPLR method and retain
at least a linear convergence rate. In numerical experiments, cuRegOT
consistently achieves faster convergence than GPU Sinkhorn baselines
(including Anderson acceleration), with larger gains at higher accuracy
targets and larger problem sizes. Future work includes further reducing
reliance on host-side symbolic steps, exploring iterative or mixed-precision
sparse solvers, and extending the approach to broader OT formulations.

\bibliographystyle{apalike}
\bibliography{ref}

\appendix

\section{Additional Experiment Details}

\subsection{Computing Environment}

The numerical experiments are conducted on a workstation with an AMD
5900X CPU and an Nvidia RTX 6000 Ada GPU. The workstation runs on
a Ubuntu 25.10 Linux operating system and the CUDA 13.2 platform.

\subsection{Sparsity Pattern of the Transport Plan Across Iterations}

\label{subsec:evolution_plan}

\begin{figure}[h]
\begin{centering}
\includegraphics[width=0.3\textwidth]{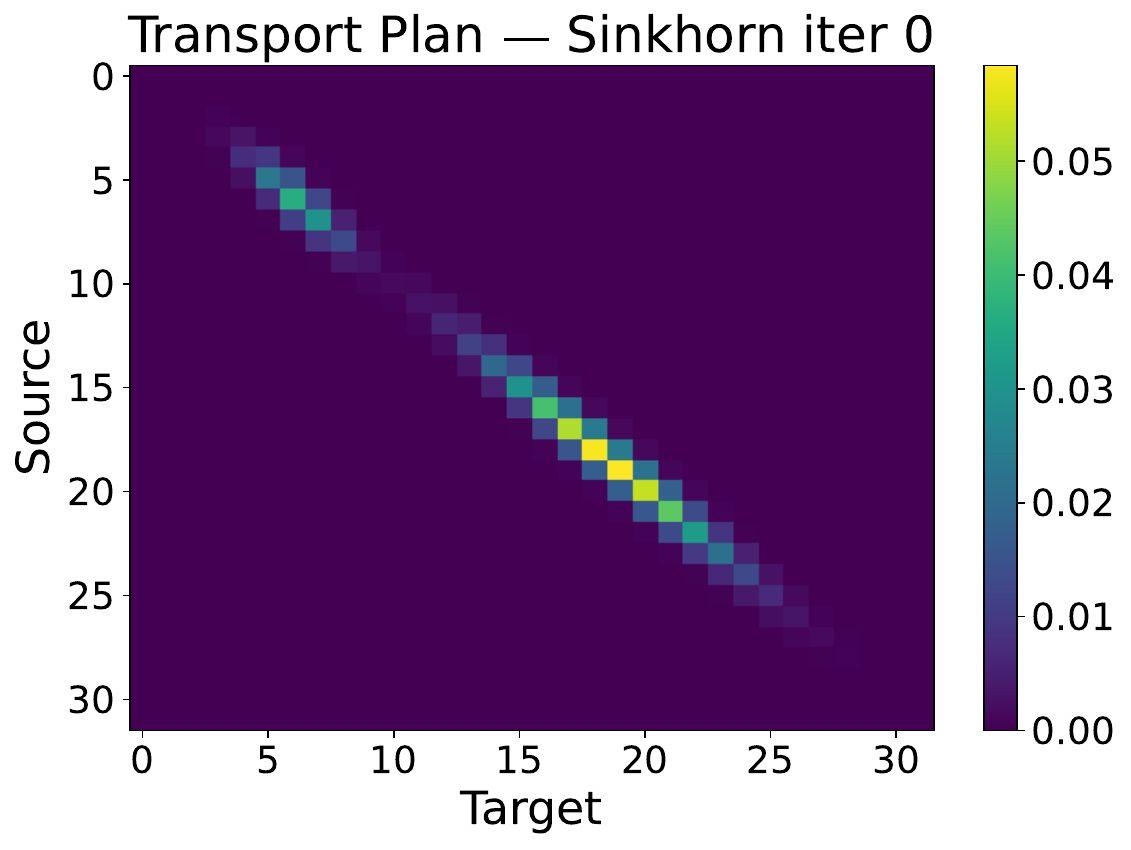}\includegraphics[width=0.3\textwidth]{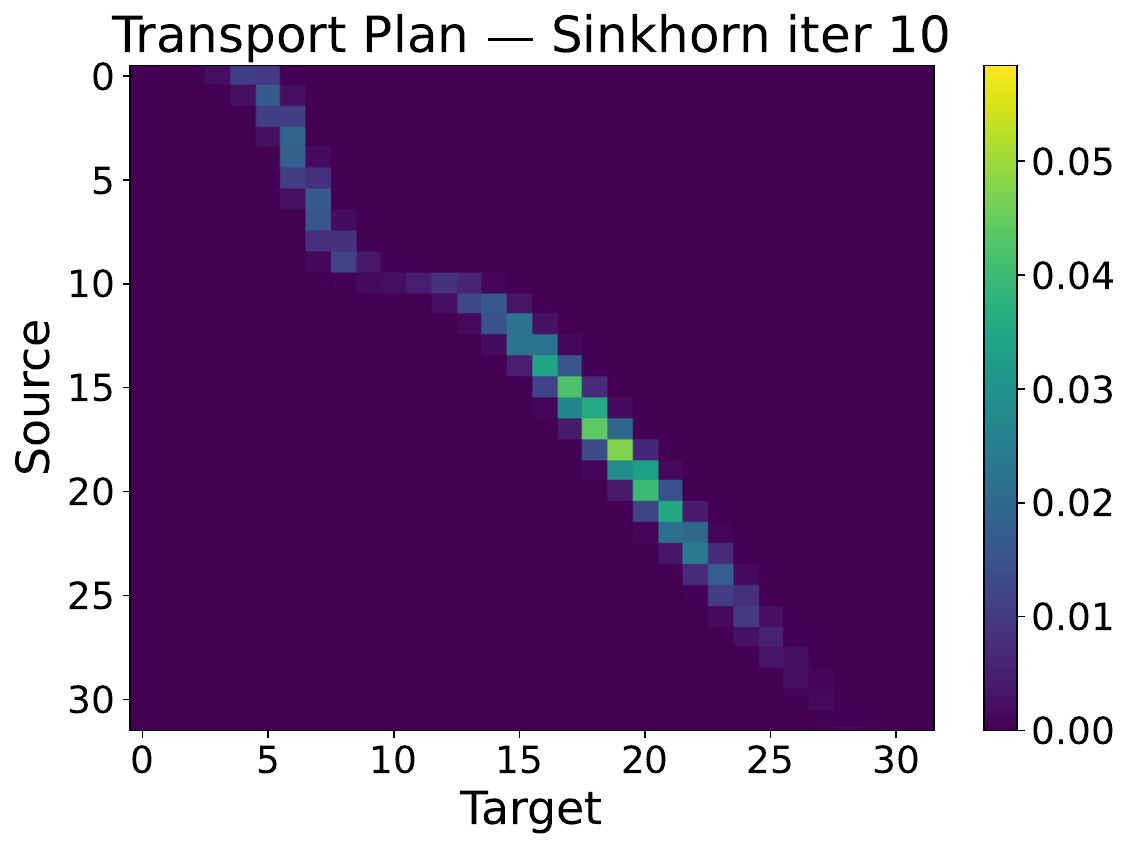}\includegraphics[width=0.3\textwidth]{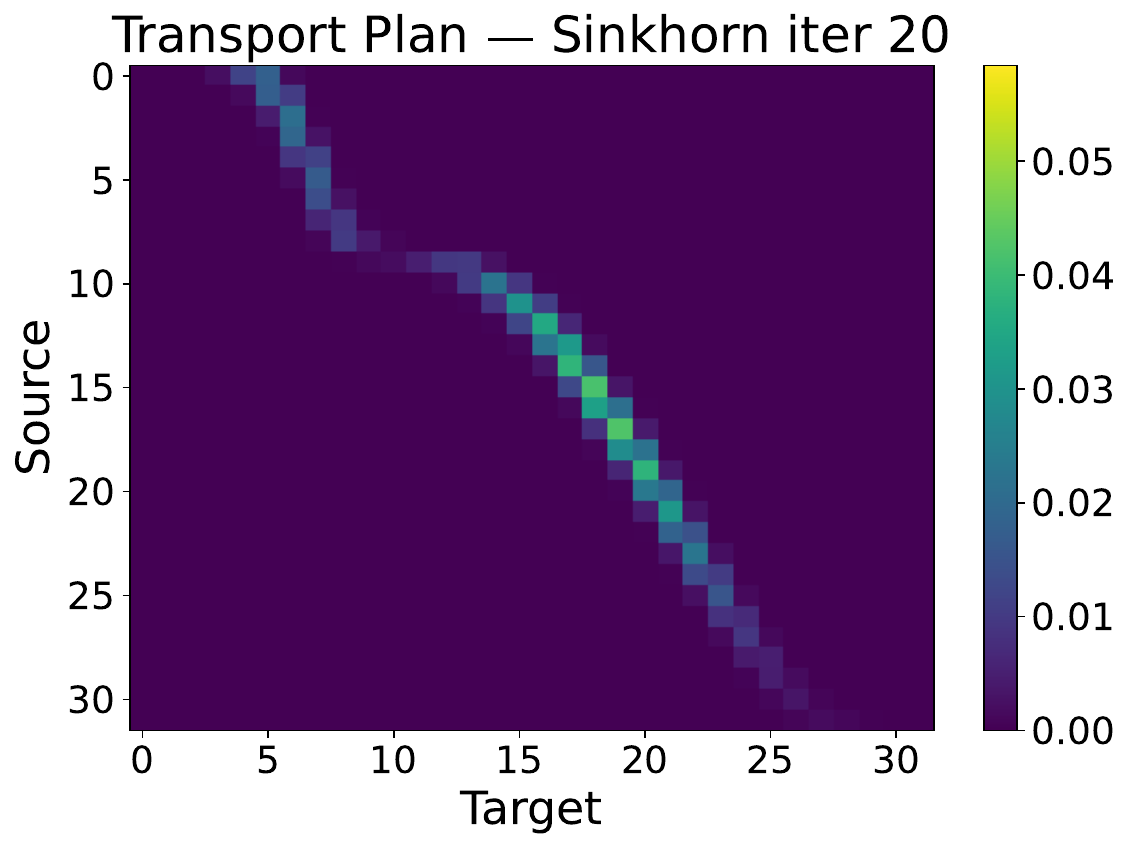}
\par\end{centering}
\begin{centering}
\includegraphics[width=0.3\textwidth]{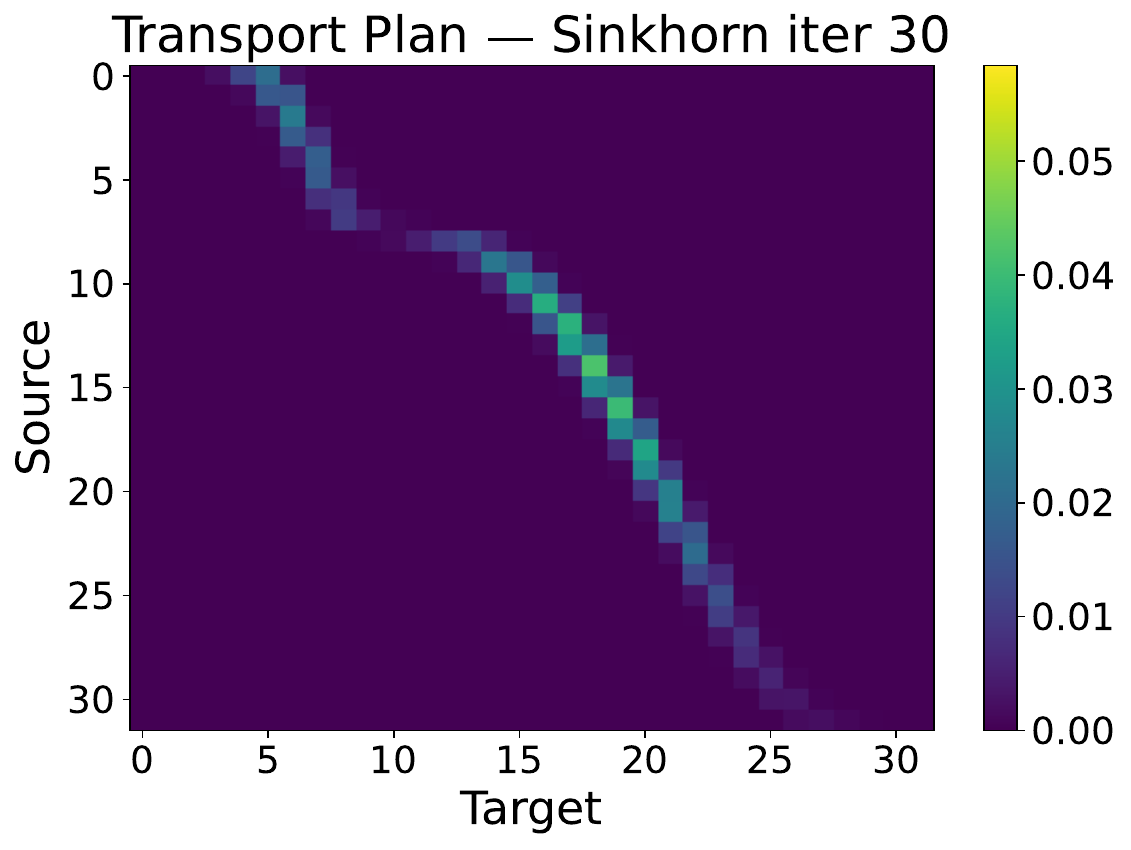}\includegraphics[width=0.3\textwidth]{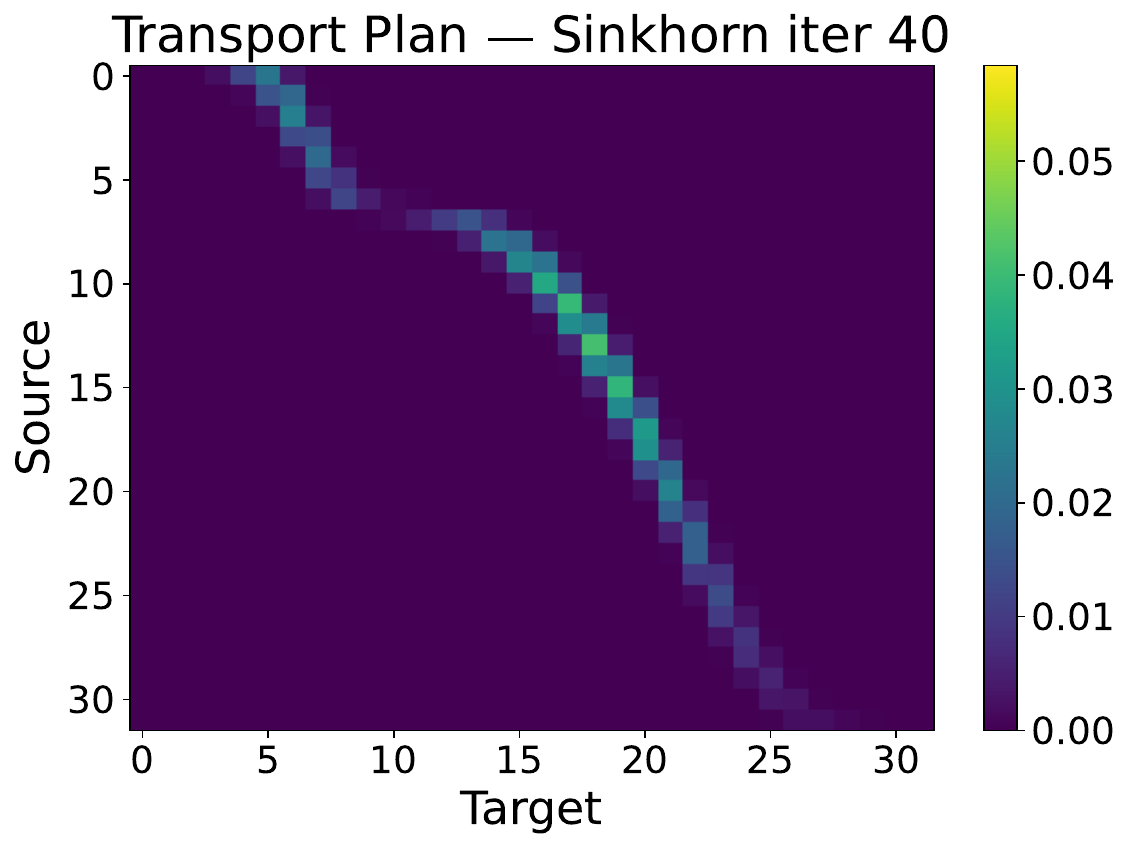}\includegraphics[width=0.3\textwidth]{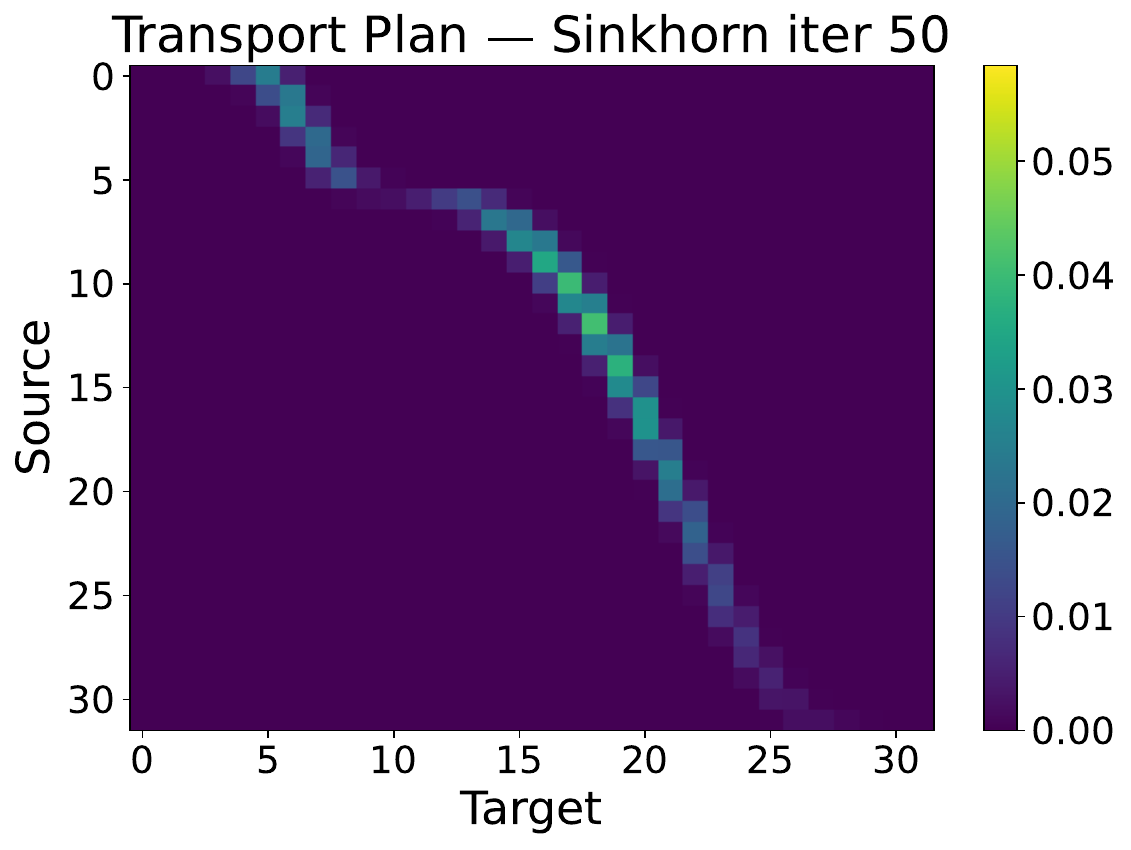}
\par\end{centering}
\begin{centering}
\includegraphics[width=0.3\textwidth]{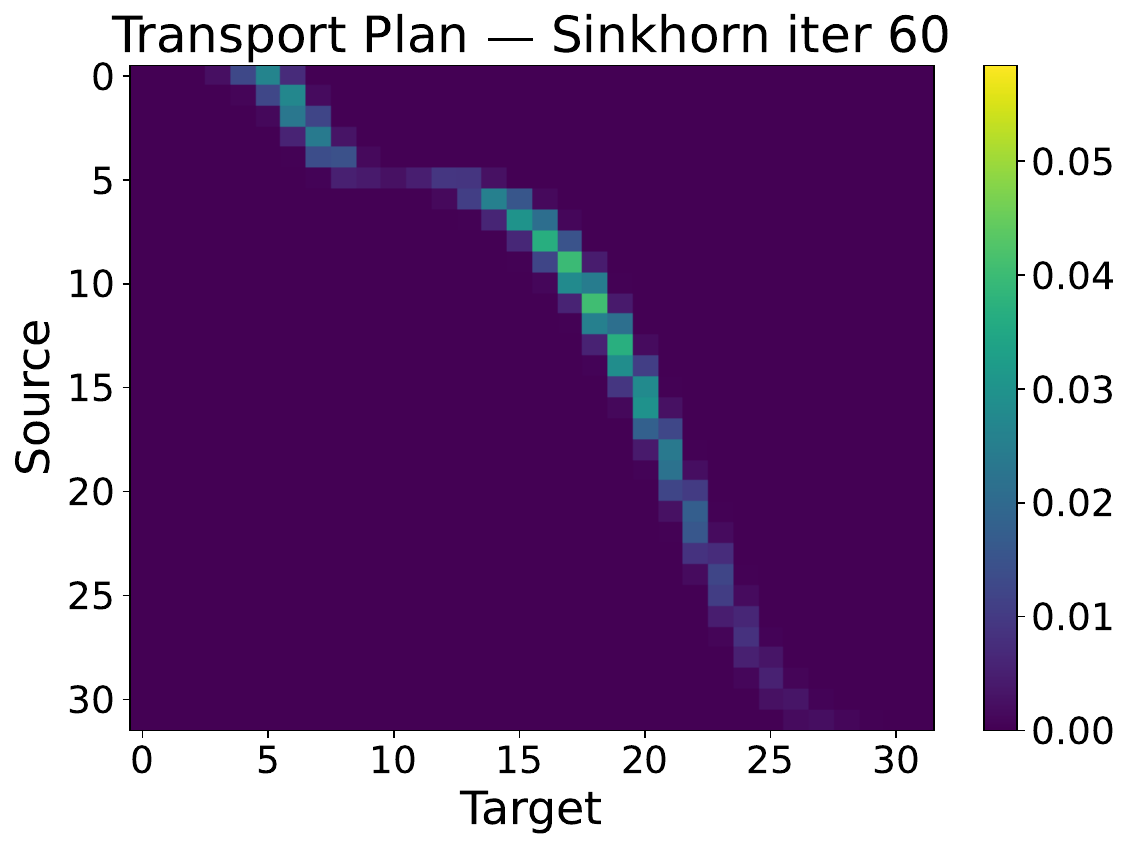}\includegraphics[width=0.3\textwidth]{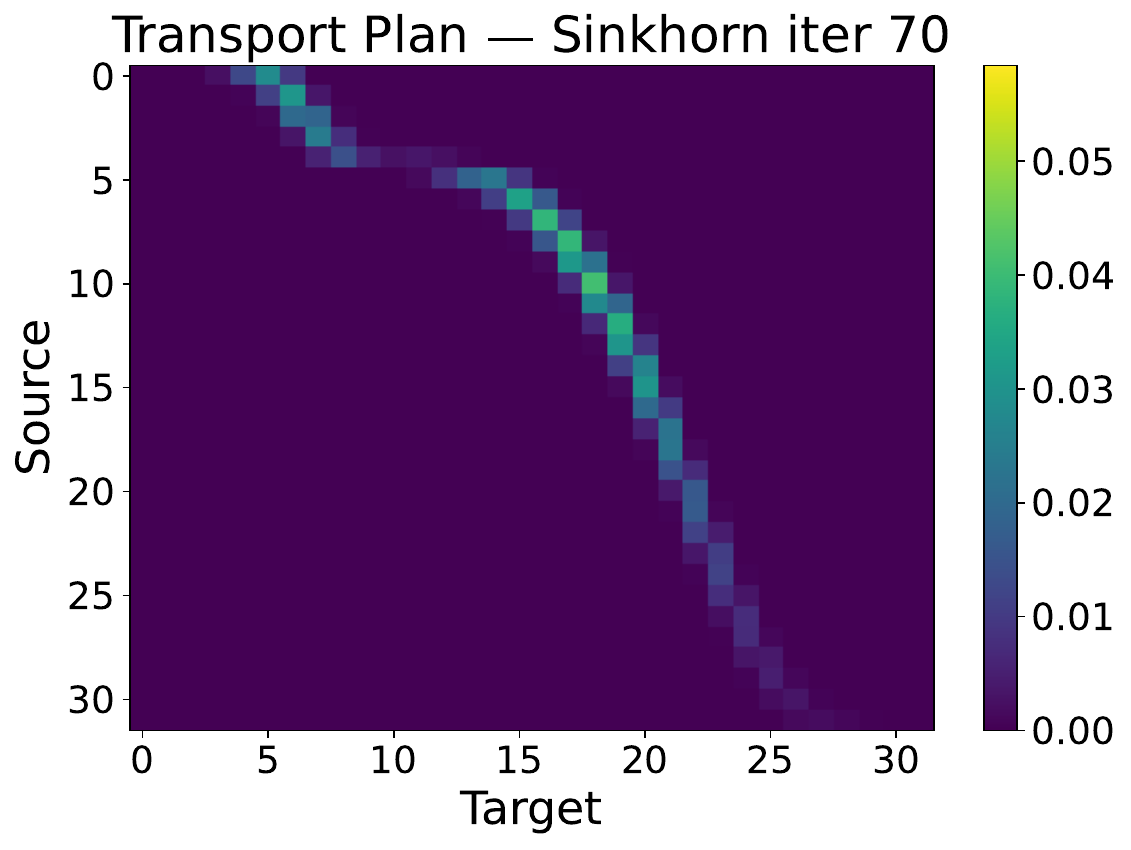}\includegraphics[width=0.3\textwidth]{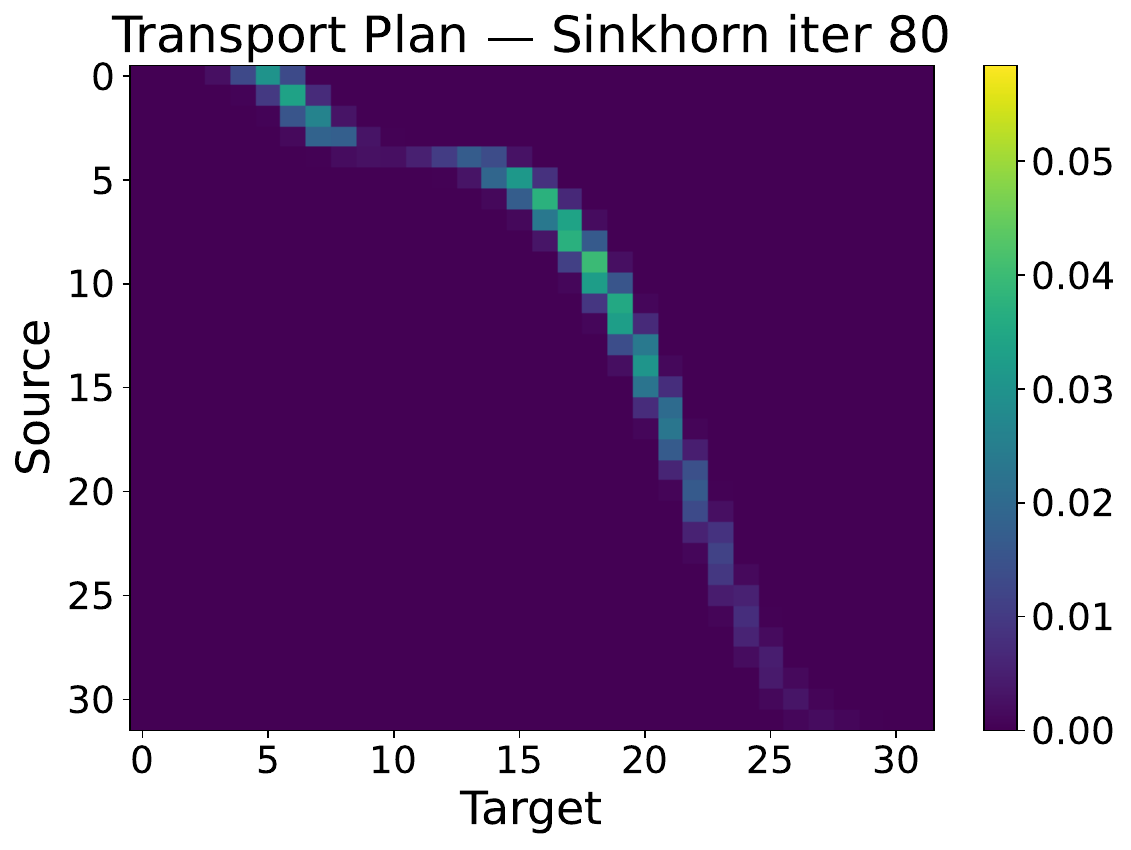}
\par\end{centering}
\begin{centering}
\includegraphics[width=0.3\textwidth]{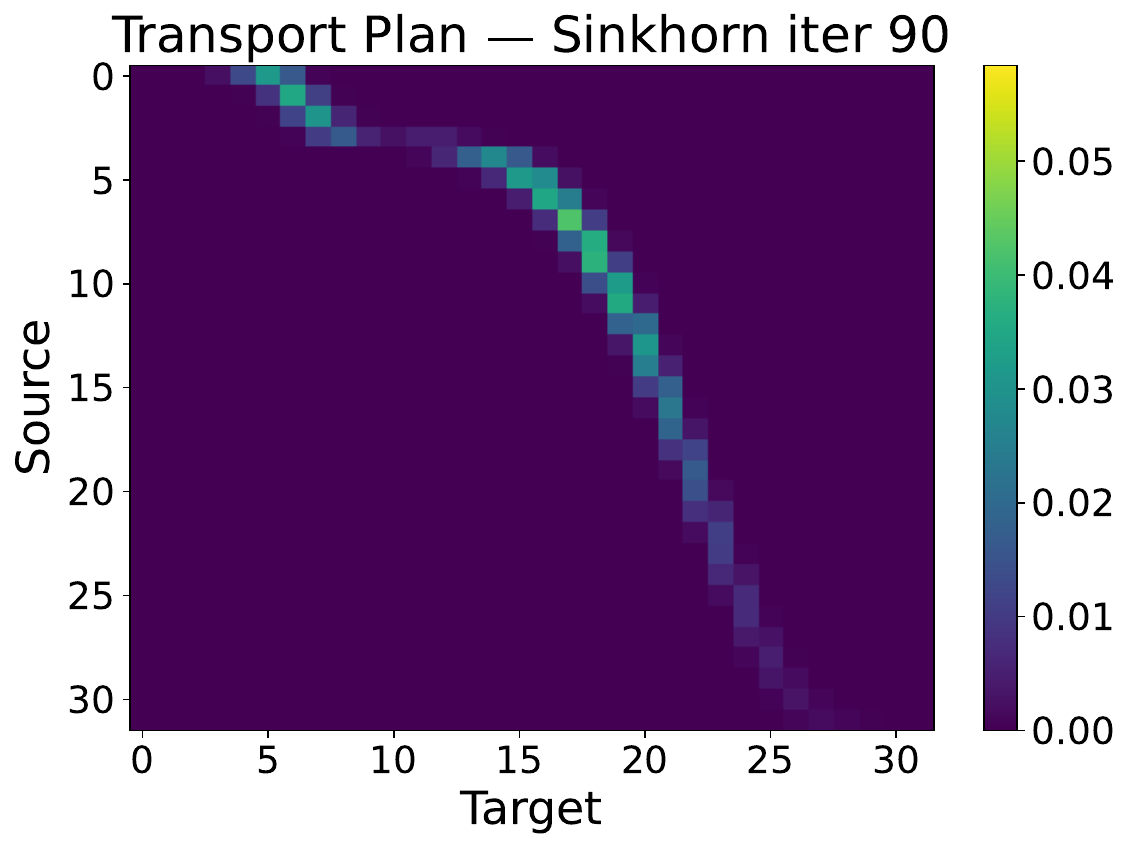}\includegraphics[width=0.3\textwidth]{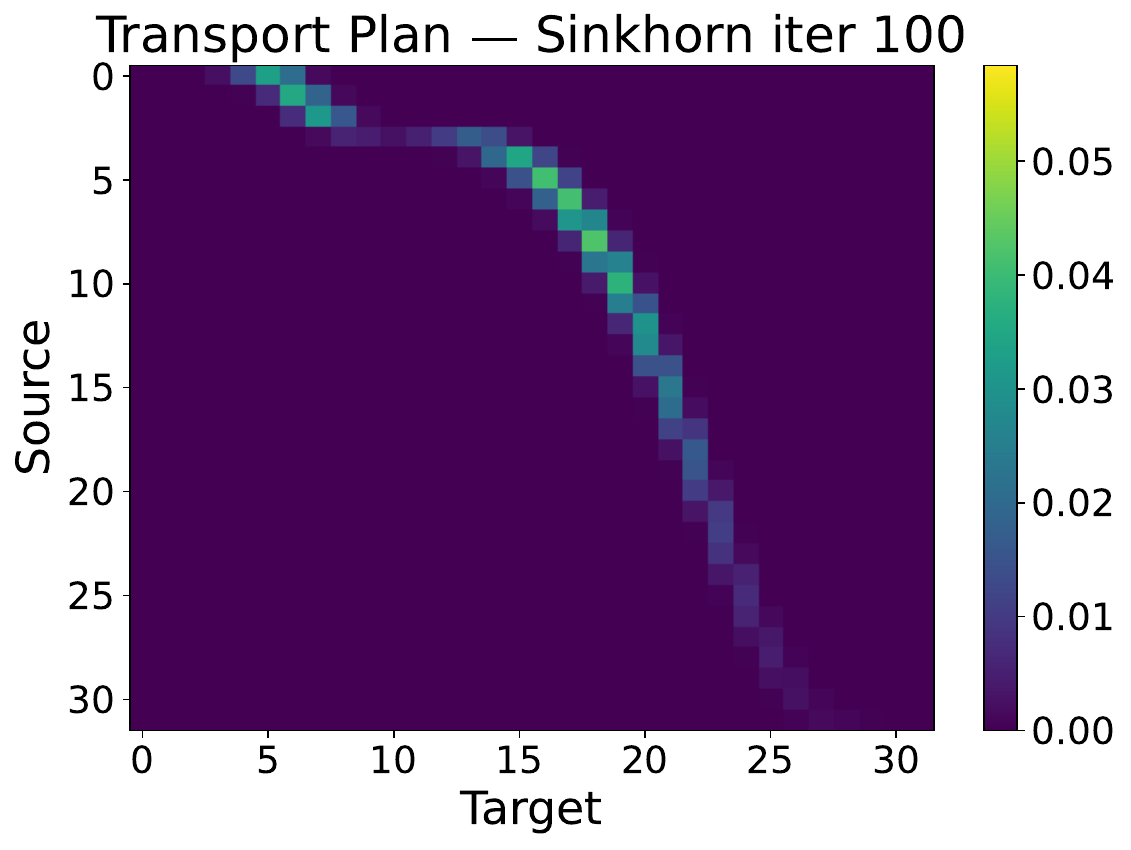}\includegraphics[width=0.3\textwidth]{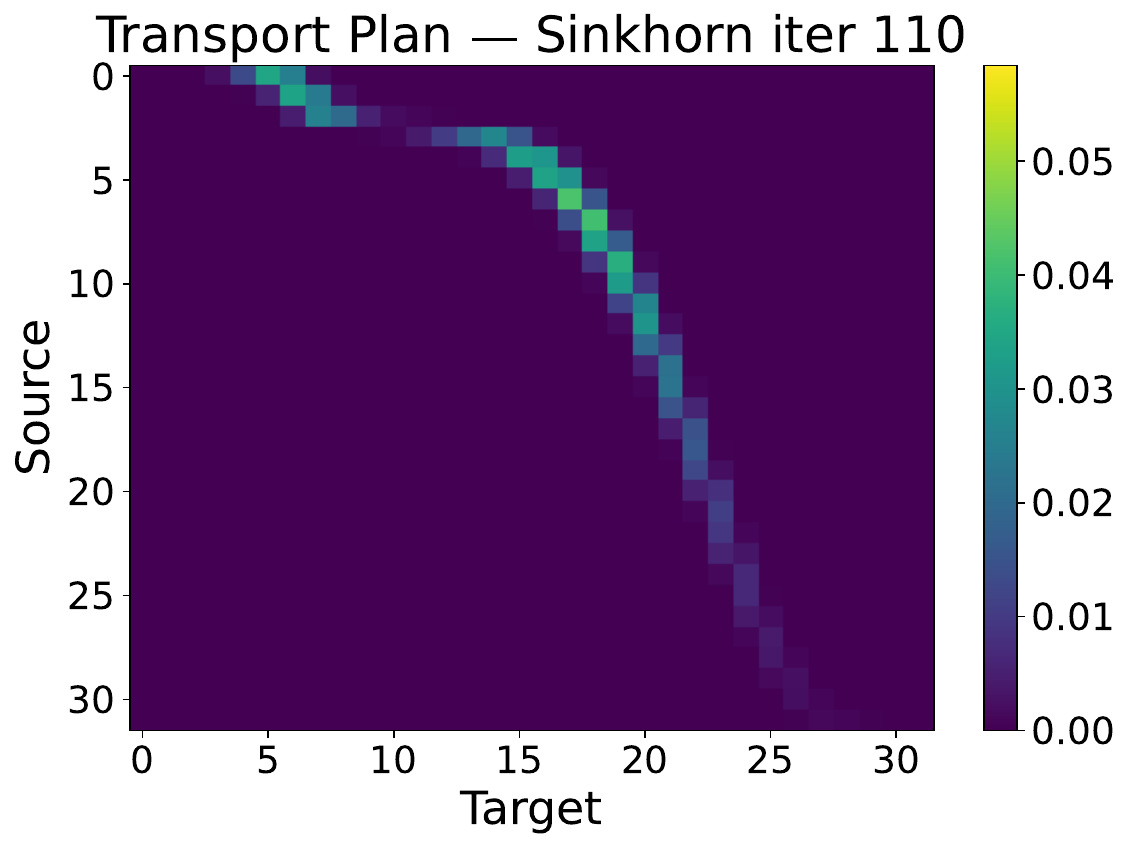}
\par\end{centering}
\caption{\label{fig:evolution_plan}Evolution of the transport plan during
Sinkhorn iterations, based on the Synthetic II problem introduced
in Section \ref{subsec:synthetic} with $n=m=32$ and $\eta=0.001$.}
\end{figure}

To support the claim in Section \ref{subsec:amortized} that the sparsity
pattern of the $T(x)$ matrix changes slowly across iterations, below
we use a motivating example to visualize the evolution of the transport
plan $T(x)$ in a realization of the Sinkhorn algorithm. The OT task
is defined by the Synthetic II problem introduced in Section \ref{subsec:synthetic},
with the problem size $n=m=32$ and a regularization parameter $\eta=0.001$.
We compute the $T(x)$ matrix every ten iterations, and visualize
it using a heatmap, as illustrated in Figure \ref{fig:evolution_plan}.
It can be observed that the top-$k$ entries of $T(x)$ show smooth
changes across iterations, suggesting that we can reuse the sparsity
pattern to reduce symbolic analysis overhead.

\subsection{Preprocessing of the CIFAR-10 Data}

\label{subsec:cifar10_preprocess}

\paragraph{Problem construction}

For each benchmark instance, we select two classes from CIFAR-10 as
the source and target distributions. For example, we transport between
class 0 (airplane) and class 1 (automobile), or class 3 (cat) and
class 5 (dog). This creates problems with $n=m=5000$, where each
point represents an image.

\paragraph{Feature extraction}

Directly using raw pixel values ($32\times32\times3=3072$ dimensions)
for distance computation is inadequate, as pixel-wise Euclidean distance
does not capture semantic similarity between images. Instead, we extract
deep features using a pretrained ResNet-18 model \citep{he2016deep}
trained on the ImageNet dataset \citep{deng2009imagenet}. Specifically:
\begin{enumerate}
\item Each CIFAR-10 image is upsampled from $32\times32\times3$ to $224\times224\times3$
to match the ImageNet input size expected by ResNet-18.
\item Images are normalized using ImageNet's mean and standard deviation.
\item We remove the final classification layer of ResNet-18 and extract
the 512-dimensional feature vectors from the penultimate layer (after
global average pooling).
\end{enumerate}
This feature extraction pipeline produces semantically meaningful
embeddings where distances reflect perceptual similarity between images.

\paragraph{Cost matrix construction}

For each pair of source and target images, we compute the squared
Euclidean distance between their feature vectors to construct the
cost matrix $M\in\mathbb{R}^{5000\times5000}$. Both marginal distributions
$a$ and $b$ are set to uniform.

\subsection{Additional Test Cases}

\begin{figure}[h]
\centering
\begin{centering}
\includegraphics[width=0.328\textwidth]{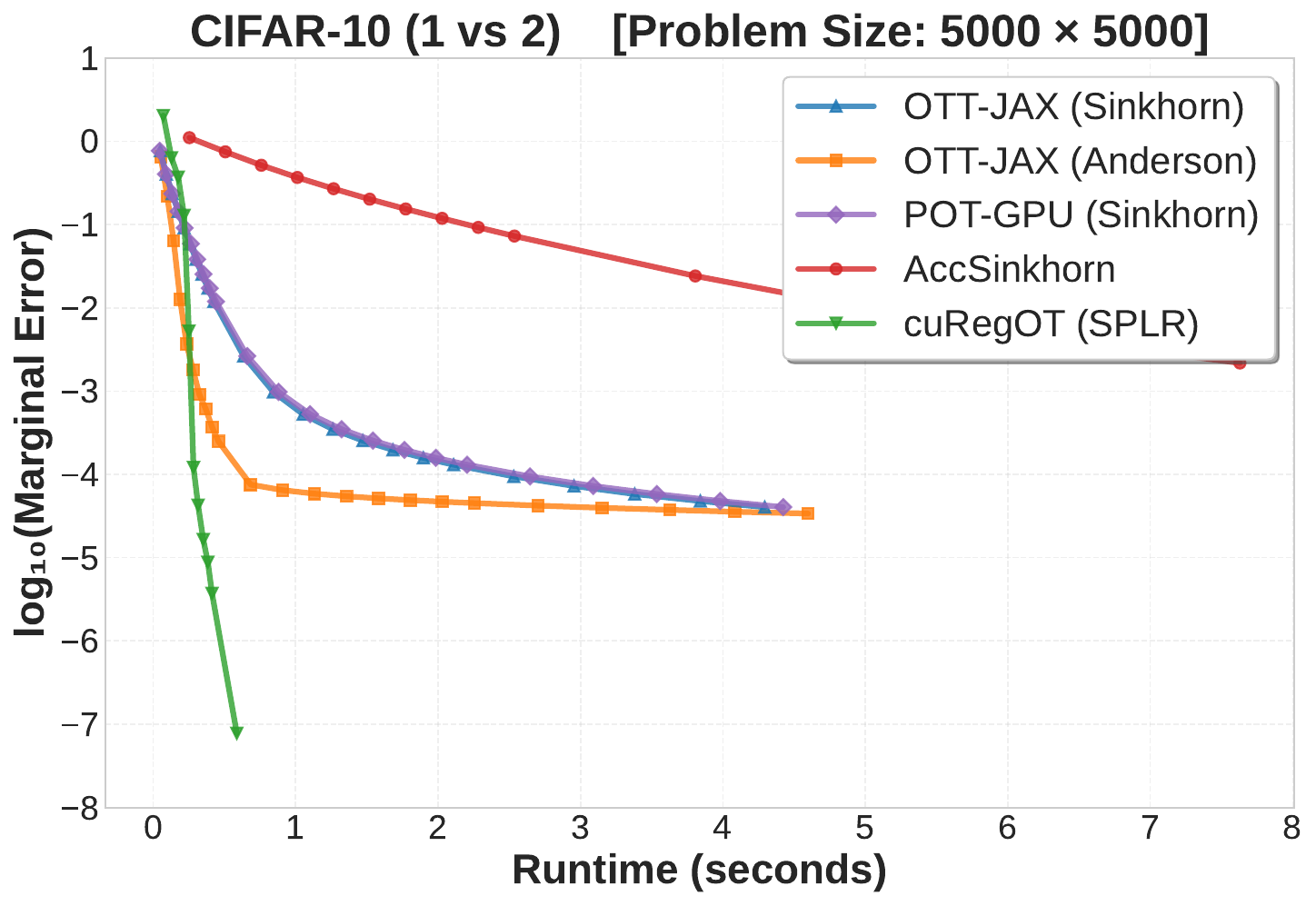}
\includegraphics[width=0.328\textwidth]{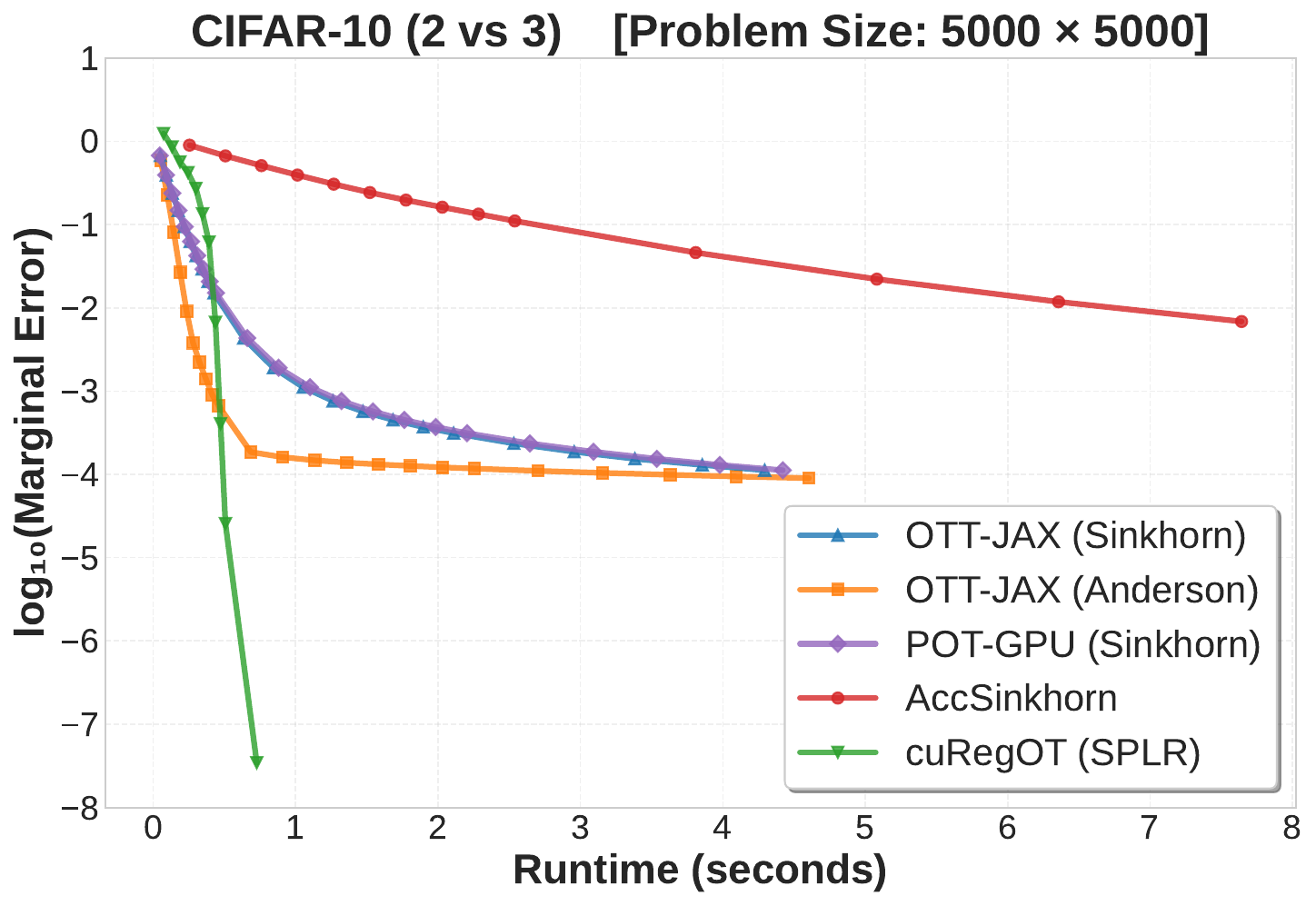}
\includegraphics[width=0.328\textwidth]{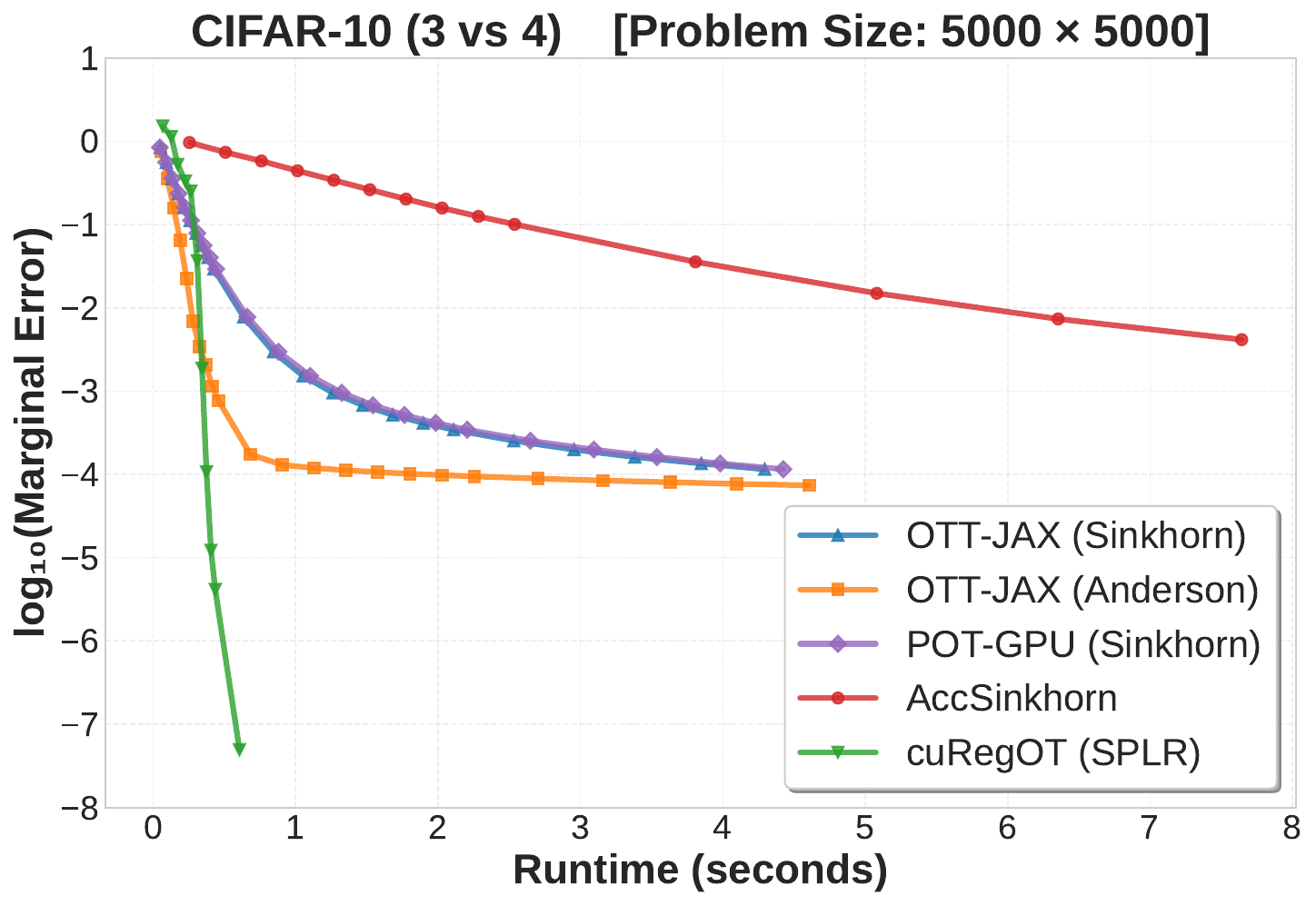}
\par\end{centering}
\begin{centering}
\includegraphics[width=0.328\textwidth]{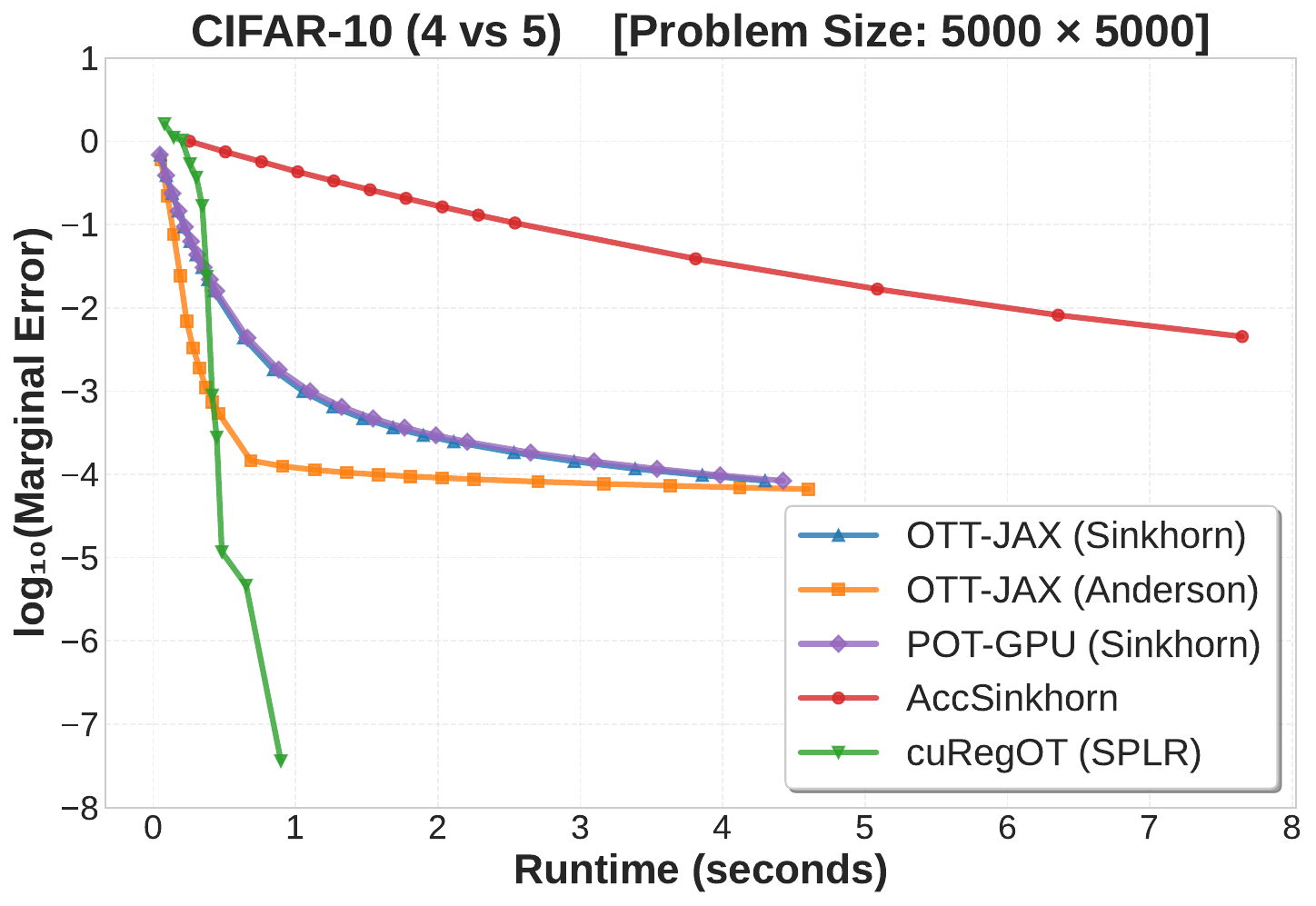}
\includegraphics[width=0.328\textwidth]{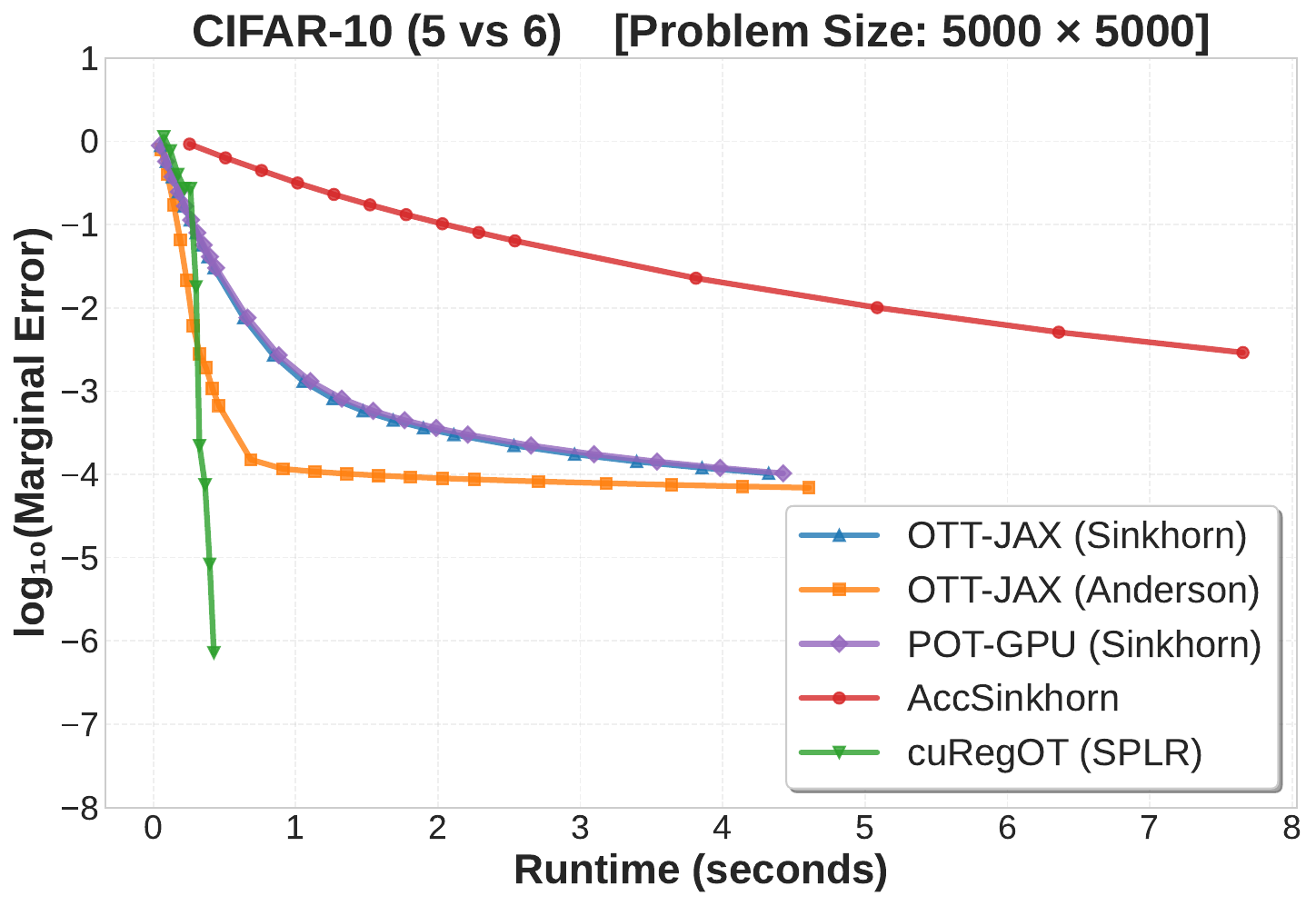}
\includegraphics[width=0.328\textwidth]{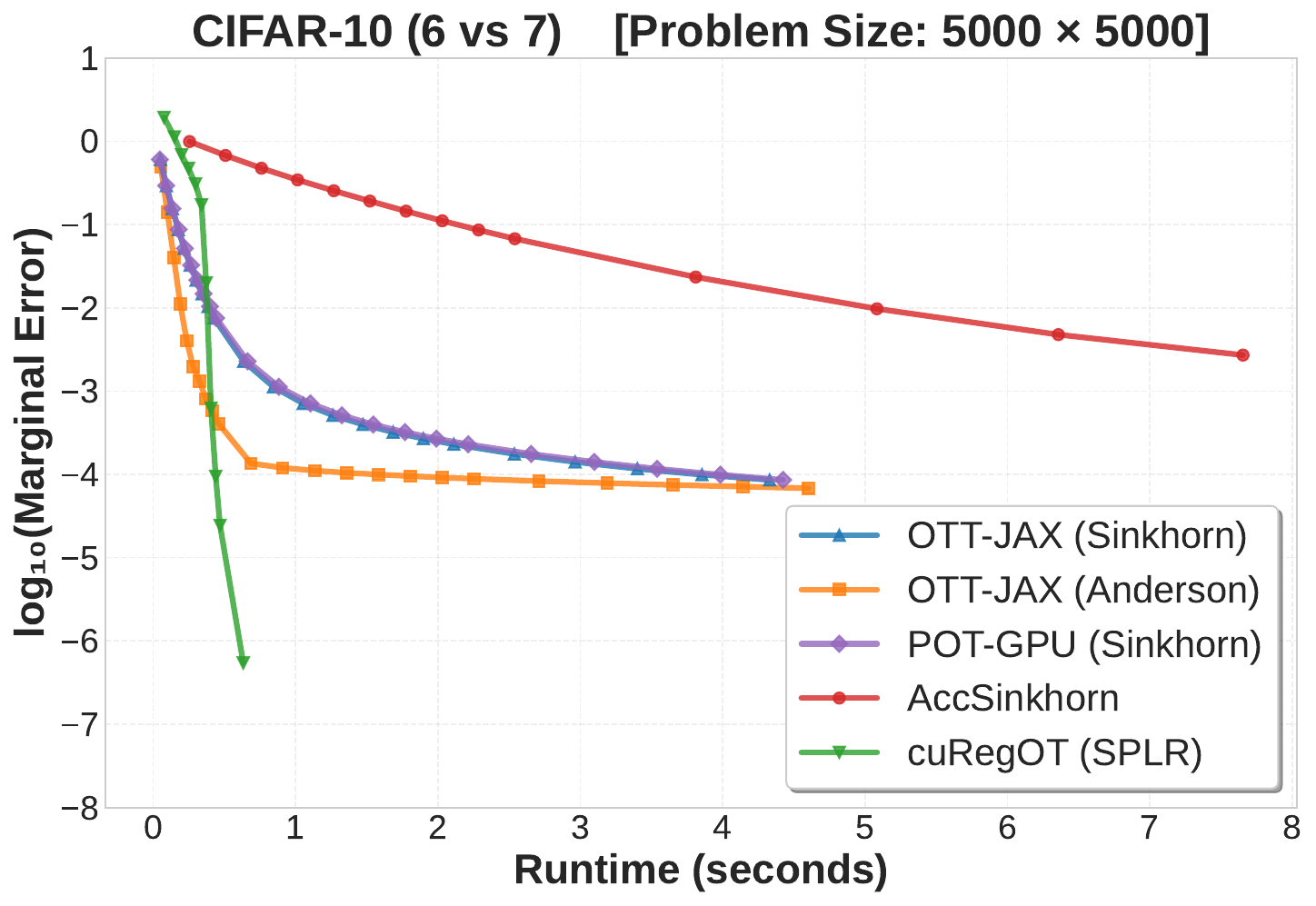}
\par\end{centering}
\begin{centering}
\includegraphics[width=0.328\textwidth]{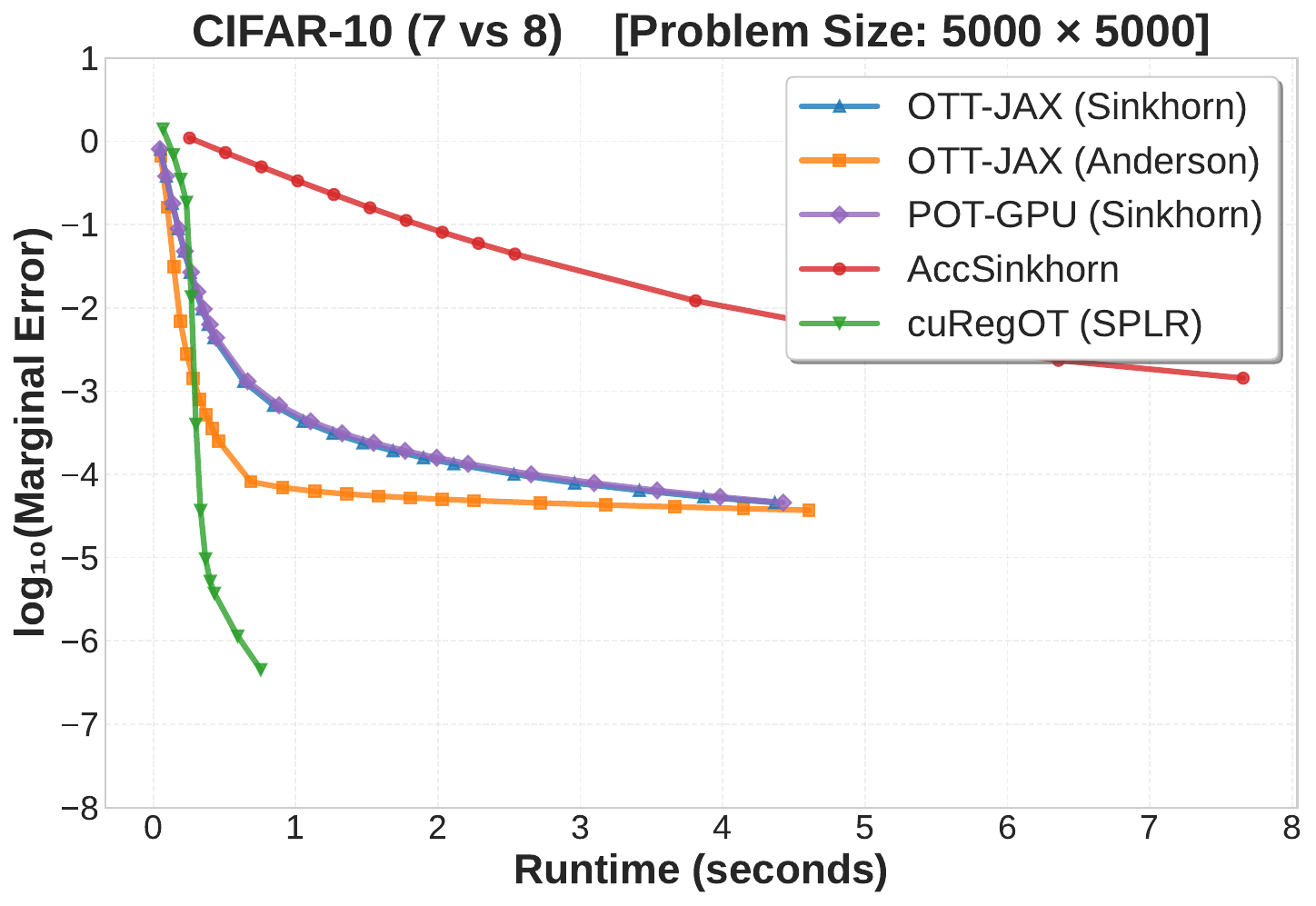}
\includegraphics[width=0.328\textwidth]{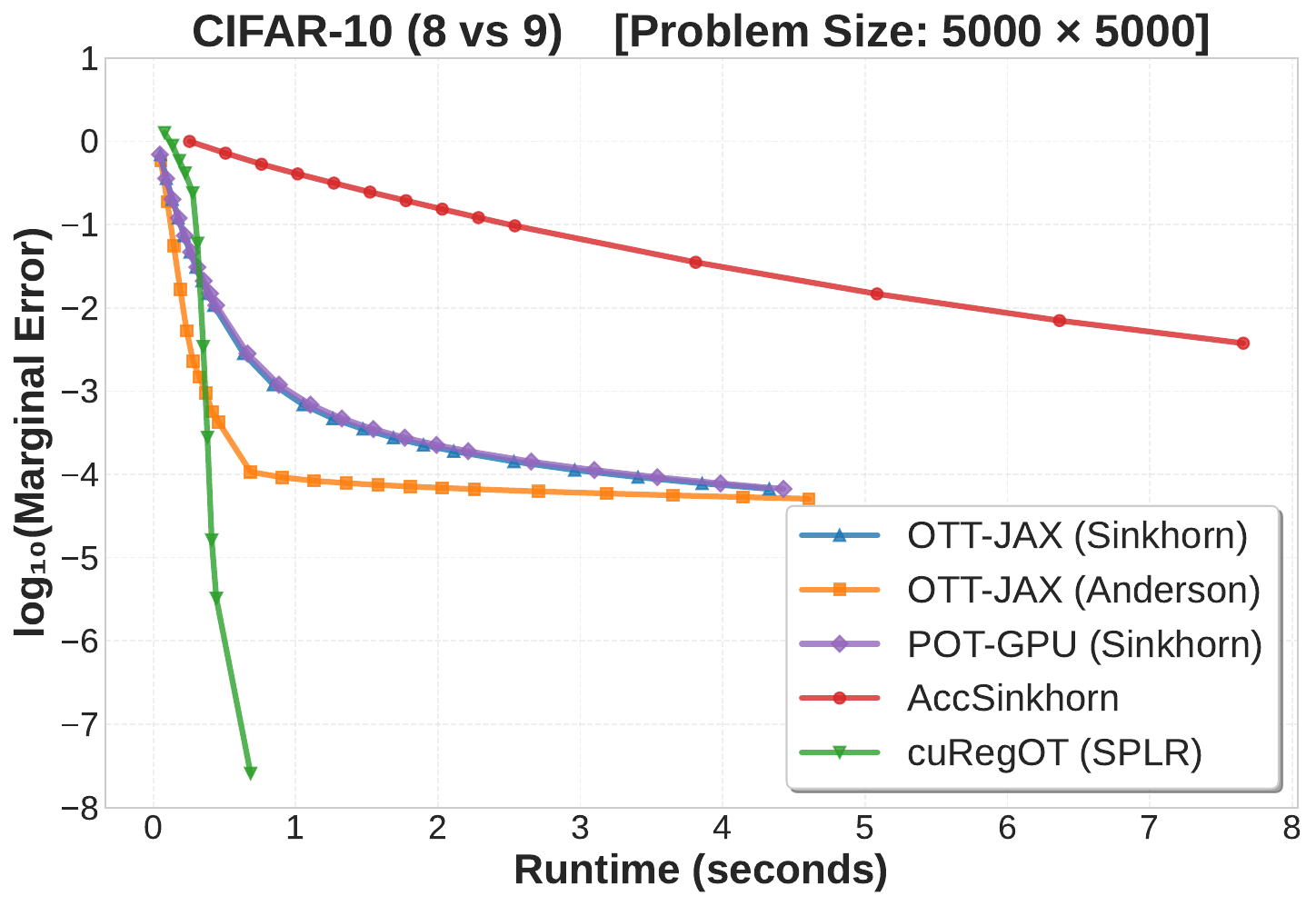}
\includegraphics[width=0.328\textwidth]{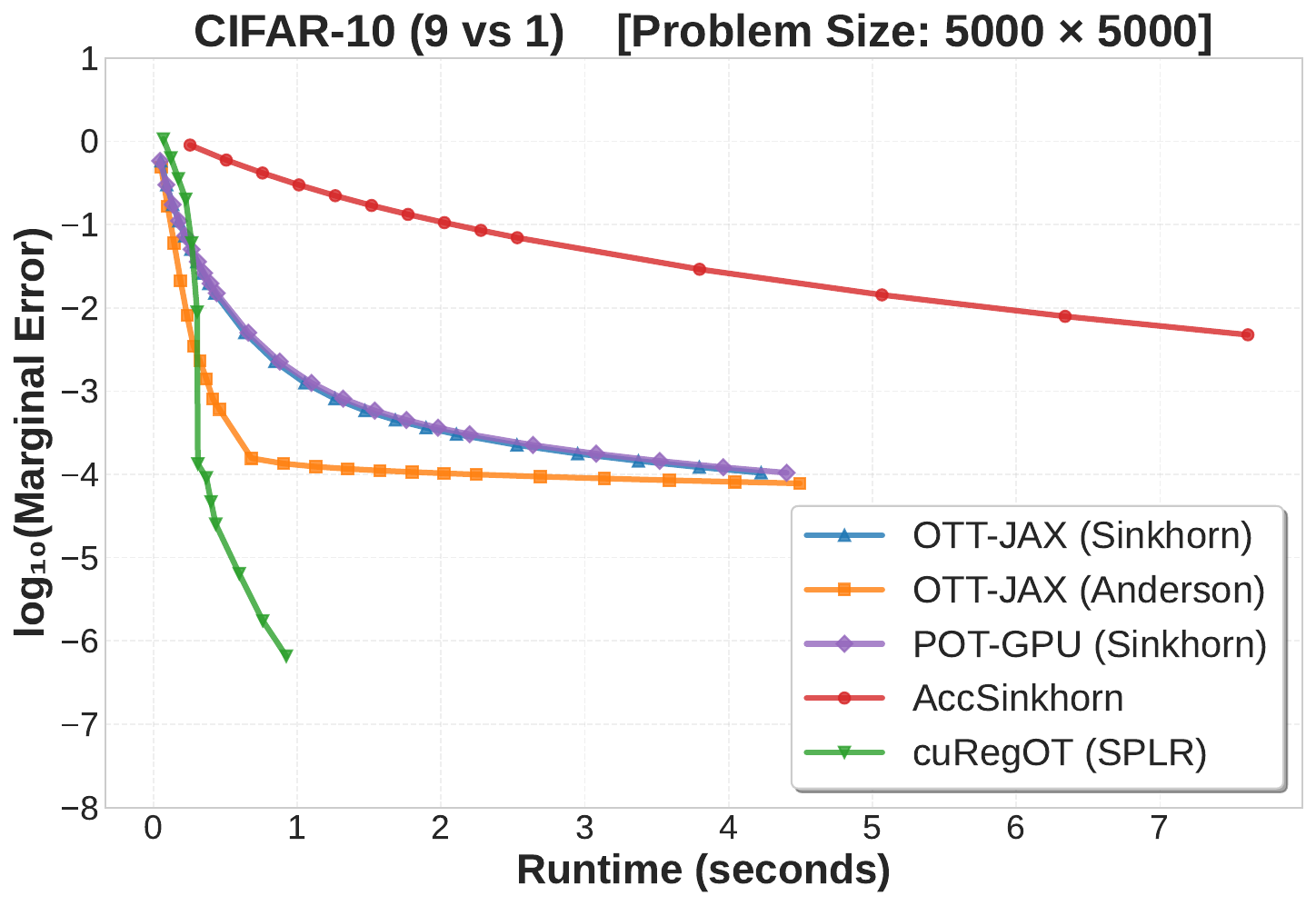}
\par\end{centering}
\caption{\label{fig:additional_experiments}Benchmark result for the CIFAR-10
dataset with additional class pairs.}
\end{figure}

Figure \ref{fig:additional_experiments} shows the benchmark result
on more test cases using the CIFAR-10 dataset.

\subsection{Using Duality Gap as the Optimization Error Metric}

\label{subsec:duality_gap}

In the numerical experiments, we have used the marginal error to quantify
the optimization error. For general algorithms, only testing the marginal
error is not sufficient, as the trivial transport plan output $ab^{T}$
results in a zero marginal error. However, the methods compared in
this paper are all based on the dual problem (\ref{eq:entropic_dual}),
meaning that the output transport plan is obtained as $T_{ij}=\exp((\alpha_{i}+\beta_{j}-M_{ij})/\eta)$.
Under this special structure, the marginal error is exactly the gradient
norm of the dual problem. Therefore, testing the gradient is a valid
indicator of the optimality.

In addition to the marginal error, we also consider the duality gap
as a metric to evaluate the performance of different solvers. By definition,
the duality gap is the difference between the current primal value
$L_{p}$ and the current dual value $L_{d}$. Given dual variables
$(\alpha,\beta)$, let $T_{ij}=\exp\{\eta^{-1}(\alpha_{i}+\beta_{j}-M_{ij})\}$,
and then
\begin{align*}
L_{p} & =\langle T,M\rangle-\eta\cdot h(T)=\langle T,M\rangle+\eta\cdot\sum_{i,j}T_{ij}\cdot\log(T_{ij})-\eta\cdot\sum_{i,j}T_{ij},\\
L_{d} & =-\eta\cdot\sum_{ij}T_{ij}+\alpha^{T}a+\beta^{T}b,\\
L_{p}-L_{d} & =\langle T,M\rangle+\sum_{i,j}T_{ij}(\alpha_{i}+\beta_{j}-M_{ij})-\alpha^{T}a-\beta^{T}b\\
 & =\sum_{i,j}T_{ij}(\alpha_{i}+\beta_{j})-\alpha^{T}a-\beta^{T}b=\alpha^{T}(T\mathbf{1}_{m}-a)+\beta^{T}(T^{T}\mathbf{1}_{n}-b).
\end{align*}

\begin{figure}[h]
\centering
\begin{centering}
\includegraphics[width=0.328\textwidth]{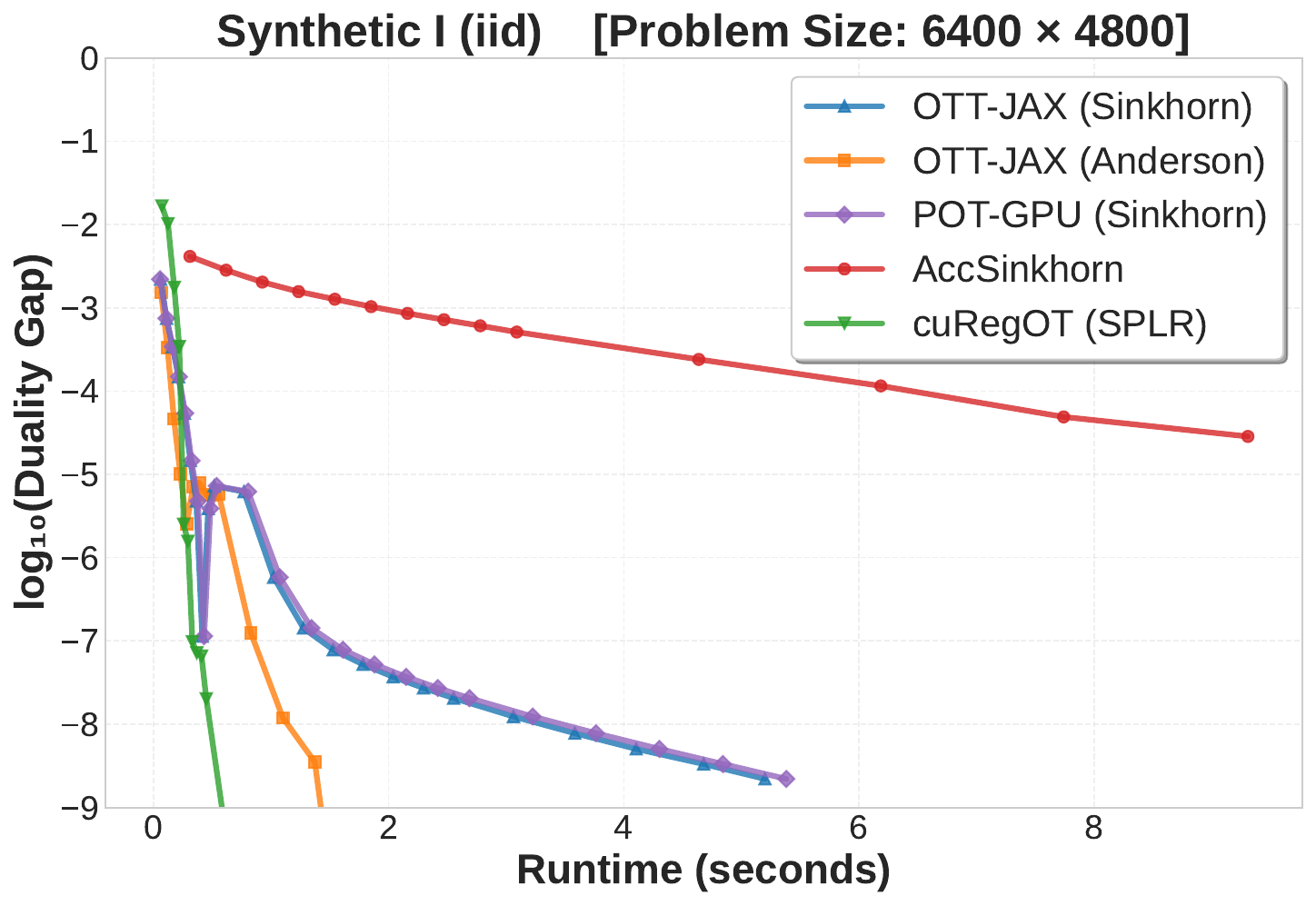}
\includegraphics[width=0.328\textwidth]{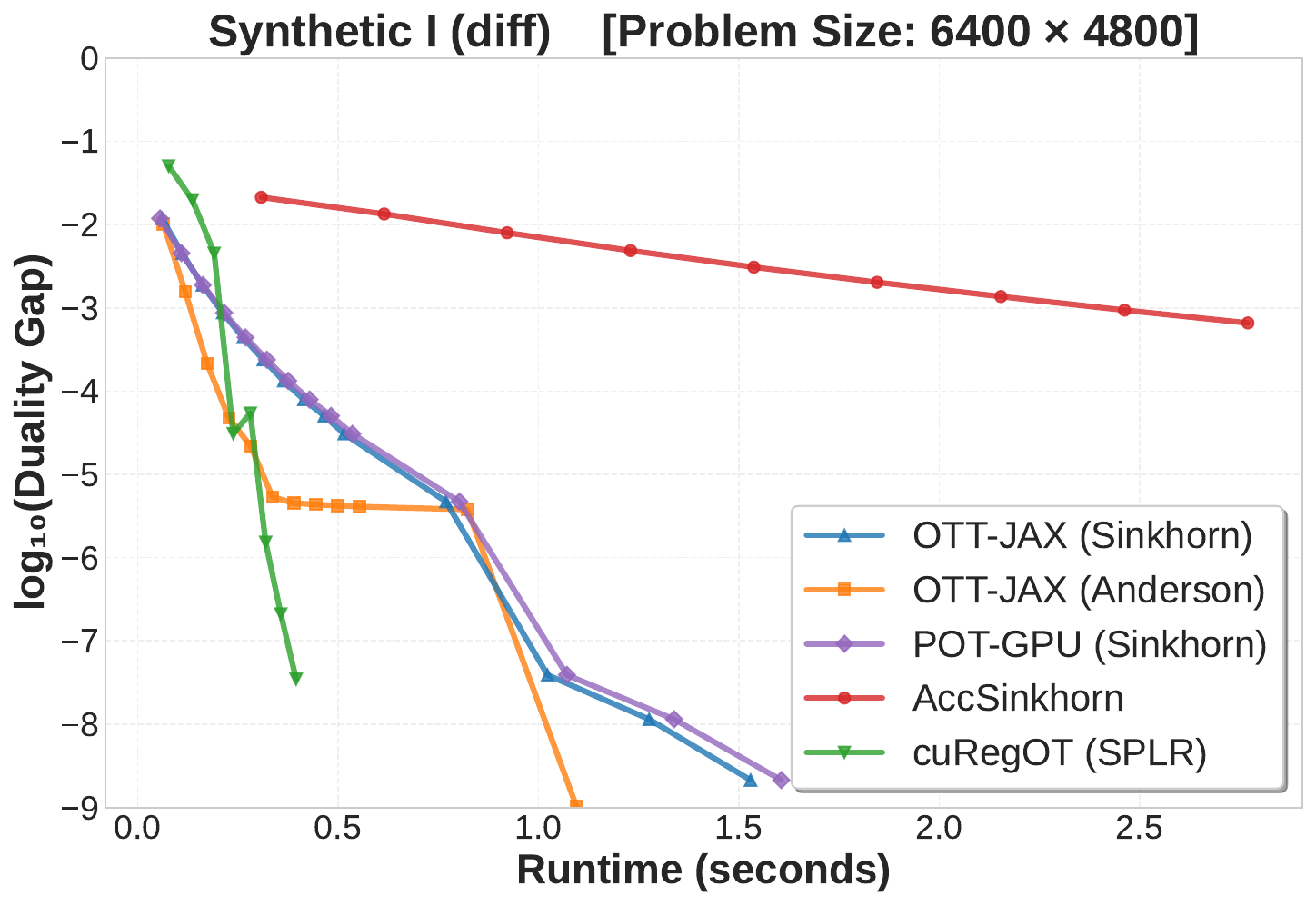}
\includegraphics[width=0.328\textwidth]{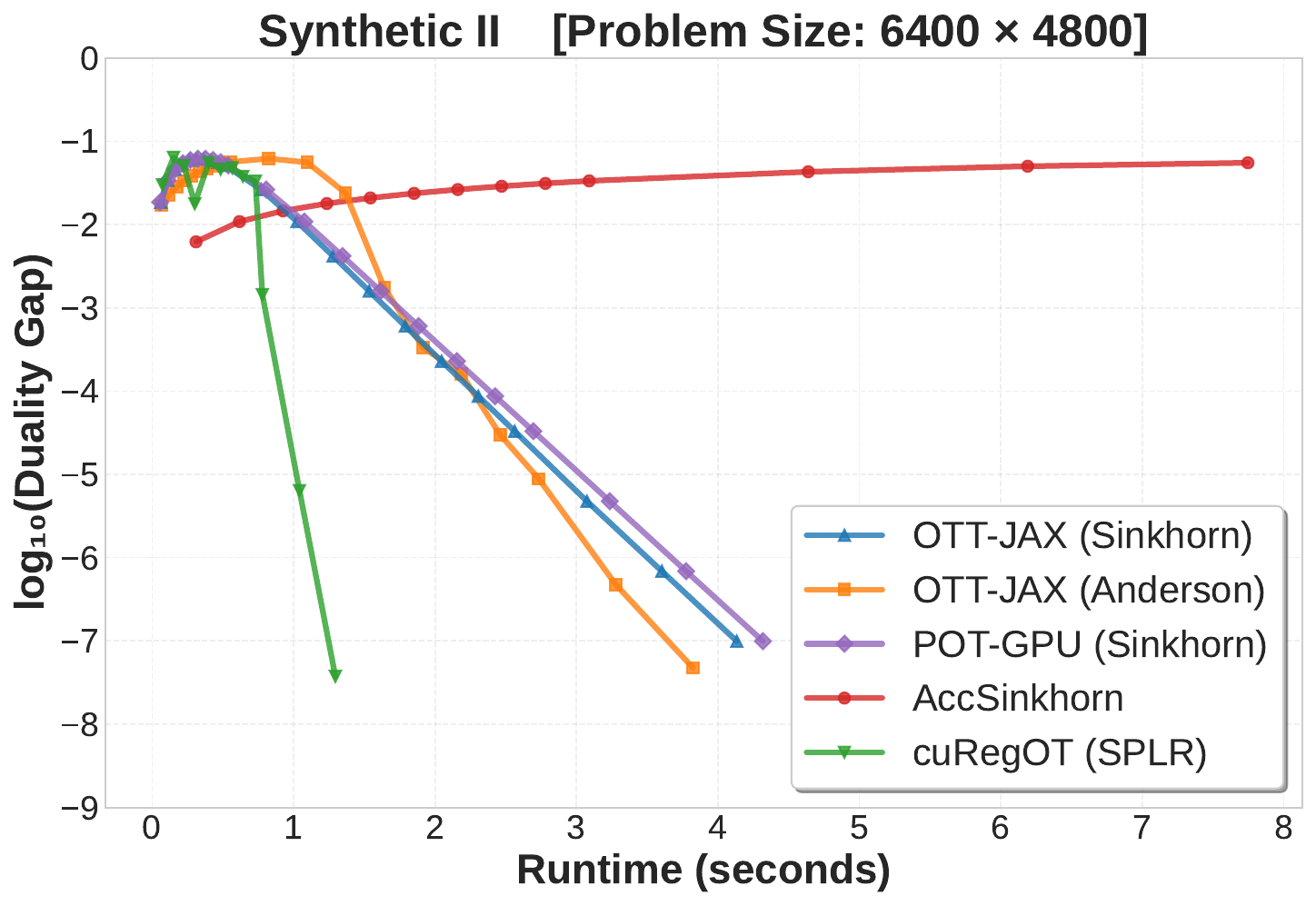}
\par\end{centering}
\begin{centering}
\includegraphics[width=0.328\textwidth]{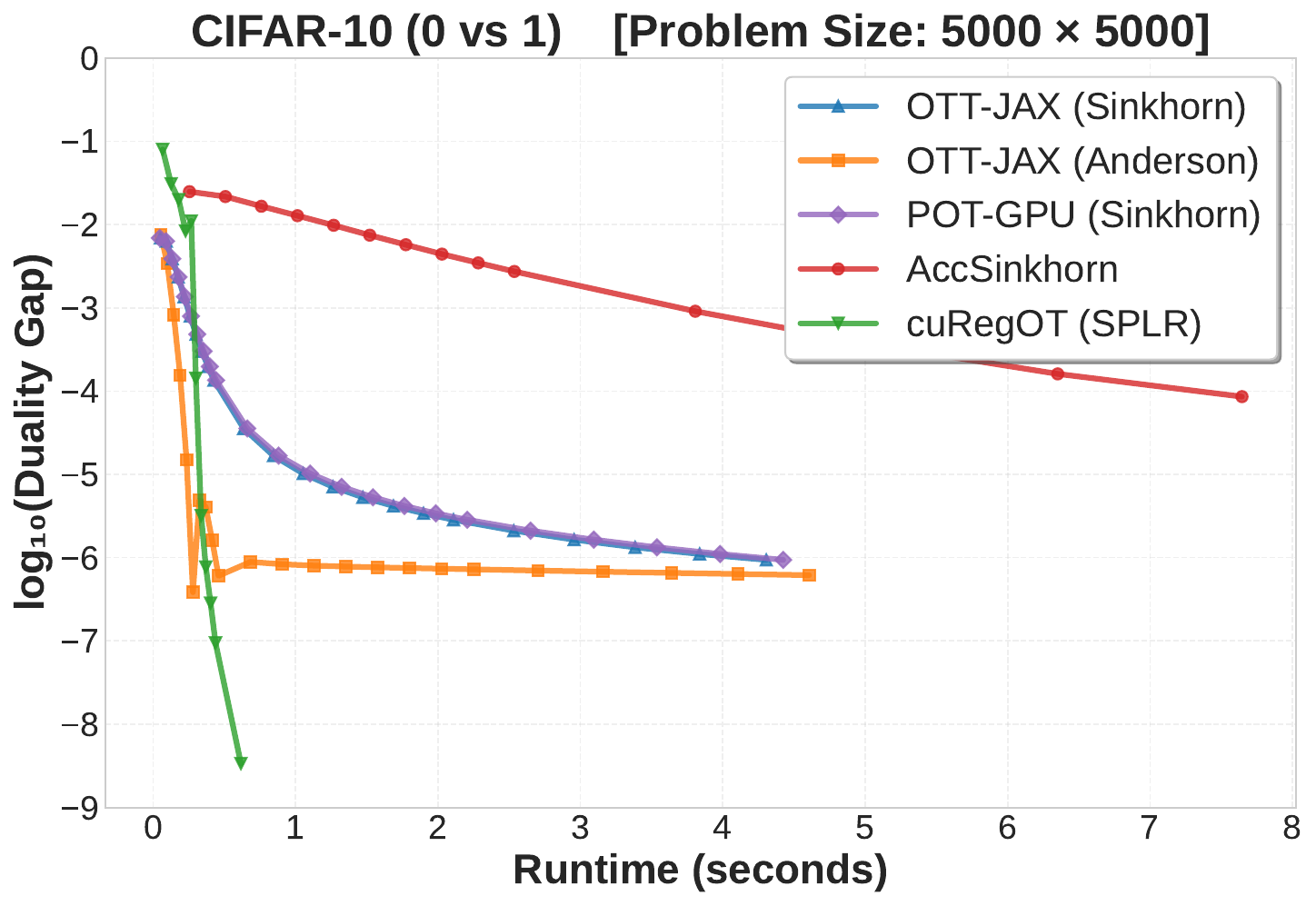}
\includegraphics[width=0.328\textwidth]{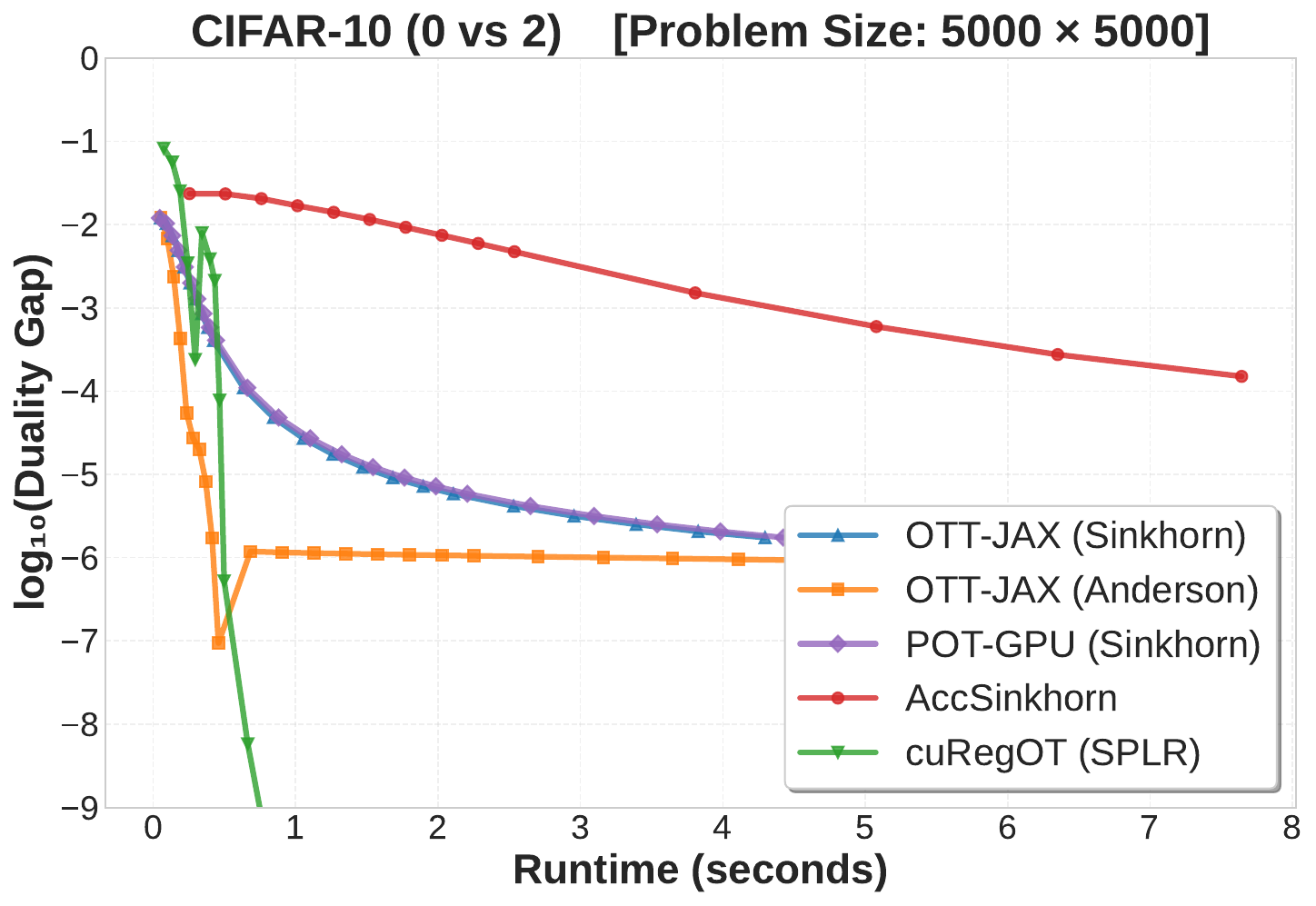}
\includegraphics[width=0.328\textwidth]{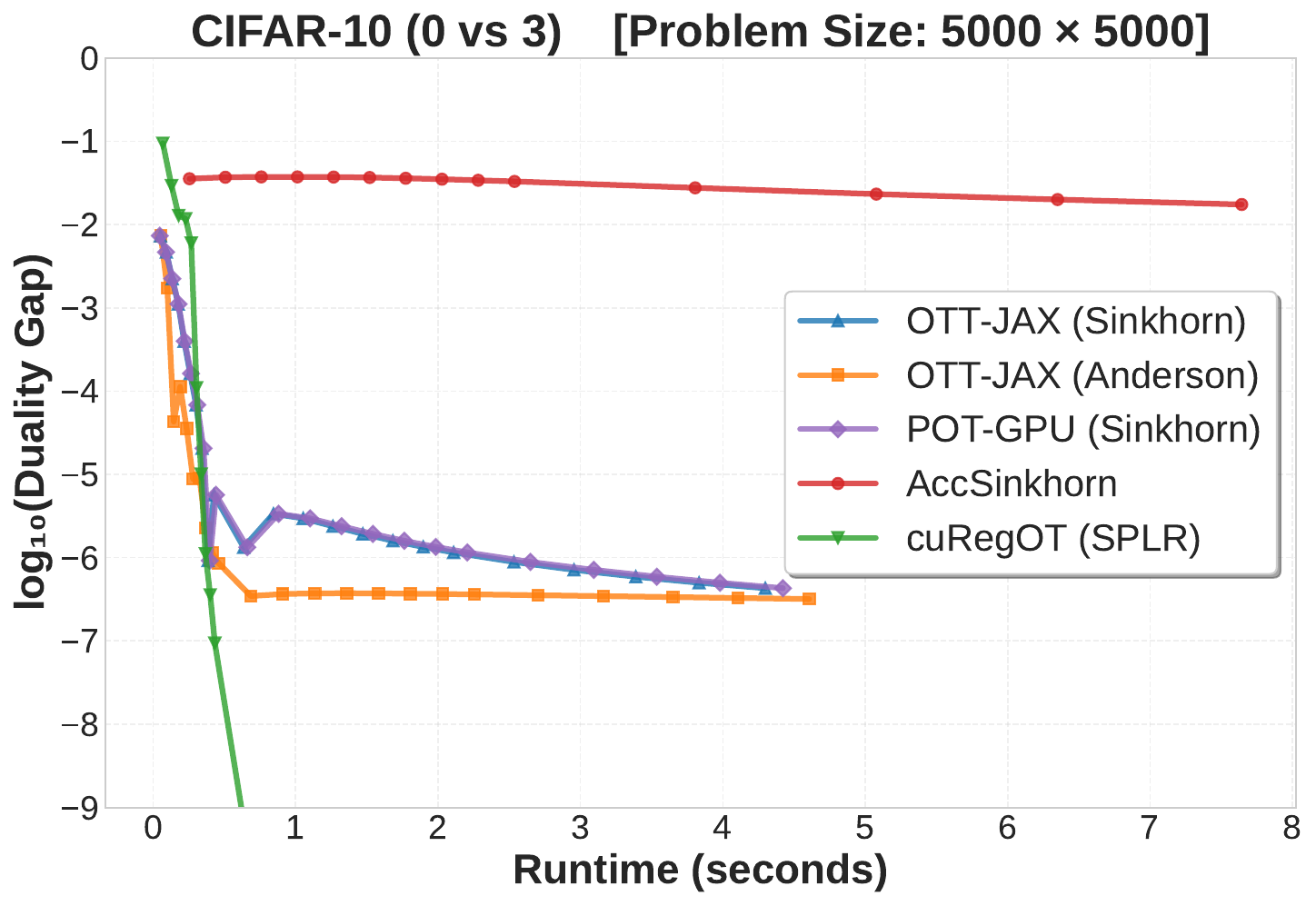}
\par\end{centering}
\caption{\label{fig:duality_gap}Benchmark result that uses the duality gap
(on a logarithmic scale) as the metric for optimization error. The
regularization parameter is set to $\eta=0.001$.}
\end{figure}

In Figure \ref{fig:duality_gap}, we use $\log_{10}|L_{p}-L_{d}|$
as the optimality metric, and plot it against the wall time. The plots
show that the duality gap demonstrates very similar patterns as the
marginal error. In fact,
\begin{align*}
|L_{p}-L_{d}| & \le\Vert\alpha\Vert_{\infty}\Vert T\mathbf{1}_{m}-a\Vert_{1}+\Vert\beta\Vert_{\infty}\Vert T^{T}\mathbf{1}_{n}-b\Vert_{1}\\
 & \le\max\{\Vert\alpha\Vert_{\infty},\Vert\beta\Vert_{\infty}\}\cdot(\Vert T\mathbf{1}_{m}-a\Vert_{1}+\Vert T^{T}\mathbf{1}_{n}-b\Vert_{1}),
\end{align*}
where $\Vert T\mathbf{1}_{m}-a\Vert_{1}+\Vert T^{T}\mathbf{1}_{n}-b\Vert_{1}$
is exactly the marginal error. Therefore, in some sense, the marginal
error provides an upper bound for the duality gap, if we assume that
$\Vert\alpha\Vert_{\infty}$ and $\Vert\beta\Vert_{\infty}$ can be
properly bounded.

\subsection{Impact of the Regularization Parameter $\eta$}

\label{subsec:different_eta}

\begin{figure}[h]
\centering
\begin{centering}
\includegraphics[width=0.328\textwidth]{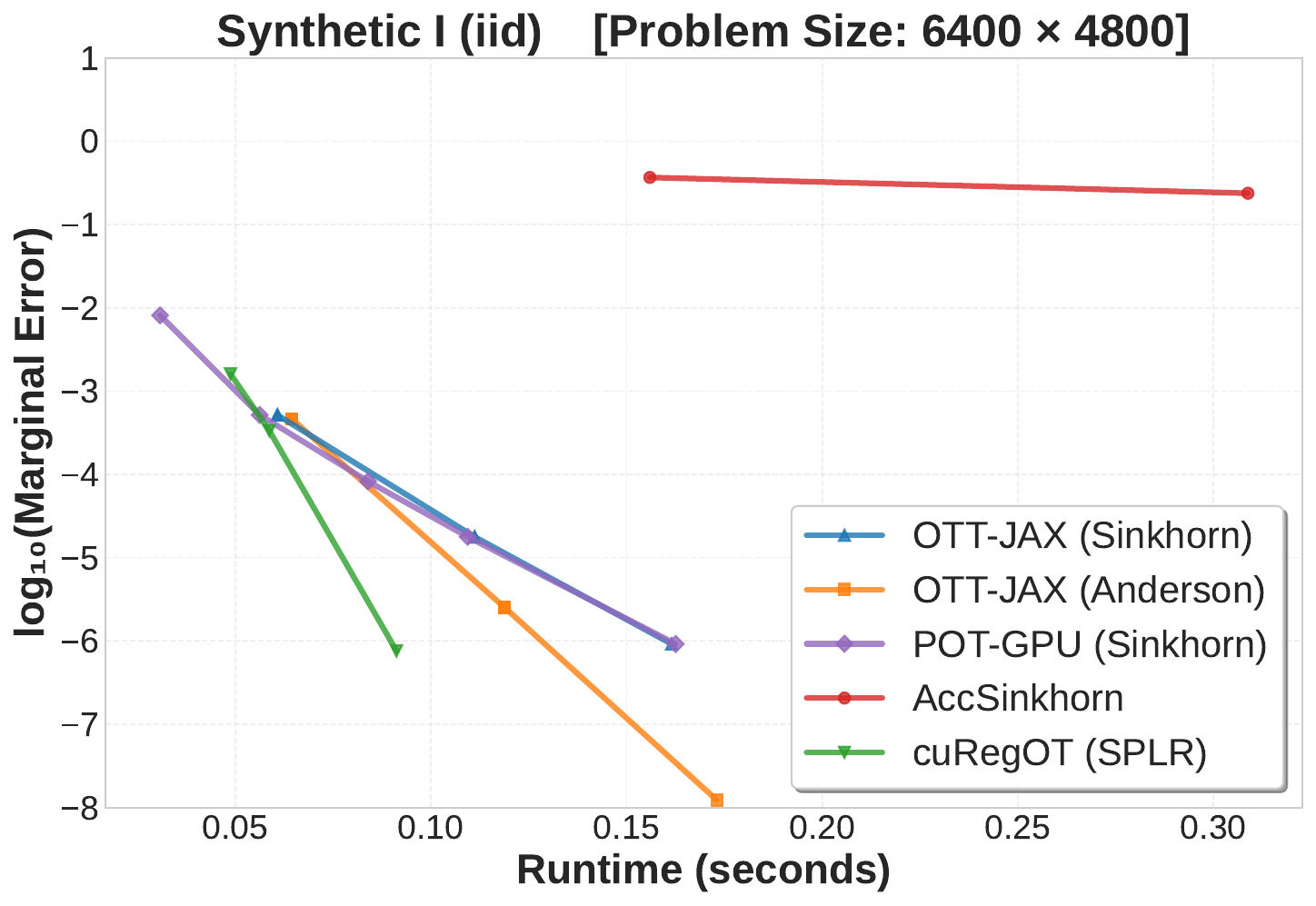}
\includegraphics[width=0.328\textwidth]{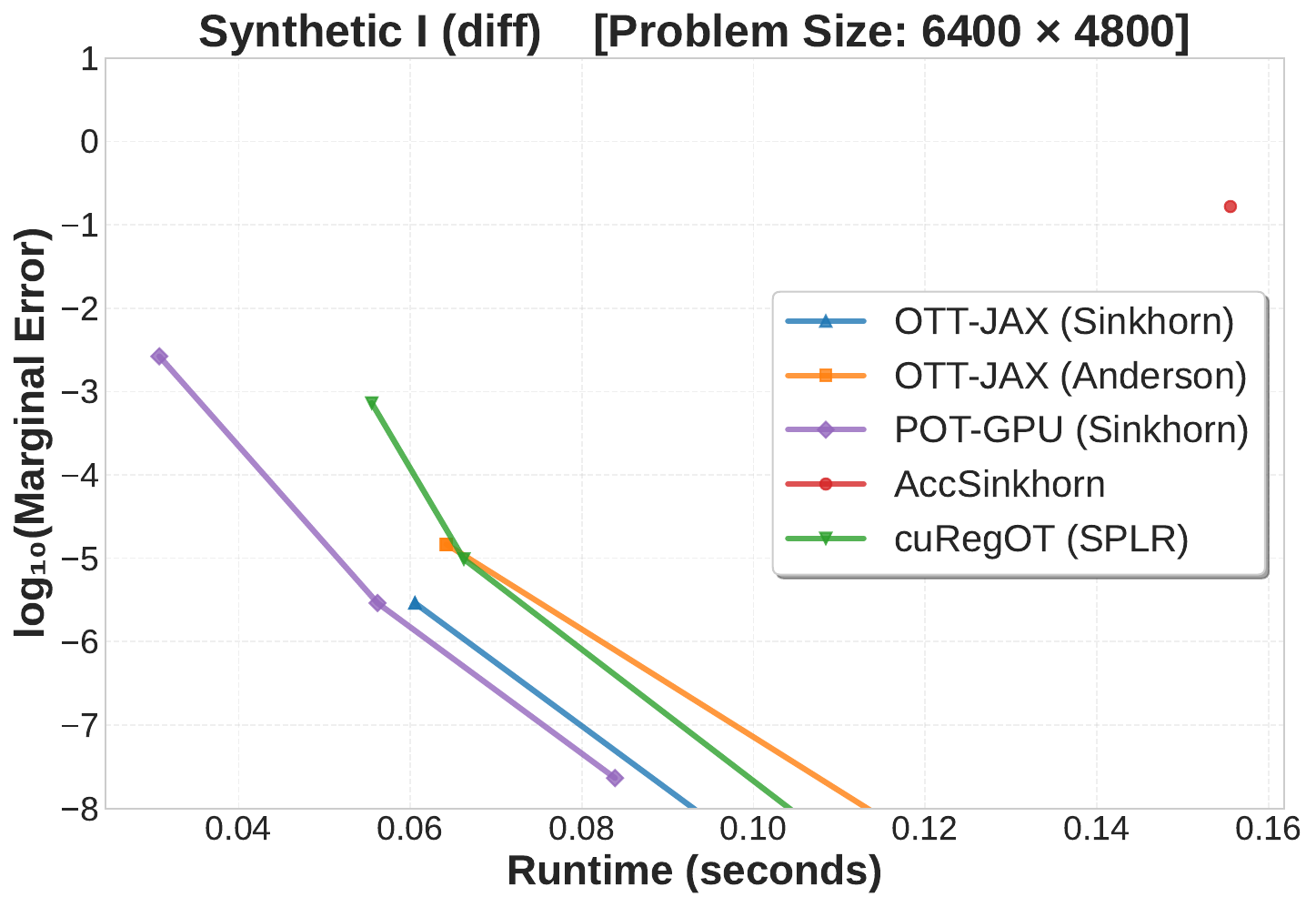}
\includegraphics[width=0.328\textwidth]{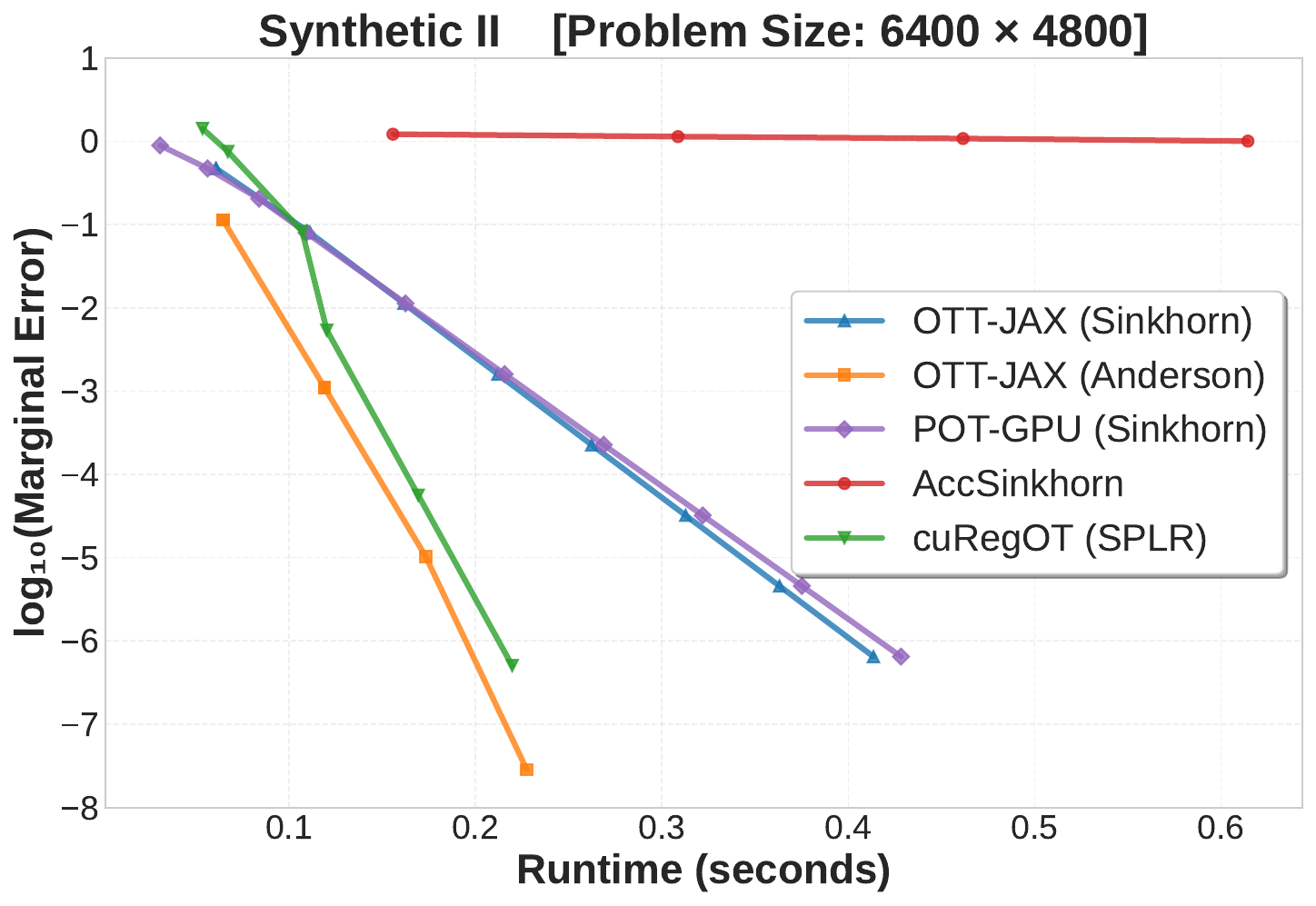}
\par\end{centering}
\begin{centering}
\includegraphics[width=0.328\textwidth]{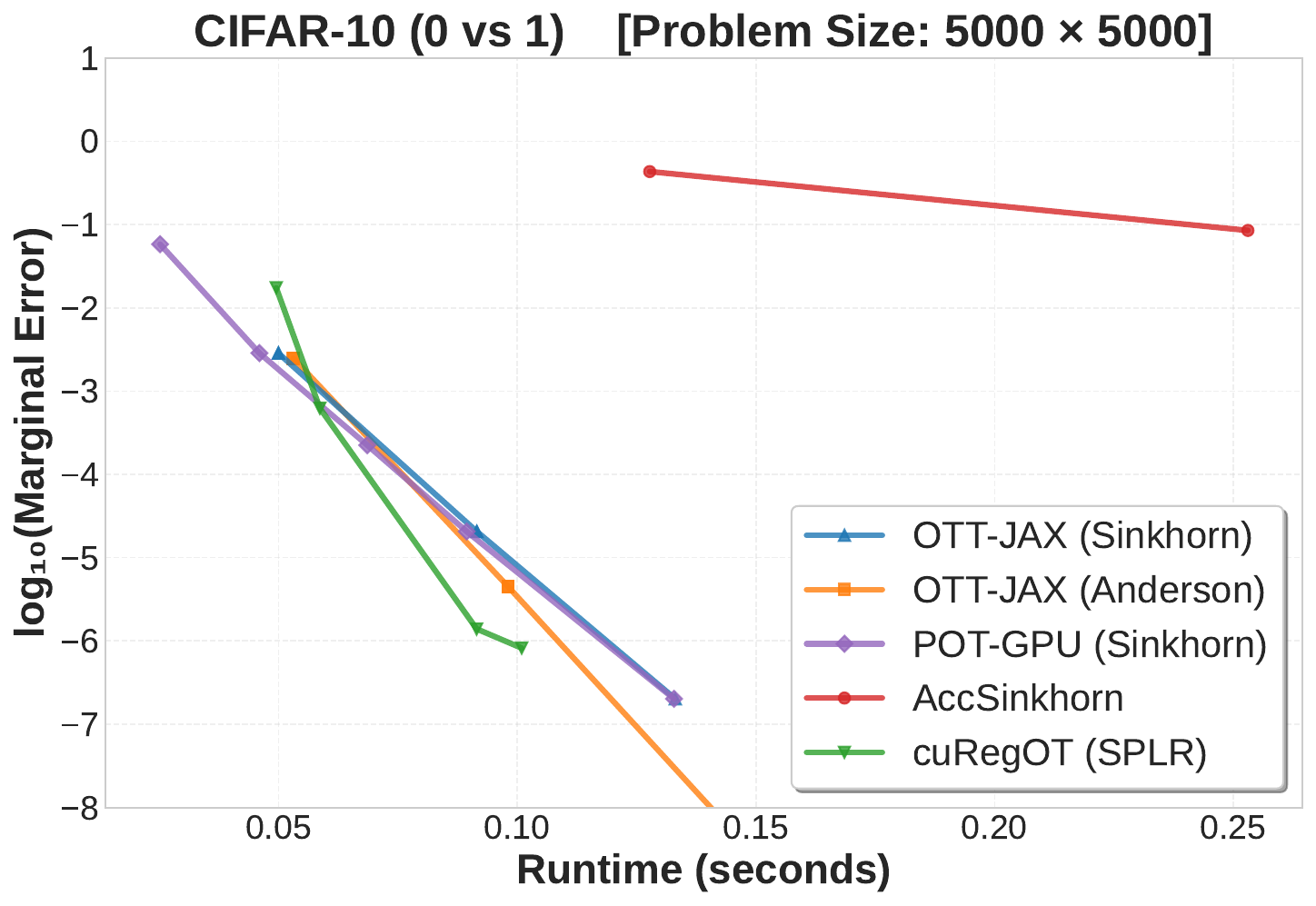}
\includegraphics[width=0.328\textwidth]{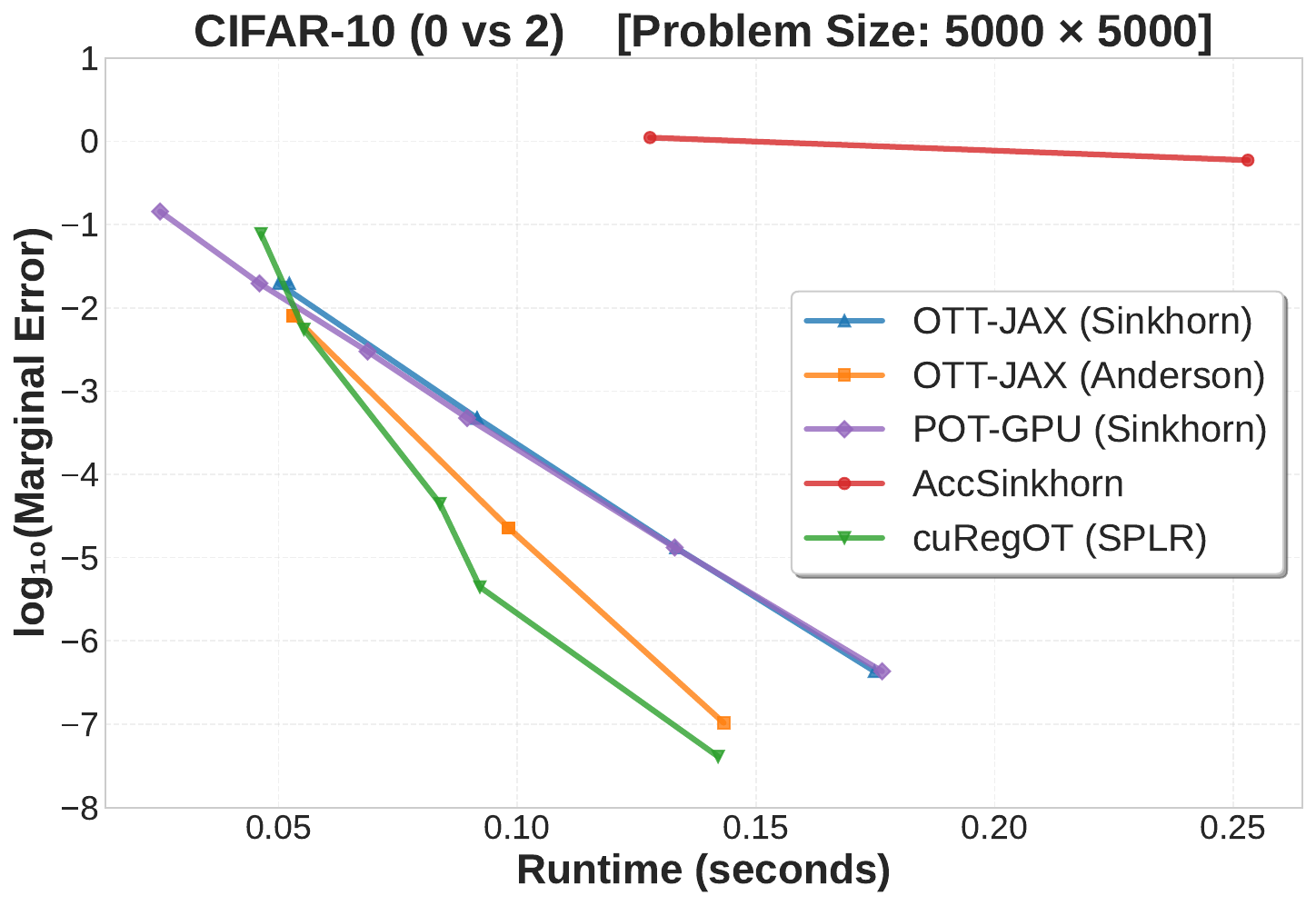}
\includegraphics[width=0.328\textwidth]{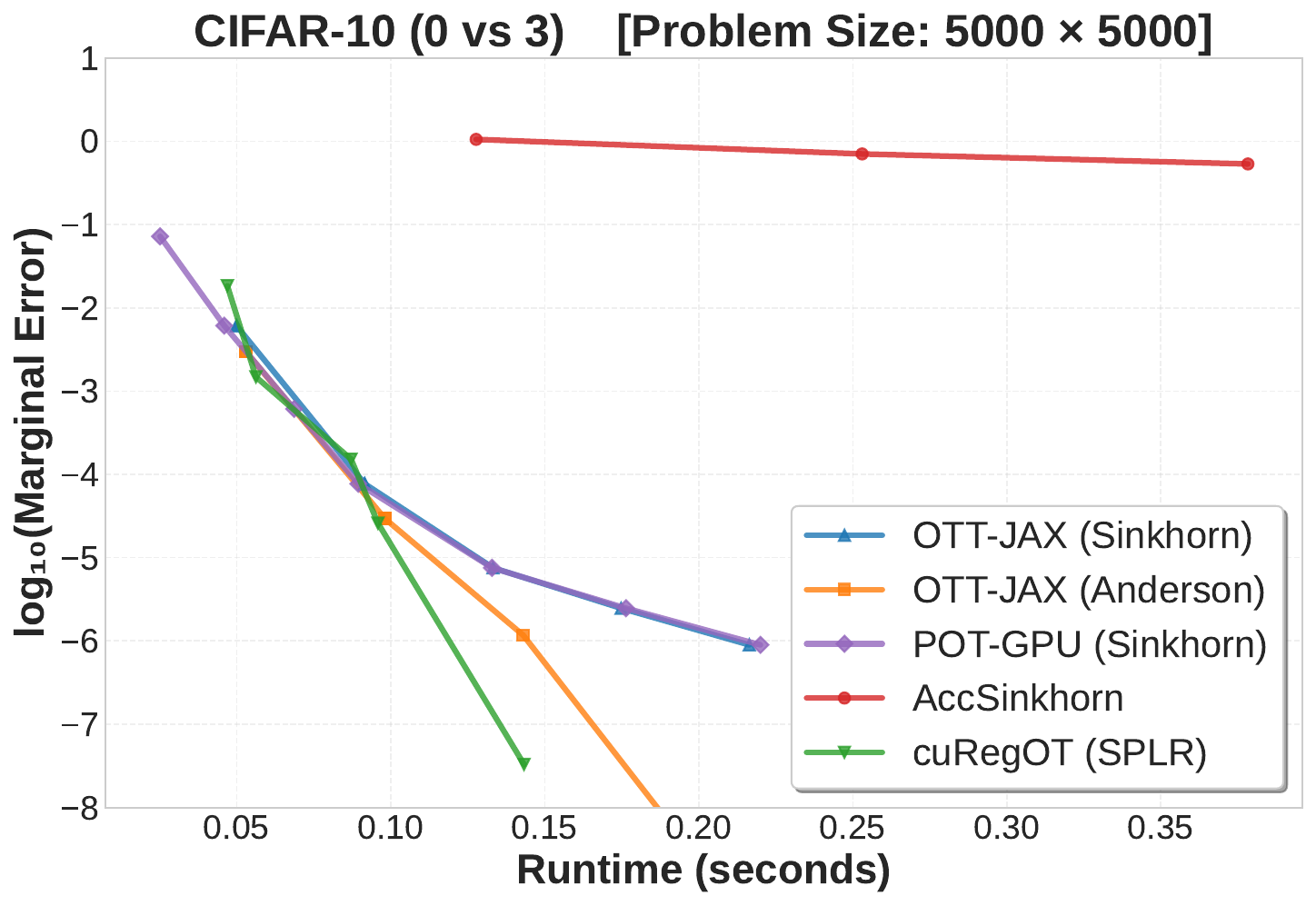}
\par\end{centering}
\caption{\label{fig:eta_0.01}Benchmark result for different solvers under
the regularization parameter $\eta=0.01$.}
\end{figure}

\begin{figure}[H]
\centering
\begin{centering}
\includegraphics[width=0.328\textwidth]{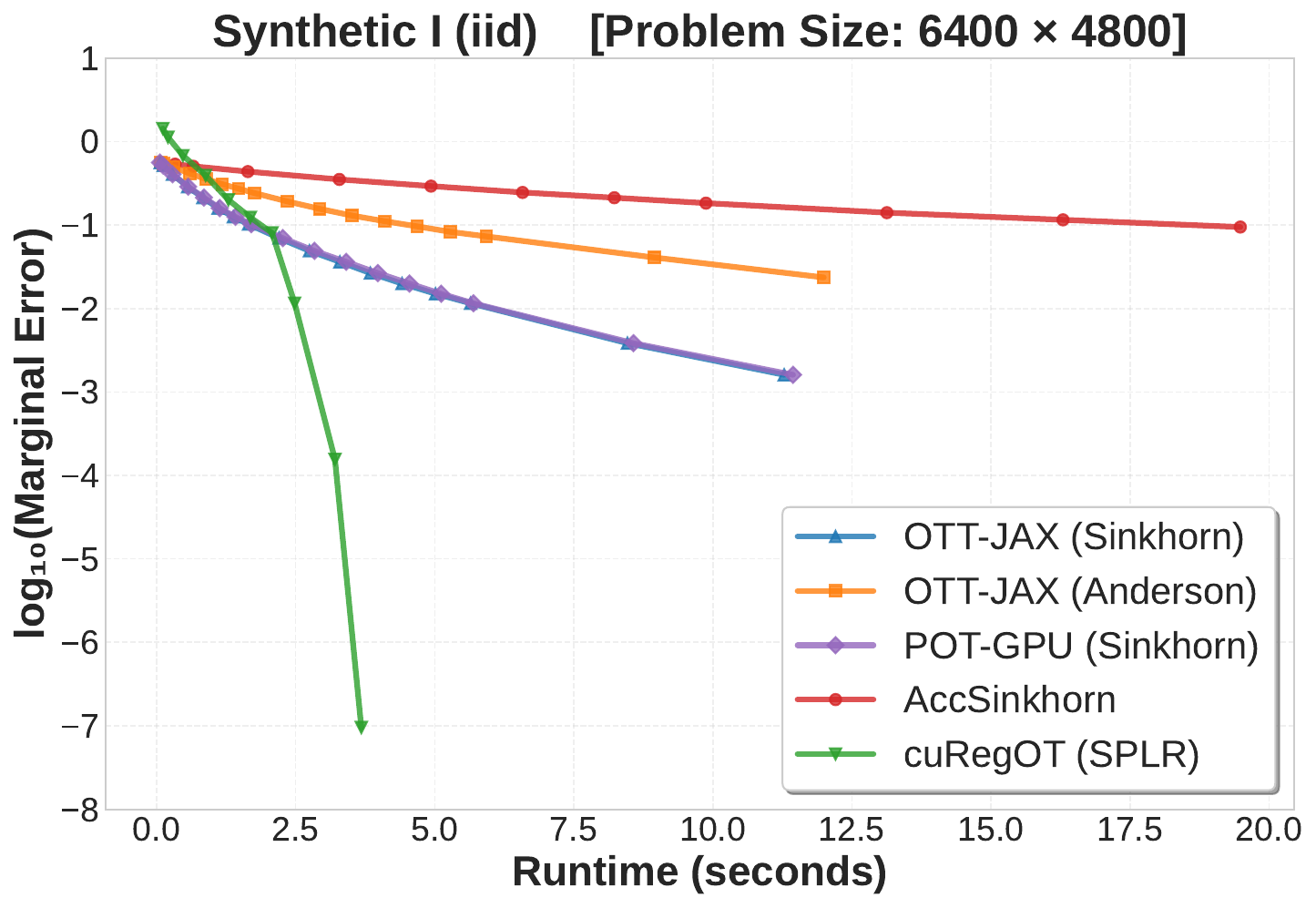}
\includegraphics[width=0.328\textwidth]{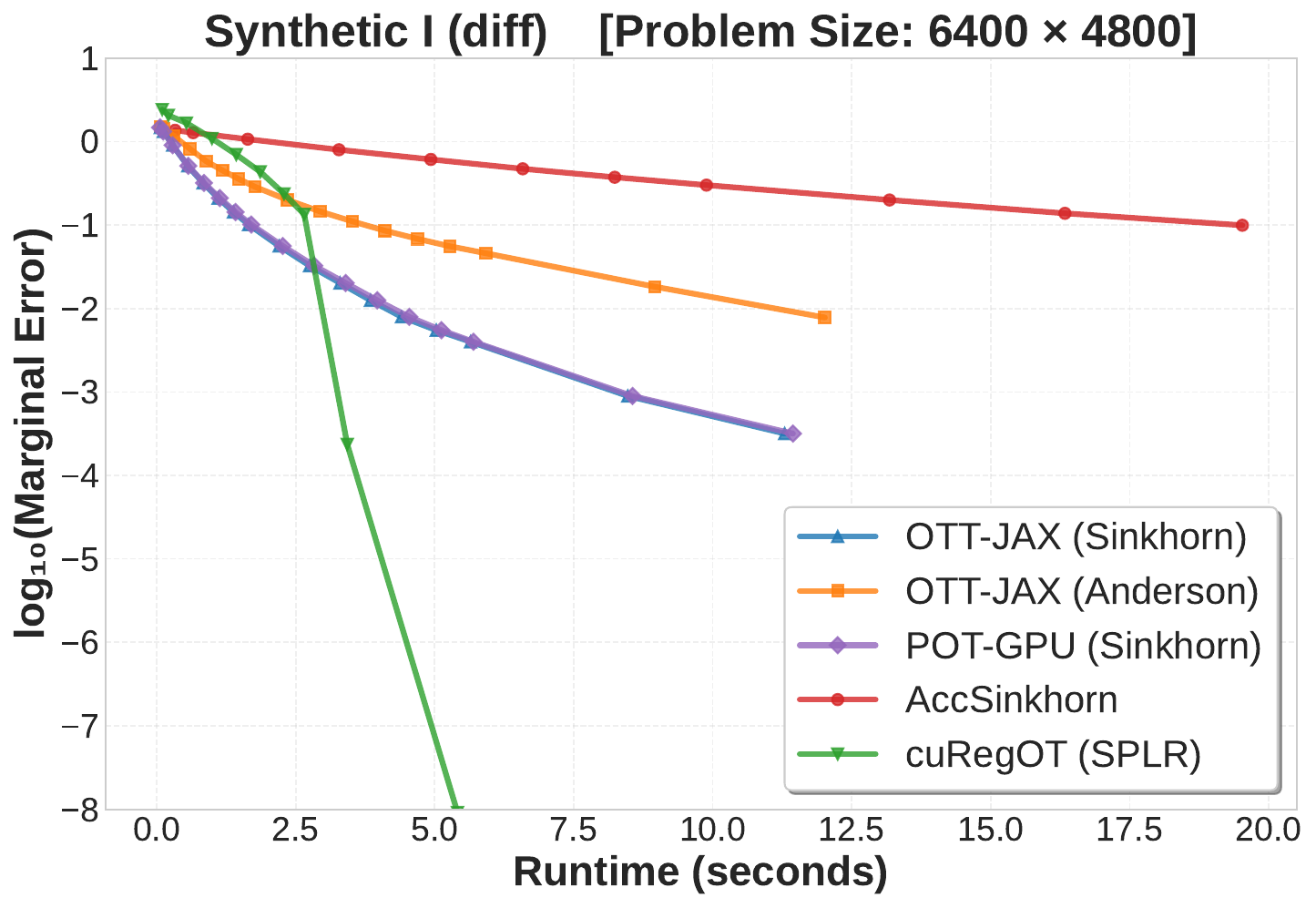}
\includegraphics[width=0.328\textwidth]{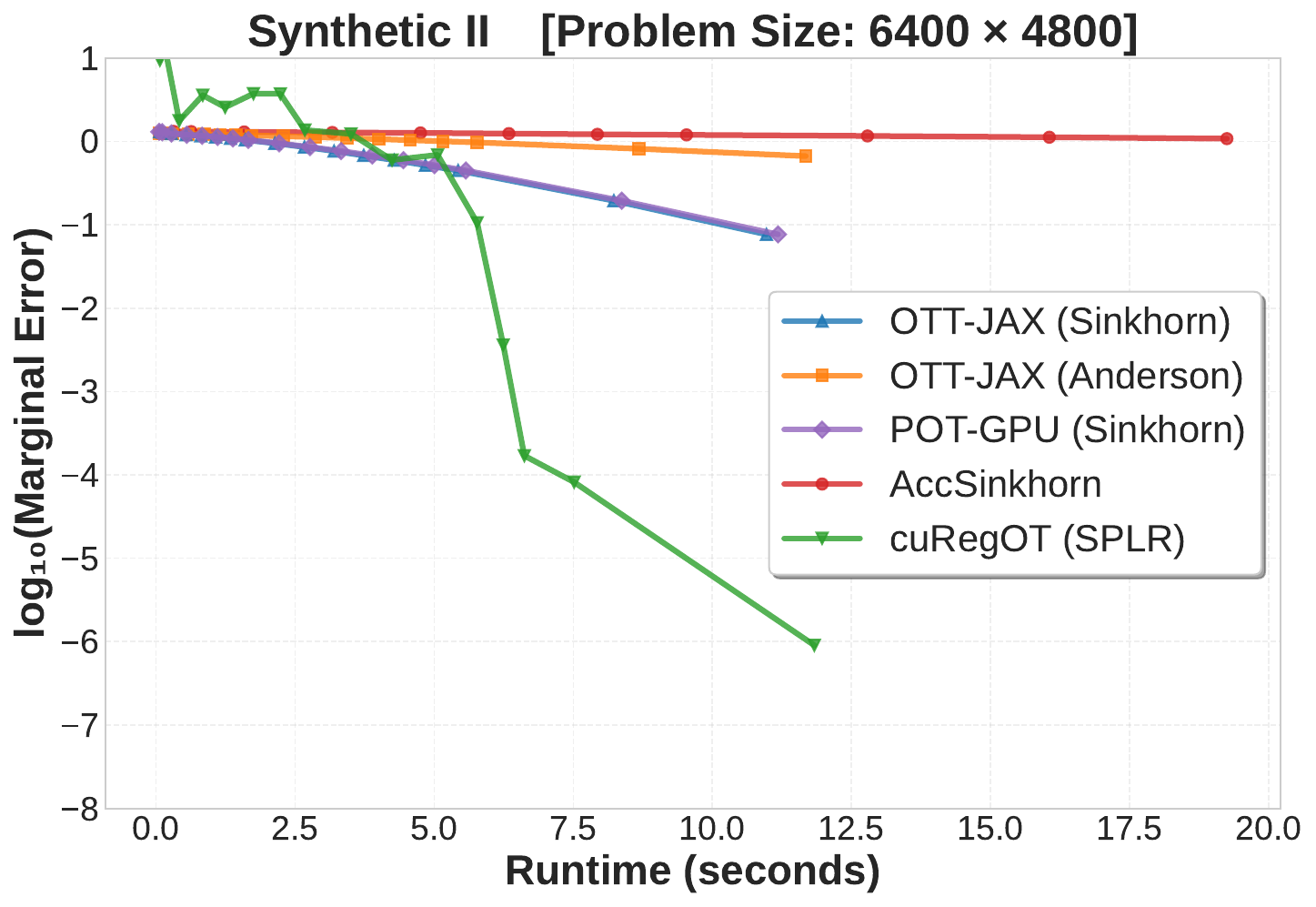}
\par\end{centering}
\begin{centering}
\includegraphics[width=0.328\textwidth]{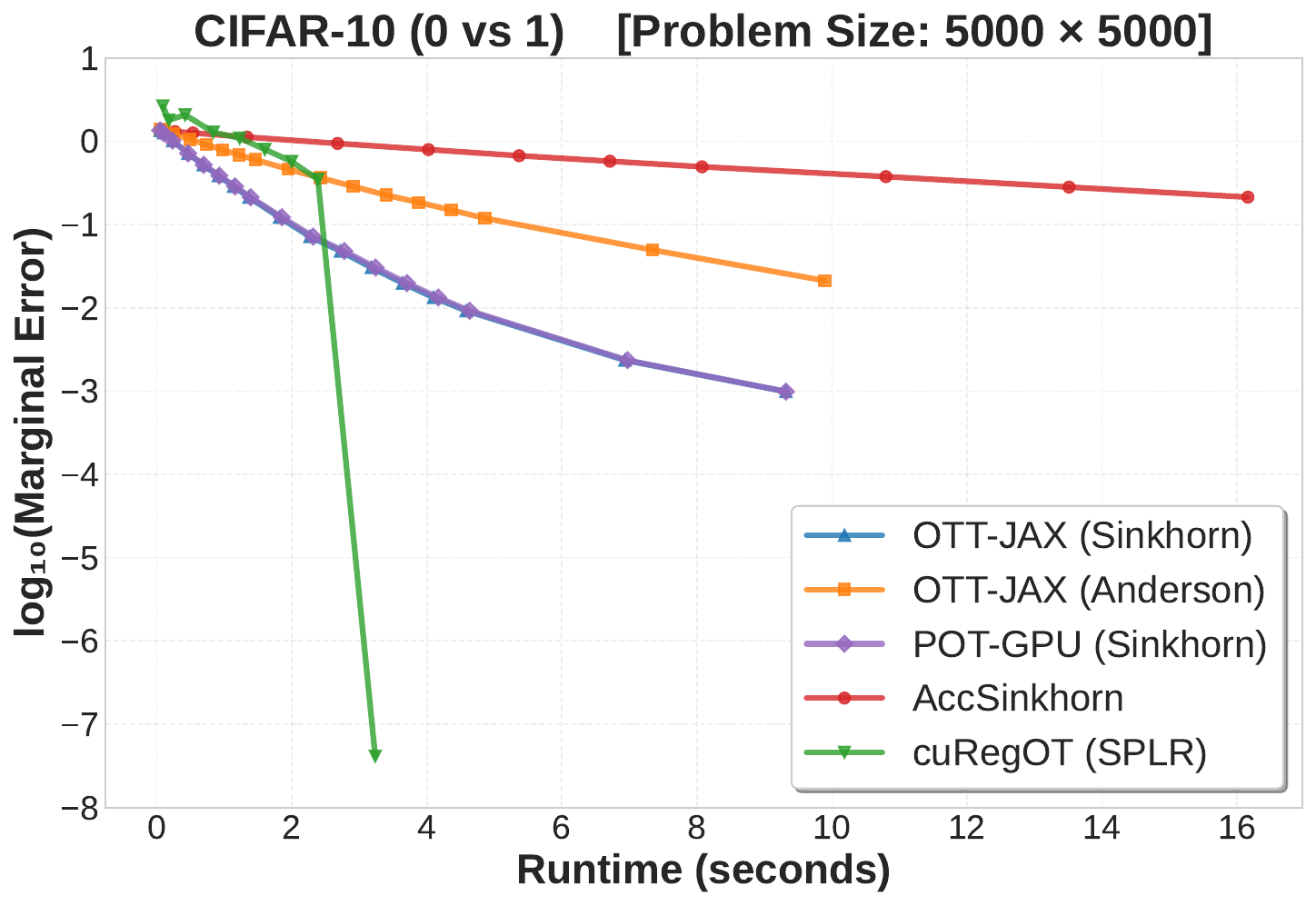}
\includegraphics[width=0.328\textwidth]{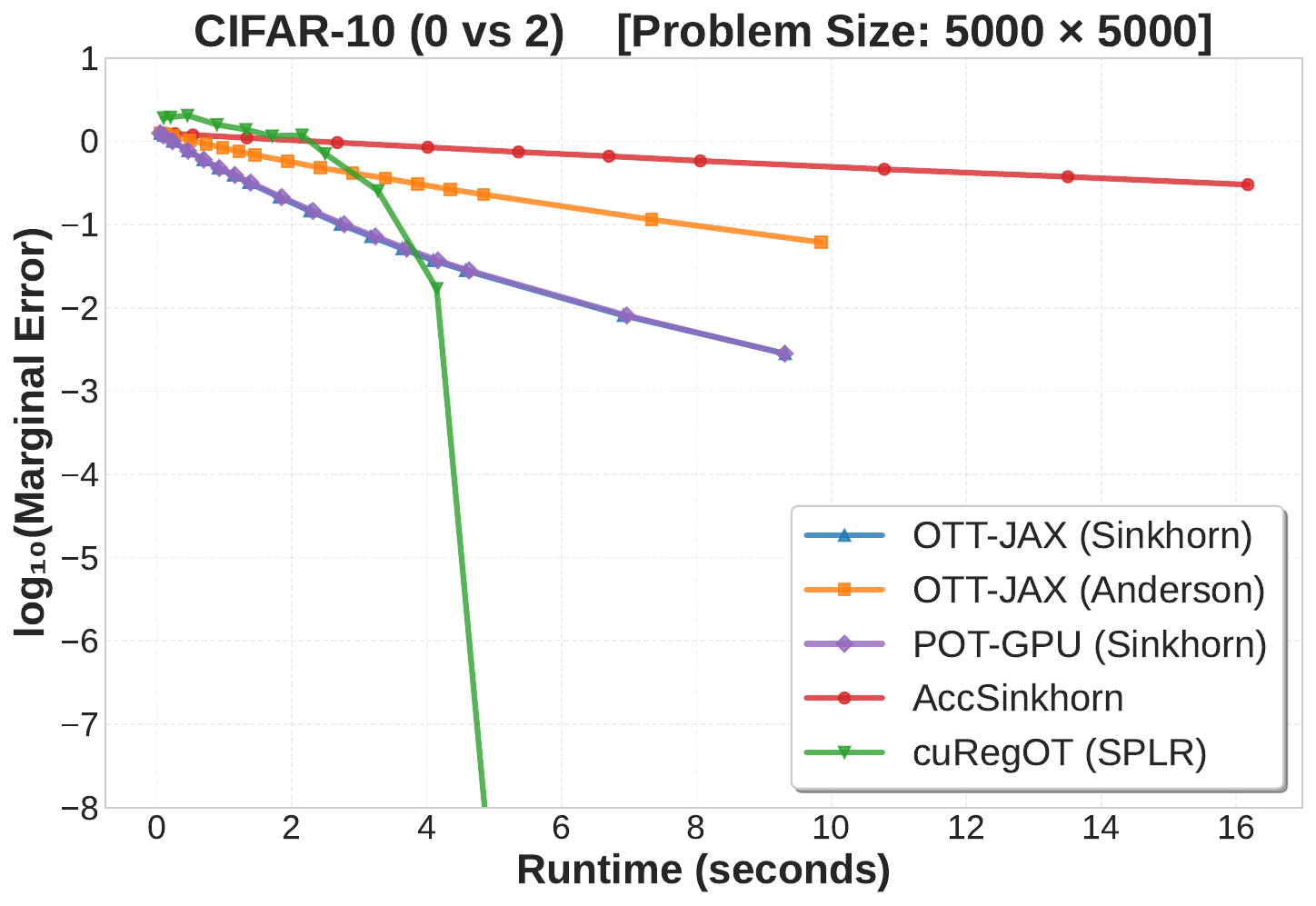}
\includegraphics[width=0.328\textwidth]{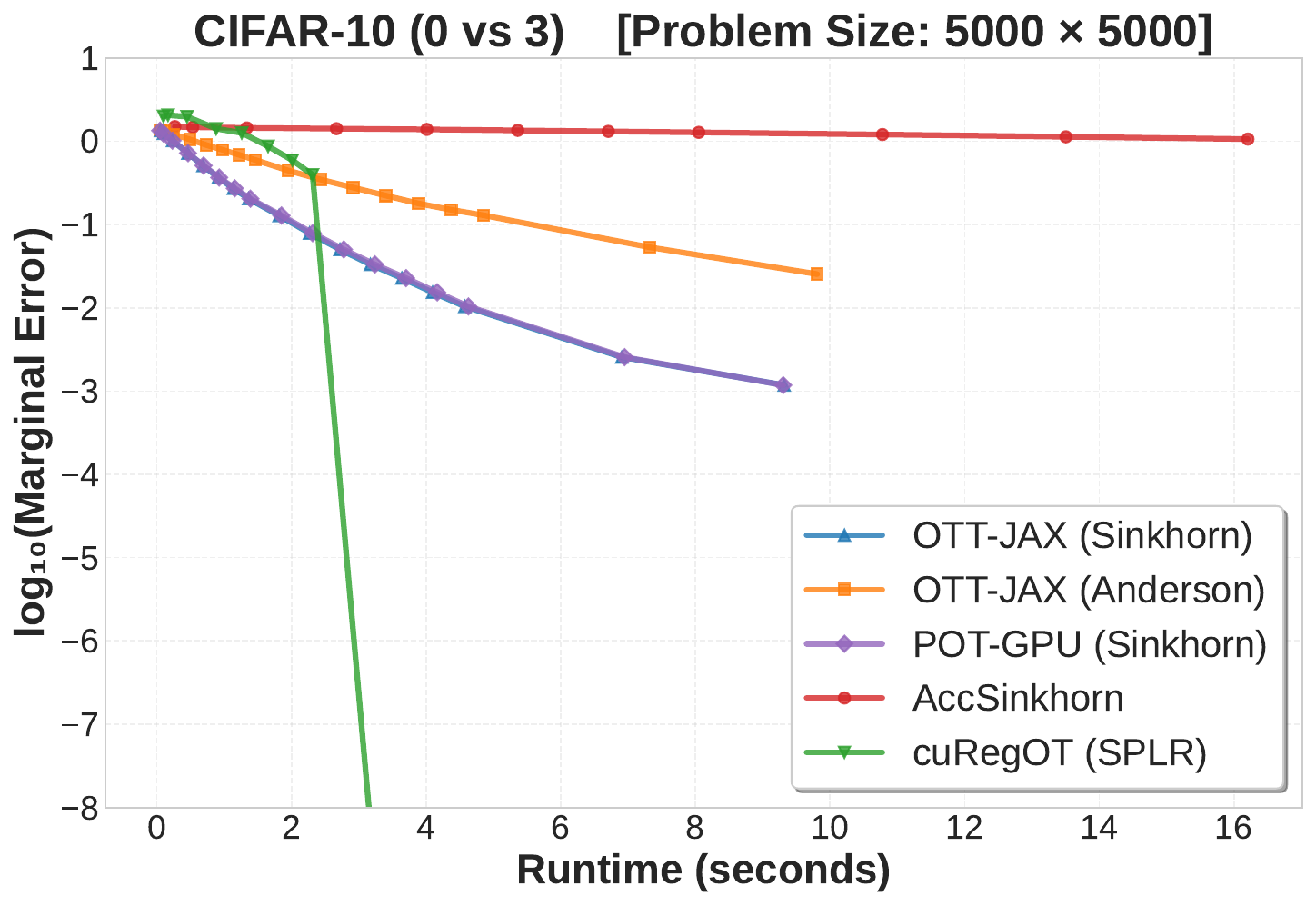}
\par\end{centering}
\caption{\label{fig:eta_0.0001}Benchmark result for different solvers under
the regularization parameter $\eta=0.0001$.}
\end{figure}

For the experiments in the main article, we have fixed the regularization
parameter to $\eta=0.001$. To test solver's performance under different
$\eta$ values, we additionally conduct experiments evaluating $\eta=0.01$
and $\eta=0.0001$, with results given in Figures \ref{fig:eta_0.01}
and \ref{fig:eta_0.0001}, respectively. It can be observed that as
$\eta$ decreases, all methods slow down due to ill-conditioning,
but cuRegOT's advantage becomes more pronounced: the plots show that
while Sinkhorn and its variants deteriorate rapidly at smaller $\eta$
values, cuRegOT maintains a fast convergence speed. This indicates
that the second-order curvature information we utilize is practically
indispensable for overcoming the ill-conditioning caused by small
$\eta$.

\subsection{Impact of the Hyperparameter $S$}

\label{subsec:hyperparameter}

\begin{figure}[h]
\centering
\begin{centering}
\includegraphics[width=0.328\textwidth]{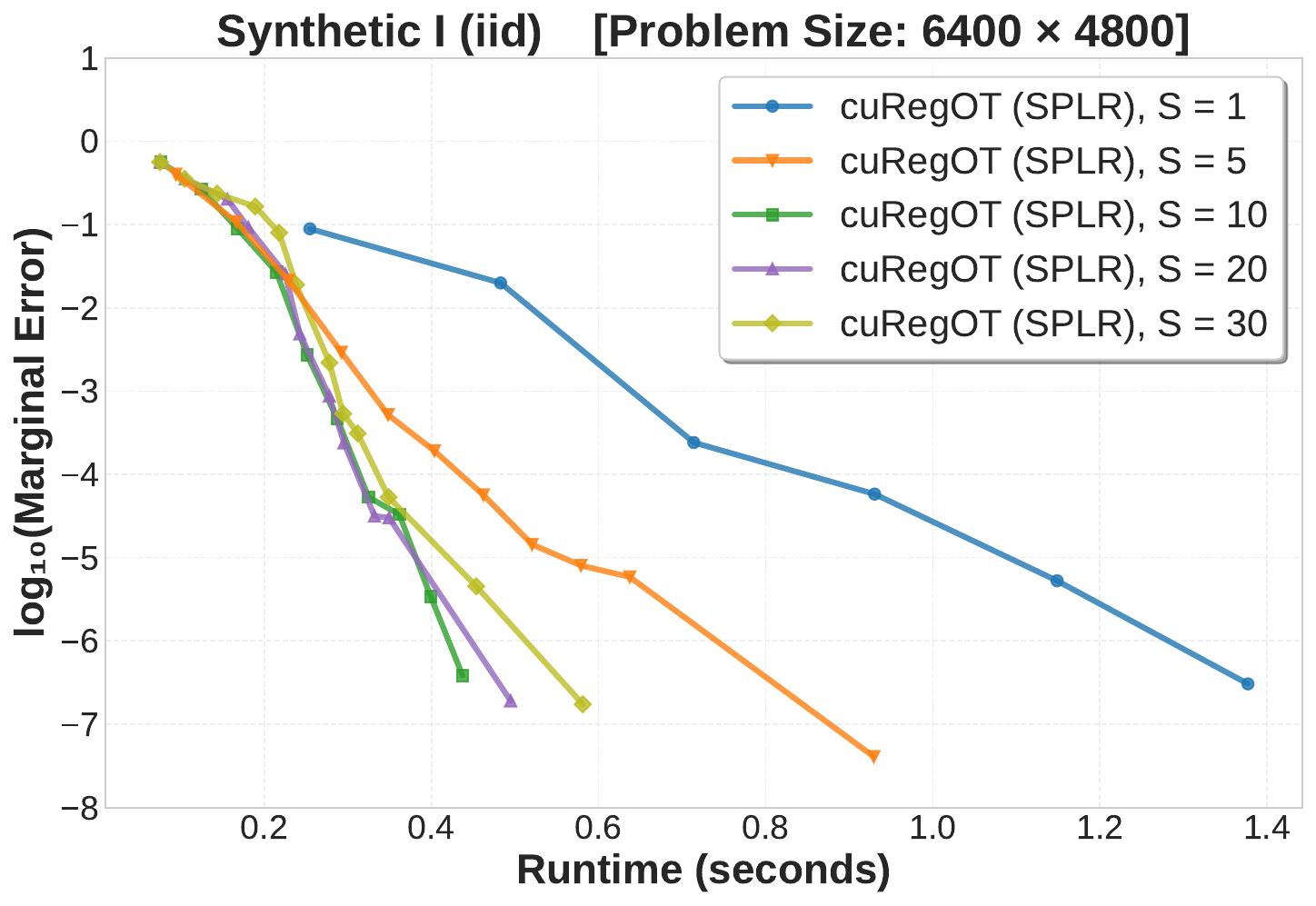}
\includegraphics[width=0.328\textwidth]{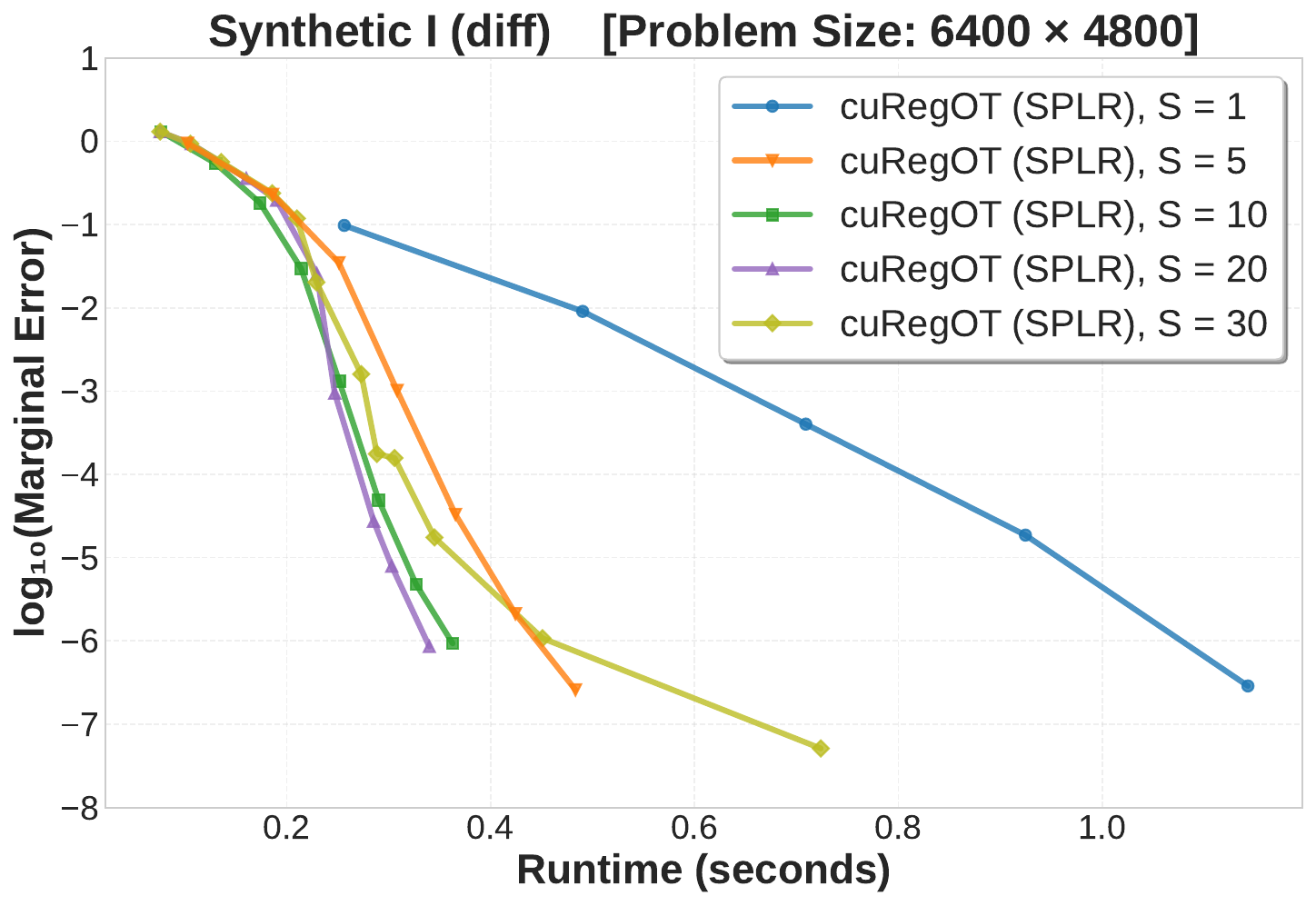}
\includegraphics[width=0.328\textwidth]{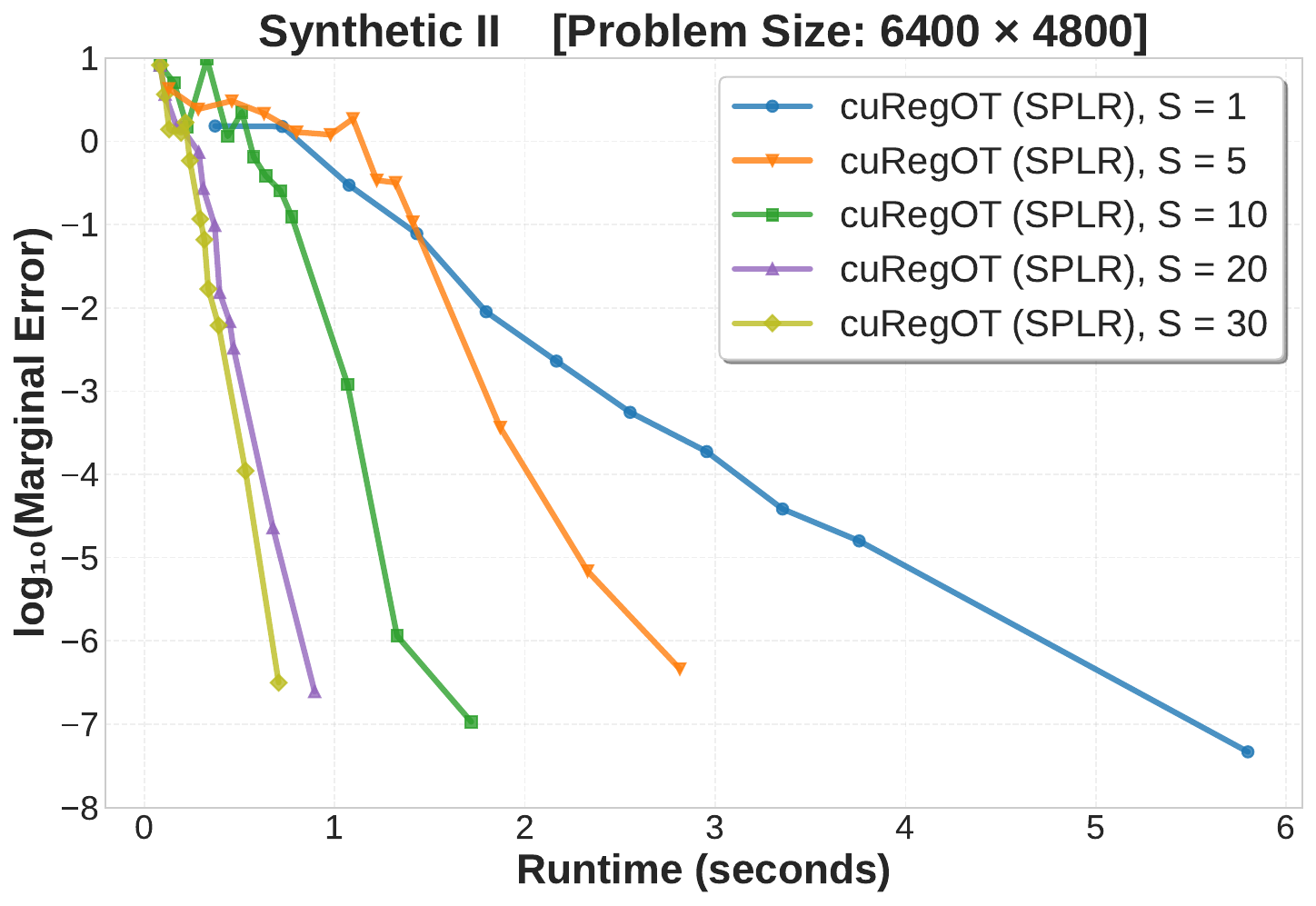}
\par\end{centering}
\begin{centering}
\includegraphics[width=0.328\textwidth]{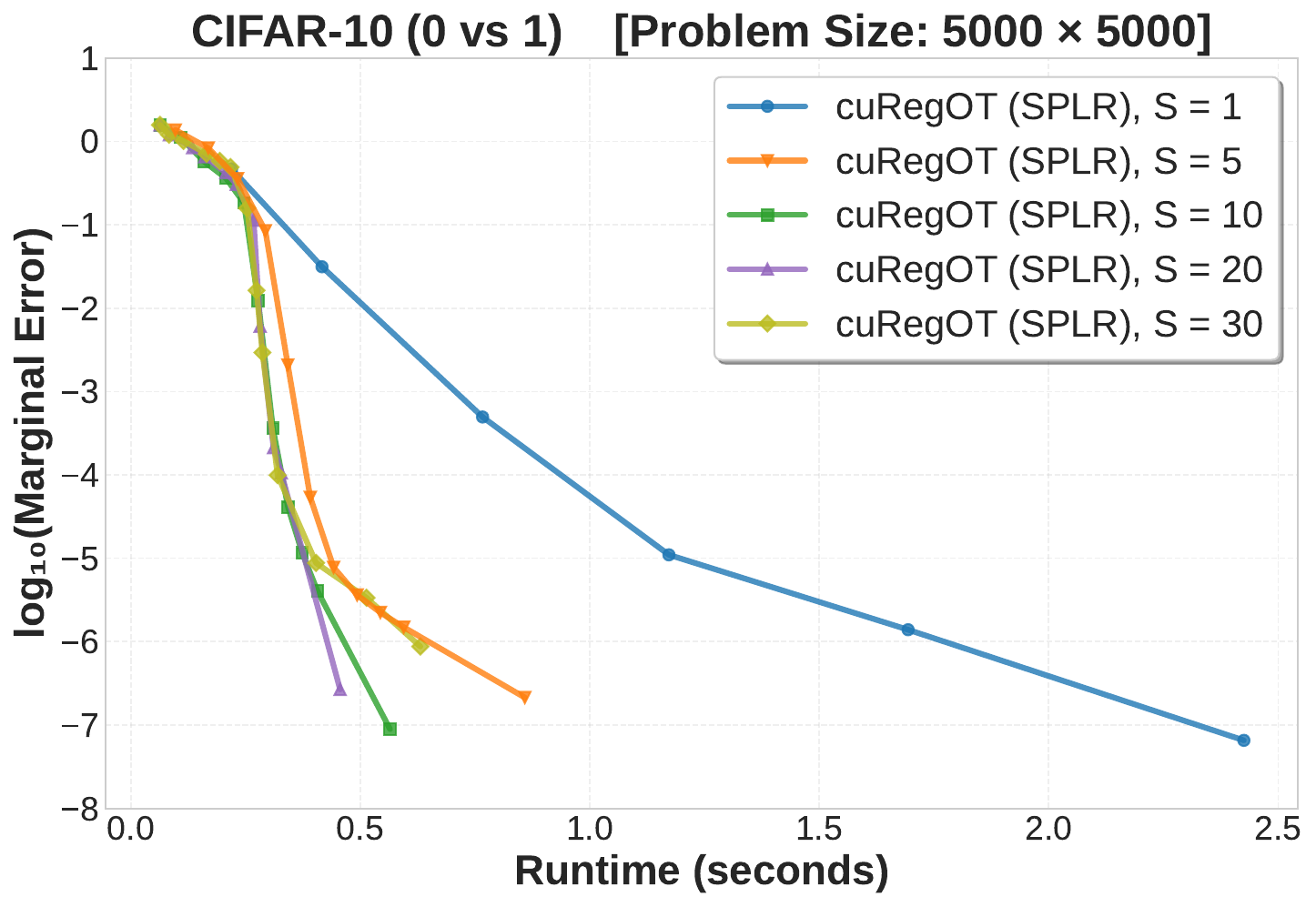}
\includegraphics[width=0.328\textwidth]{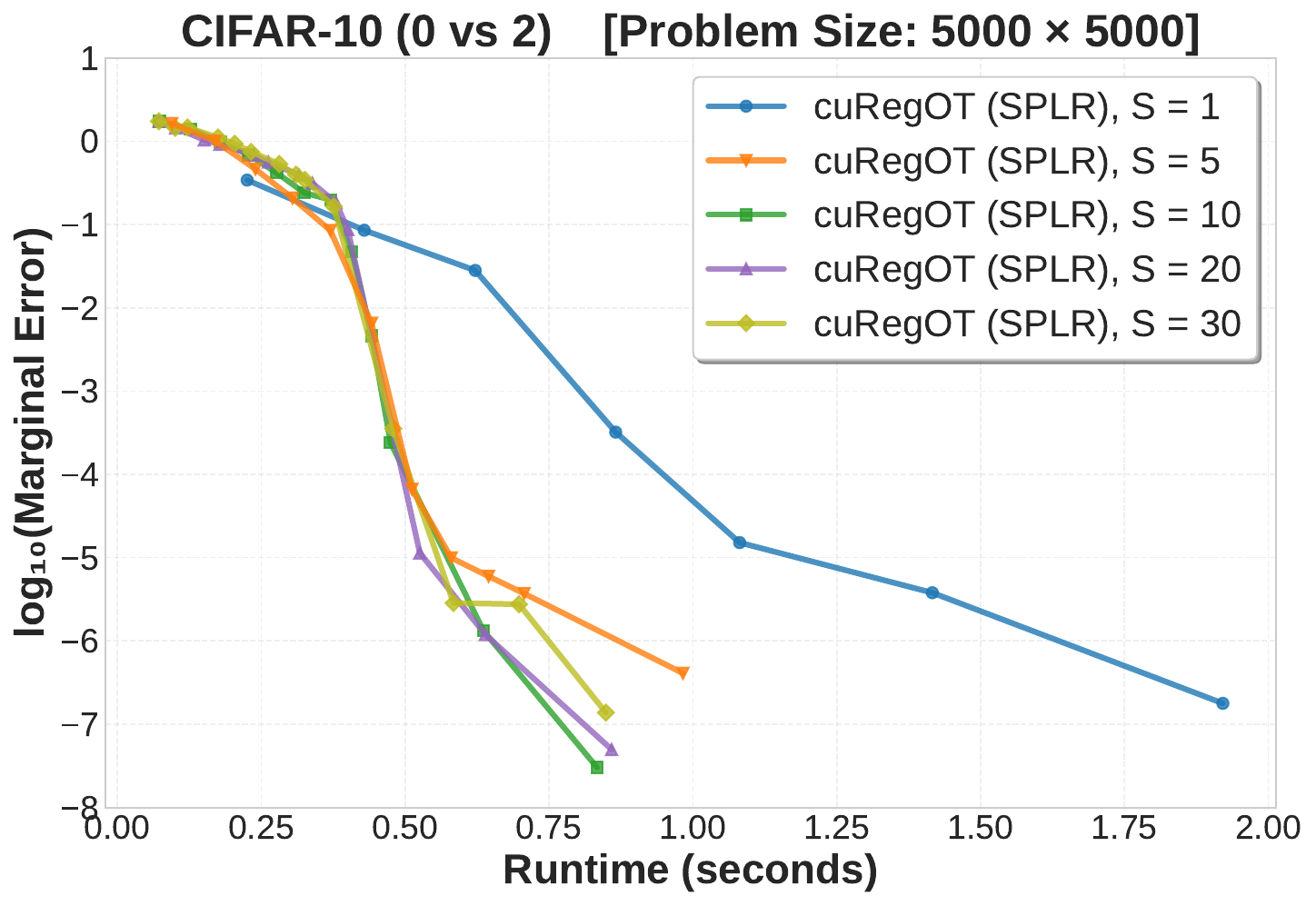}
\includegraphics[width=0.328\textwidth]{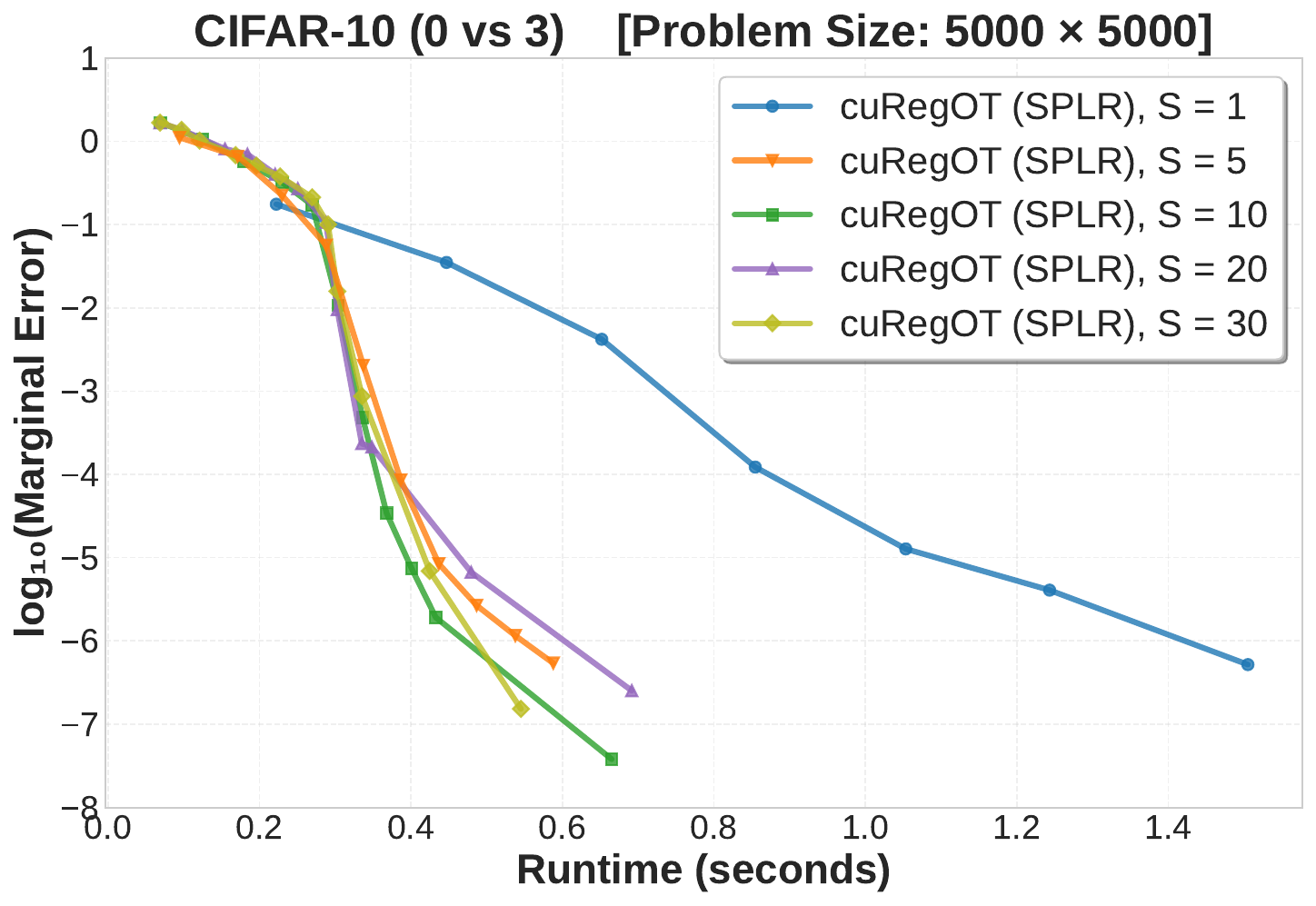}
\par\end{centering}
\caption{\label{fig:hyperparameter}Benchmark result for the hyperparameter
$S$ in the cuRegOT solver. The regularization parameter is set to
$\eta=0.001$.}
\end{figure}

The main hyperparameter in the cuRegOT solver is $S$, the number
of iterations for which we reuse the sparsity pattern (and thus reuse
symbolic analysis). In Figure \ref{fig:hyperparameter} we conduct
an $S$-sensitivity study to examine the impact of $S$ on the solver
performance. We find that the method is relatively robust: $S\in[5,30]$
consistently yields good accuracy-time curves across problems, while
$S=1$ (no reuse) can be slower due to CPU overhead.

\section{Theoretical Analysis Details}

\label{sec:proof}

\subsection{Complexity Analysis for Table \ref{tab:complexity}}

\label{subsec:complexity_details}

Each row of the $D\times32$ thread block forms a warp, and it can
compute the block row sum using 5 parallel additions. Moreover, each
thread block needs to process $nm/K$ tiles of the matrix, so computing
the block-level row sums for the whole $T$ matrix costs $O(5nm/K)$
wall time.

Each warp produces one partial row sum for a 32-column segment and
atomically accumulates it into the corresponding row sum entry in
the global memory. For each row, there are roughly $m/32$ such partial
sums. Although different column blocks are launched in parallel, their
atomic additions target the same row sum address, and we use a conservative
model in which these updates are serialized. Since $g_{r}D$ rows
can be processed simultaneously, the global memory writing time for
row sums is
\[
O\left(\frac{n}{g_{r}D}\cdot\frac{m}{32}\right)=O\left(\frac{g_{c}nm}{K}\right).
\]

Computing the column sums consists of two stages. In the first stage,
each block adds the $T_{ij}$ elements in the same column to the shared
memory, and in the second stage, the block partial column sums are
added to the global memory. In the first stage, each thread sequentially
processes $n/(g_{r}D)$ rows of $T$, leading to $O(n/(g_{r}D))$
register-level additions. Then within the block, adding thread-owned
values to the shared memory costs $O(D)$ time. Next, for each column,
$g_{r}$ block-owned partial column sums are atomically accumulated
into the same column sum entry in the global memory, yielding the
conservative estimate of $O(g_{r})$ time. Since the $K$ threads
can process $32g_{c}$ columns simultaneously, in total, computing
the column sums requires $O(m/(32g_{c})\cdot n/(g_{r}D))=O(mn/K)$
additions, $O(Dm/(32g_{c}))$ shared memory writes, and $O(g_{r}m/(32g_{c}))$
global memory writes.

\subsection{Proof of Theorem \ref{thm:global_convergence}}

Let $x_{0}$ be the initial value of the iterates $\{x_{k}\}$, and
define the level set $\mathsf{L}=\{x:f(x)\le f(x_{0})\}$. Clearly,
$\mathsf{L}$ is a convex and compact set. Since $f(x)$ is twice
differentiable on $\mathsf{L}$ and $H(x)\coloneqq\nabla^{2}f(x)$
is positive definite, we have that the eigenvalues of $H(x)$ must
be bounded on $\mathsf{L}$. That is, there exists constants $0<\mu\le L<\infty$
such that $\mu\le\lambda_{\min}(H(x))\le\lambda_{\max}(H(x))\le L$
for all $x\in\mathsf{L}$, where $\lambda_{\min}(A)$ and $\lambda_{\max}(A)$
denote the smallest and largest eigenvalues of a symmetric matrix
$A$, respectively. This implies that $\nabla f$ is $L$-Lipschitz,
\emph{i.e.},
\begin{equation}
\Vert\nabla f(x)-\nabla f(y)\Vert\le L\Vert x-y\Vert,\quad\forall x,y\in\mathsf{L},\label{eq:smoothness}
\end{equation}
 and that the Polyak-\L ojasiewicz condition holds:
\begin{equation}
\Vert\nabla f(x)\Vert^{2}\ge2\mu[f(x)-f^{*}],\quad\forall x\in\mathsf{L},\label{eq:pl_condition}
\end{equation}
where $f^{*}$ is the optimal value of $f(x)$.

Let $x^{-},x\in\mathsf{L}$ be two arbitrary points in the level
set $\mathsf{L}$, and let $x^{+}$ be the output of Algorithm \ref{alg:splr}
applied to $x$. Clearly, $x^{+}=x-\gamma B^{-1}g$, where $g=\nabla f(x)$
and $B=H_{\Omega}+\xi uu^{T}+\zeta vv^{T}+\tau I$ as defined in (\ref{eq:approx_hessian})
and (\ref{eq:low_rank}). We show next that the eigenvalues of $B$
are also bounded on $\mathsf{L}$, uniformly on the values of $x^{-}$
and $x$.

To start with, Corollary 3.4 of \citet{wang2025sparse} shows that
$\lambda_{\min}(H)\le\lambda_{\min}(H_{\Omega})\le\lambda_{\max}(H_{\Omega})\le\lambda_{\max}(H)$,
$H=H(x)$, so we have $\mu\le\lambda_{\max}(H_{\Omega})\le L$. In
case $(y^{-})^{T}s^{-}\le0$, Algorithm \ref{alg:splr} forces $B=H_{\Omega}+\tau I$,
so we easily obtain
\[
\mu\le\lambda_{\min}(H_{\Omega})\le\lambda_{\min}(B)\le\lambda_{\max}(H_{\Omega})+\tau\le L+\tau_{\max}.
\]
For $(y^{-})^{T}s^{-}>0$, the proof of Theorem 5.1 of \citet{wang2025sparse}
applies here, so by that theorem, we have
\[
m_{B}\coloneqq(2+3L/\mu)^{-1}\mu\le\lambda_{\min}(B)\le\lambda_{\max}(B)\le M_{B}\coloneqq2L+\tau_{\max}.
\]
Clearly, the bounds $(m_{B},M_{B})$ also hold for the $(y^{-})^{T}s^{-}\le0$
case, and they do not depend on the values of $x^{-}$ and $x$.

Next, we prove that there exists a constant $C>0$ such that
\begin{equation}
f(x^{+})\le f(x)-C\Vert g\Vert^{2},\label{eq:objfn_diff_grad}
\end{equation}
and $C$ does not depend on the values of $x^{-}$ and $x$.

To see this, define $\phi(\gamma)=f(x+\gamma d)$, where $d=-B^{-1}g$.
Since $\nabla f$ is $L$-Lipschitz by (\ref{eq:smoothness}), we
have
\[
|\phi'(\gamma)-\phi'(0)|=|[\nabla f(x+\gamma d)-\nabla f(x)]^{T}d|\le\Vert\nabla f(x+\gamma d)-\nabla f(x)\Vert\cdot\Vert d\Vert\le L\gamma\Vert d\Vert^{2}
\]
using the Cauchy--Schwarz inequality. The Wolfe conditions (\ref{eq:wolfe})
guarantees that
\[
\phi'(\gamma)-\phi'(0)=(g^{+}-g)^{T}d\ge(c_{2}-1)g^{T}d=(1-c_{2})|\phi'(0)|,
\]
where $g^{+}=\nabla f(x^{+})$. Then by combining the two inequalities
above, we have
\[
\gamma\ge\frac{(1-c_{2})|\phi'(0)|}{L\Vert d\Vert^{2}}.
\]
Also by the Wolfe conditions, we have
\[
f(x^{+})\le f(x)+c_{1}\gamma g^{T}d=f(x)-c_{1}\gamma|\phi'(0)|,
\]
so we get
\[
f(x^{+})\le f(x)-\frac{c_{1}(1-c_{2})|\phi'(0)|^{2}}{L\Vert d\Vert^{2}}.
\]

Notice that $|\phi'(0)|=-g^{T}d=g^{T}B^{-1}g\ge M^{-1}_{B}\Vert g\Vert^{2}$
and $\Vert d\Vert=\Vert B^{-1}g\Vert\le m^{-1}_{B}\Vert g\Vert$,
so we then obtain
\[
\frac{|\phi'(0)|}{\Vert d\Vert}\ge\frac{M^{-1}_{B}\Vert g\Vert^{2}}{m^{-1}_{B}\Vert g\Vert}=\frac{m_{B}}{M_{B}}\Vert g\Vert,
\]
which immediately gives (\ref{eq:objfn_diff_grad}) with
\[
C=\frac{c_{1}(1-c_{2})m^{2}_{B}}{LM^{2}_{B}}.
\]

Then we are ready to prove the main theorem. For $k=iS$, $i=1,2,\ldots$,
we first note that $f(x^{q}_{k})\le f(x_{k-1})$, since $x^{q}_{k}$
is generated by the line search procedure that guarantees the decrease
on objective function value. Then by the design of the algorithm,
we can show that $f(x_{k})$ is non-increasing on $k$. This implies
that all the iterates $\{x_{k}\}$ stay in the level set $\mathsf{L}$.
Then by (\ref{eq:objfn_diff_grad}), we have that for all $k\ge1$,
\begin{equation}
f(x_{k})\le f(x^{q}_{k})\le f(x_{k-1})-C\Vert\nabla f(x_{k-1})\Vert^{2}.\label{eq:telescope}
\end{equation}
Summing (\ref{eq:telescope}) over $k$, and then we have
\[
\sum^{\infty}_{k=0}\Vert\nabla f(x_{k})\Vert\le\frac{f(x_{0})-f^{*}}{C}<\infty,
\]
which gives the desired conclusion that $\Vert\nabla f(x_{k})\Vert\rightarrow0$
as $k\rightarrow\infty$.

\subsection{Proof of Theorem \ref{thm:linear_convergence}}

\label{subsec:proof_rate}

Using the same definitions for $x^{-}$, $x$, and $x^{+}$ as in
the proof of Theorem \ref{thm:global_convergence}, we will prove
that there exists a constant $0<r<1$ such that
\begin{equation}
f(x^{+})-f^{*}\le r[f(x)-f^{*}],\label{eq:linear_convergence_qn}
\end{equation}
and $r$ does not depend on the values of $x^{-}$ and $x$.

Indeed, in the proof of Theorem \ref{thm:global_convergence} we prove
that
\[
f(x^{+})-f^{*}\le f(x)-f^{*}-C\Vert\nabla f(x)\Vert^{2}.
\]
By the Polyak-\L ojasiewicz condition (\ref{eq:pl_condition}), we
have
\[
\Vert\nabla f(x)\Vert^{2}\ge2\mu[f(x)-f^{*}].
\]
Therefore, it is easy to get
\[
f(x^{+})-f^{*}\le f(x)-f^{*}-2C\mu[f(x)-f^{*}]=(1-2C\mu)[f(x)-f^{*}],
\]
which implies (\ref{eq:linear_convergence_qn}) with $r=1-2C\mu$.

For $k=iS$, $i=1,2\ldots$, we already have
\[
f(x^{q}_{k})-f^{*}\le r[f(x_{k-1})-f^{*}].
\]
Since by construction we have $f(x_{k})\le f(x^{q}_{k})$, we immediately
get
\begin{equation}
f(x_{k})-f^{*}\le r[f(x_{k-1})-f^{*}].\label{eq:linear_convergence_actual}
\end{equation}
The iterates $\{x_{k}\}$ for $k\neq iS$, $i=1,2,\ldots$ are generated
by Algorithm \ref{alg:splr}, so they also satisfy (\ref{eq:linear_convergence_actual}).
Overall, the inequality (\ref{eq:linear_convergence_actual}) holds
for all iterates.
\end{document}